\documentclass[prd,superscriptaddress,nofootinbib,amsmath,amssymb,aps,11pt]{revtex4-2}

\usepackage[normalem]{ulem}
\usepackage{bm}
\usepackage{amsfonts}
\usepackage{latexsym}
\usepackage[latin1]{inputenc}
\usepackage{amsmath}
\usepackage{tabularx}
\usepackage{array}
\usepackage{palatino}
\usepackage{mathpazo}
\usepackage{graphicx} 
\usepackage{appendix}
\usepackage{booktabs}
\usepackage{dcolumn}
\usepackage[dvipsnames]{xcolor}
\usepackage{multirow}
\usepackage{hyperref}

\begin{document}

\title{Images of the Thin Accretion Disk Around Kerr Black Holes coupled to time periodic scalar fields}

\author{Galin N. Gyulchev}
	\email{gyulchev@phys.uni-sofia.bg}
	\affiliation{Department of Theoretical Physics, Faculty of Physics, Sofia University, Sofia 1164, Bulgaria}
  
\author{Daniela D. Doneva}
\email{daniela.doneva@uv.es}
\affiliation{Departamento de Astronom\'ia y Astrof\'isica, Universitat de Val\`encia,
	Dr. Moliner 50, 46100, Burjassot (Val\`encia), Spain}
\affiliation{Theoretical Astrophysics, Eberhard Karls University of T\"ubingen, 72076 T\"ubingen, Germany}

\author{Valentin O. Deliyski}
	\email{valentin.deliyski@phys.uni-sofia.bg}
	\affiliation{Department of Theoretical Physics, Faculty of Physics, Sofia University, Sofia 1164, Bulgaria}

\author{Petya G. Nedkova}
    \email{pnedkova@phys.uni-sofia.bg}
    \affiliation{Department of Theoretical Physics, Faculty of Physics, Sofia University, Sofia 1164, Bulgaria}

\author{Stoytcho S. Yazadjiev}
	\email{yazad@phys.uni-sofia.bg}	
	\affiliation{Department of Theoretical Physics, Faculty of Physics, Sofia University, Sofia 1164, Bulgaria}
	\affiliation{Institute of Mathematics and Informatics, 	Bulgarian Academy of Sciences, 	Acad. G. Bonchev St. 8, Sofia 1113, Bulgaria}

\begin{abstract}
We investigate the orbital structure and observable appearance of rotating Kerr black holes endowed with synchronized scalar hair described by two time-periodic scalar fields with a flat target-space geometry. The presence of scalar hair enriches the geodesic structure of the spacetime relative to the Kerr case and significantly modifies the emission properties of geometrically thin Novikov-Thorne accretion disks. Combining an analysis of timelike circular orbits with backward ray tracing, we show that the normalized scalar charge governs the morphology and luminosity of both prograde and counter-rotating disks. In the strongly scalarized regime, additional light rings and modified circular-orbit regions produce multiple inner emitting zones and strongly enhanced redshift patterns that depart markedly from the Kerr prediction. The most pronounced deviations occur in the counter-rotating sector, where scalar hair generates inner retrograde radiative rings with substantially enhanced luminosity and distinctive frequency-shift signatures. Even when the spacetime approaches the Kerr geometry at weaker scalarization, the retrograde disk remains highly sensitive to the presence of scalar hair. Our results demonstrate that geometrically thin accretion disks can provide robust observational diagnostics of synchronized scalar hair and may offer a promising avenue for testing tensor-multi-scalar gravity with future horizon-scale black-hole imaging observations.
\end{abstract}

\maketitle
\vspace{-0.5mm}

\section{Introduction}

A remarkable feature of rotating Kerr black holes is that, under a synchronization condition, they can support stationary configurations of massive, time-periodic scalar fields. In the test-field approximation this phenomenon was first identified in \cite{hairysol1}. Fully nonlinear and self-consistent solutions describing Kerr black holes with synchronized scalar hair were subsequently constructed in \cite{herdeiro2014, herdeiro2015, herdeiro2016a, brihaye2016, delgado2016}. In these solutions the scalar-field frequency is locked to the angular velocity of the event horizon, ensuring regularity at the horizon and preventing scalar flux into the black hole, thus allowing for stationary hairy configurations.

These solutions can be naturally embedded within tensor-multi-scalar theories of gravity, in which the scalar sector is described by multiple real fields \cite{Damour:1992we,horbatsch2015tensor}. In the present work we focus on the flat target space case, corresponding to vanishing Gaussian curvature, which reduces the theory to the original model of Kerr black holes with synchronized scalar hair \cite{collodel2020rotating}. This restriction allows us to isolate the impact of synchronized scalar hair without additional geometric effects associated with target-space curvature. Such solutions exist across a broad mass range, from stellar-mass to supermassive black holes, and thus provide a useful theoretical framework for exploring potential observational signatures of scalar hair. Recent horizon-scale observations by the Event Horizon Telescope (EHT) collaboration have demonstrated the capability to resolve ring-like emission structures around supermassive black holes such as M87* and Sgr~A*, opening the possibility of testing strong-field deviations from the Kerr geometry through direct imaging \cite{event2019first,event2022first}. We note, however, that the nonlinear dynamical stability of these configurations remains under investigation, and recent analyses indicate that certain branches may be unstable \cite{herdeiro2025}.

The presence of synchronized scalar hair deforms the Kerr geometry in the strong-field region and modifies the geodesic structure of the spacetime. In particular, the configuration of timelike circular orbits and light rings can differ substantially from the Kerr case. These deviations affect strong gravitational lensing and the appearance of black-hole shadows. In idealized isotropic illumination scenarios, the shadows of Kerr black holes with synchronized scalar hair have been studied in detail for flat target spaces in~\cite{cunha2015shadows,cunha2016shadows}, while the role of target space curvature has been addressed separately in~\cite{Gyulchev2024}. 

Astrophysical black holes are not illuminated isotropically, but rather by surrounding accretion flows. In many astrophysically relevant systems, such as active galactic nuclei and X-ray binaries in the thin-disk regime, the dominant source of illumination is a geometrically thin accretion disk. In this case, the observed emission is governed by the properties of circular equatorial geodesics, in particular by the existence and stability of timelike circular orbits and by the associated light-ring structure, as described within the Novikov-Thorne framework in~\cite{Novikov1973,Page1974}. In long-lived accretion systems the disk is expected to align with the black hole spin through the Bardeen--Petterson effect \cite{Bardeen1975,Scheuer1996}. Nevertheless, counter-rotating disks may arise in several astrophysical situations, including chaotic accretion episodes, misaligned gas inflows during galaxy mergers, or tidal disruption events \cite{King2006Chaotic,Nixon2012Retrograde,Stone2019TDE}. Such scenarios motivate the investigation of both co-rotating and counter-rotating disk configurations when exploring observational signatures of modified black hole geometries.

The optical appearance of thin accretion disks around compact objects has a long history, beginning with the seminal work of Luminet \cite{Luminet_1979}, who first constructed images of a thin disk around a Schwarzschild black hole. Subsequent studies extended this analysis to rotating black holes, incorporating relativistic ray tracing, transfer functions, and detailed modeling of disk emission \cite{CunninghamBardeen1972,Cunningham1975,Fanton1997}. More recently, thin-disk imaging has emerged as a powerful tool for probing deviations from the Kerr geometry. It has been applied to black holes with scalar hair \cite{Collodel_2021,ThinDiskHorndeski2024}, compact scalar-field configurations such as boson stars \cite{BosonStarThinDisk2025}, and horizonless spacetimes including traversable wormholes supported by scalar fields \cite{Ishkaeva_2023,WormholeDiskImaging2023}.

For Kerr black holes with synchronized scalar hair, the scalar field assumes a toroidal distribution outside the event horizon, inducing nontrivial modifications of the spacetime geometry in the strong-field region. As a consequence, the structure of timelike circular orbits may depart significantly from that of the Kerr spacetime, and additional light rings, including stable ones, can emerge \cite{Collodel_2021}. These features have a direct impact on the morphology and radiative properties of geometrically thin accretion disks, potentially leading to inner emitting rings, fragmented disk configurations, or enhanced relativistic effects that have no counterpart in the Kerr case. In this work we investigate how these modifications of the strong-field orbital structure manifest in the observable appearance of geometrically thin accretion disks.

The aim of the present paper is to investigate the observable properties of geometrically thin Novikov-Thorne accretion disks around Kerr black holes with synchronized scalar hair. We first analyze the existence and stability of timelike circular orbits for both co-rotating and counter-rotating motion and study their relation to the light-ring structure. We then compute the apparent images and observed fluxes by backward ray tracing in~\cite{Luminet_1979,lora2022osiris}. Particular emphasis is placed on the role of the normalized scalar charge and the horizon radius in shaping the emission morphology and frequency-shift patterns.

The paper is organized as follows. In Sec. II we summarize the theoretical framework of Kerr black holes with synchronized scalar hair. In Sec. III we present the analysis of timelike circular orbits and light rings. The numerical ray-tracing setup and the redshift formalism are described in Secs. IV and V. Our main results for prograde and retrograde thin accretion disks are discussed in Sec. VI, followed by our conclusions in Sec. VII.

\section{Kerr black holes with synchronized scalar hair }

We consider Einstein gravity  minimally coupled to two dynamic scalar fields, $\varphi=(\varphi^{1},\varphi^{2})$. The scalar fields 
can be considered as generalized coordinates on an abstract 2-dimensional Riemmanian space $({\cal E}_2, \gamma_{ab}(\varphi))$, the so-called target space, with a positive definite metric $\gamma_{ab}(\varphi)$. The action for this theory is given by
\begin{equation}
    \label{TMST action}
    S = \frac{1}{4 \pi G} \int \sqrt{-g}\left(\frac{R}{4} - \frac{1}{2}g^{\mu \nu}\gamma_{ab}(\varphi)\partial_\mu \varphi^a \partial_\nu \varphi^b - V(\varphi) \right)\,d^4x ,
\end{equation}
where $V(\varphi)$ denotes the potential associated with the scalar fields.  By varying this action with respect to the spacetime metric and the scalar fields, we obtain the following field equations:
\begin{eqnarray}
  \label{lucas fe metric}
    R_{\mu \nu} &=& 2 \gamma_{ab}(\varphi)\partial_\mu \varphi^a \partial_\nu \varphi^b + 2V(\varphi)g_{\mu \nu}, \\
  \label{lucas fe field}
    \square \varphi^a &=& - \gamma^{a}_{bc}(\varphi)\nabla_{\mu}\varphi^b\nabla^{\mu}\varphi^c    +  \gamma^{ab}(\varphi)\frac{\partial V(\varphi)}{\partial \varphi^b}.
\end{eqnarray}
In this context, $\square\equiv g^{\mu\nu}\nabla_{\mu}\nabla_{\nu}$ represents the d'Alembert operator linked to the spacetime metric, and $\gamma^{a}_{bc}(\varphi)$  are the 
Christofell symbols for the target space metric  $\gamma_{ab}(\varphi)$. 

We shall restrict ourselves to maximally symmetric target spaces with a metric given in isotropic coordinates by  
\begin{equation}
    \gamma_{ab}(\varphi)= \Omega^2(\varphi) \delta_{ab} \,,
\end{equation}
where $\delta_{ab}$ is the Kronecker delta, and the conformal factor is 
\begin{equation}
\Omega^2(\varphi)= \frac{1}{\left(1+ \frac{\kappa}{4}\psi^2\right)^2}     
\end{equation}
with $\psi^2 = \delta \varphi^a\varphi^b$ and $\kappa$ being the Gaussian curvature of the target space. 

For the scalar field potential, we assume a standard massive form
\begin{equation}
    V(\psi) = \frac{1}{2}\mu^2 \psi^2,
\end{equation}
where  $\mu$ is the mass of the scalar fields. In the spacial case of zero Gaussian curvature ($\kappa=0$) the model reduces to the original one of~\cite{herdeiro2014} with one complex scalar field $\Psi=\varphi^1 + i \varphi^2$. 

In order to study rotating black holes, we employ the following ansatz for a stationary and axisymmetric spacetime line element:
\begin{equation} 
\label{lucas line element}
    ds^2 = -\mathcal{N}e^{2F_0}dt^2 +e^{2F_1}\left(\frac{dr^2}{\mathcal{N}} + r^2 d\theta^2\right) +e^{2F_2}r^2 \sin^2\theta \left(d \phi - \frac{\omega}{r}dt \right)^2,
\end{equation}
where $\mathcal{N} = 1 - \frac{r_\mathrm{H}}{r}$, with $r_\mathrm{H}$ denoting the location of the event horizon in these coordinates. The functions $F_0, F_1, F_2$, and $\omega$ depend only on the variables $r$ and $\theta$. We note that the radial coordinate $r$ is related to the Boyer-Lindquist radial coordinate $r_{\mathrm{BL}}$ by $r=r_{\mathrm{BL}}-a^{2}/r_{\mathrm{H,\,BL}}$ in the Kerr limit, where $a=J/M$ stands for the spin of the black hole and $r_{H,\,BL}$ is the location of the horizon in Boyer-Lindquist coordinates.

To avoid the no-scalar hair theorems, it is essential that the scalar fields are time-dependent. We employ the following ansatz:
\begin{equation}
\label{lucas scalar field ansatz}
    \varphi^1 = \psi(r, \theta)\cos(\omega_s t + m \phi), \quad \varphi^2 = \psi(r, \theta)\sin(\omega_s t +m \phi),
\end{equation}
which is consistent with the circular symmetry of the metric in equation (\ref{lucas line element}) and ensures that the field equations in (\ref{lucas fe metric}) and (\ref{lucas fe field}) remain stationary. Here, $\omega_s$ is a real parameter and $m$ is an integer. Further details can be found in \cite{collodel2020rotating}.

Assuming the above ansatze for the scalar field fields and metric, plus the appropriate boundary conditions that are regularity at the event horizon and the axes, and asymptotic flatness at infinity, stationary and axisymmetric black hole solutions with synchronized hair  were numerically constructed for our theory in \cite{collodel2020rotating}. It is  worth mentioning that the regularity at the horizon leads to the synchronization condition $\omega_s = - m \Omega_{H}$
which also ensures that there is no scalar flux into the black hole horizon. 

The black hole solutions are characterized by three global conserved charges, namely the ADM mass $M$, the total angular momentum $J$, and the Noether charge $Q$. It is convenient to introduce the normalized charge $q = \frac{mQ}{J}$ as a measure of hairiness. Since the angular momentum of the scalar field is quantized according to $J_{\psi} = mQ$, one may equivalently write $q = \frac{J_{\psi}}{J}$, which directly measures the fraction of the total angular momentum stored in the scalar field. In the limit $q \to 0$, the solutions reduce to the linear scalar cloud configurations in the test-field approximation, which do not backreact on the spacetime geometry~\cite{hairysol1}. These clouds arise at the onset of the superradiant instability and constitute the linear configurations from which the fully nonlinear Kerr black holes with synchronized scalar hair bifurcate~\cite{herdeiro2014}. In contrast, the limit $q \to 1$ corresponds to the horizonless boson star branch, where the scalar field carries essentially the entire angular momentum of the system~\cite{doneva1, doneva2}. 

\begin{figure}[!t]
    \centering
    \includegraphics[width=0.6\textwidth]{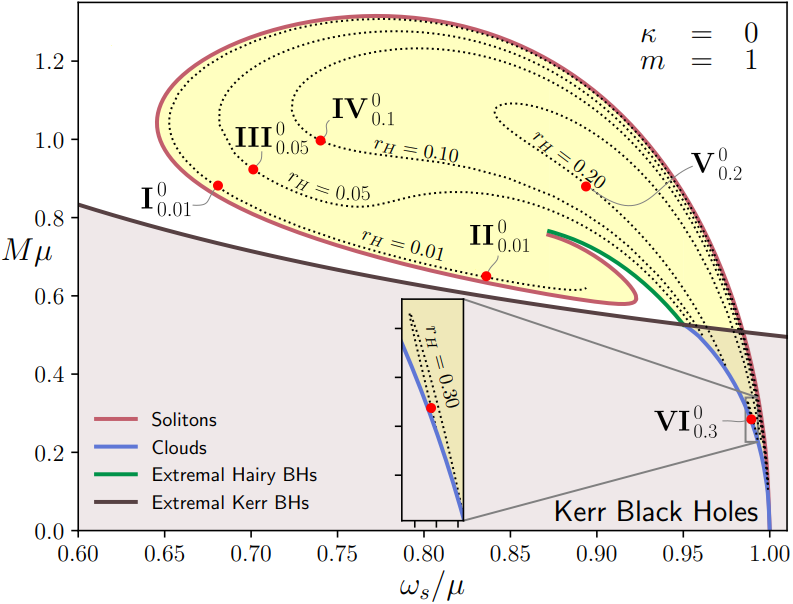}    
    \caption{\small In the $M-\omega_s$ plane, we present solution curves for fixed $r_H$ from \cite{collodel2020rotating} with $\kappa = 0$. Kerr black holes exists below the thick black line which represents extremal solutions, while hairy black holes exist in the yellow region. Red points indicate solutions with varying values of the normalized charge $q = \frac{mQ}{J}$. Configurations used for calculation of accretion disk images are labeled as $\textbf{X}^{\,0}_{\,v}$, where $0$ corresponds to $\kappa$ and $v$ represents the value of $r_\mathrm{H}$. Further details of these solutions can be found in Table \ref{tab_7} of the Appendix.}\label{fig:M-Omega Space}
\end{figure}

In the present paper we shall restrict ourselves to the case $\kappa=0$. The cases with nontrivial Gaussian curvature $\kappa\ne 0$ will be considered in \cite{GCandHairyKerr}. The existence $ M - \omega_s $ diagram for hairy black holes with $\kappa=0$ and $ m=1 $ is shown in Fig. \ref{fig:M-Omega Space} (see \cite{herdeiro2014} and \cite{collodel2020rotating}). The extremal Kerr limit with $ a=M$, depicted by the thick black line in the figure, indicates that Kerr black holes exist only below this line in the grey region. The yellow region represents hairy Kerr black hole solutions, bounded by the red line (solitonic limit with $ q=1 $ and $ r_\mathrm{H}=0 $), the green line (extremal hairy black holes with $ q \ne 0 $), and blue lines for cloud solutions ($ q=0 $).

Dotted lines represent sequences of solutions with constant horizon radii, while red dots indicate specific black hole solutions used in the construction of accretion disks in the sections below. A broader region of existence in $ \omega_s $ is observed for black holes with smaller $ r_\mathrm{H} $ values, which approaches the solitonic and extremal limits as the horizon radius tends to zero. For each fixed $ r_\mathrm{H} $, the sequence begins at the Minkowski limit ($ \omega_s/\mu=1 $) and ends at the cloud line.

\section{Calculating accretion disk images}

\subsection{Existence of Timelike Circular Orbits on the Equatorial Plane}

The study of dynamically stable timelike circular orbits (TCOs) near black holes is crucial to revealing their observational characteristics. The innermost stable circular orbit (ISCO) represents the minimum radius at which a stable orbit can exist and is often interpreted as the boundary of the inner edge of the accretion disk. As charged particles move in orbit around the black hole, they emit synchrotron radiation, which is linked to the frequency of the TC geodesics. This allows for the determination of the physical properties of the black hole through observations of the accretion disk.

We start by considering an arbitrary stationary and axisymmetric line element in the form
\begin{equation} 
\label{line element2}
    ds^2 = g_{tt}dt^2 + g_{rr}dr^2 + g_{\theta\theta}d\theta^2 + g_{\phi\phi}d \phi^{2} + 2g_{t\phi}dt d\phi.
\end{equation}

Due to the stationarity and axisymmetry of the spacetime, both $p_t = -E$ and $p_\phi = L$ are conserved quantities of the geodesic motion for a particle with four-velocity $u^\mu$. Therefore, we have
\begin{equation}\label{four_velocity}
    u^{t}=\frac{g_{t\phi}L+g_{\phi\phi}E}{g_{t\phi}^{2}-g_{tt}g_{\phi\phi}}, \quad u^{\phi}=-\frac{g_{tt}L+g_{t\phi}E}{g_{t\phi}^{2}-g_{tt}g_{\phi\phi}}.
\end{equation}
Taking into account that, for the particle's motion, the four-velocity norm is $u^\mu u_\mu = -\epsilon$, with $\epsilon = \{0,1\}$ for a massless and massive particle, respectively, one can derive the radial equation of motion, which in the equatorial plane ($\theta = \pi/2$) reduces to
\begin{equation}\label{Veff}
  \dot{r}^2=V_{\rm eff}=\frac{1}{g_{rr}}\bigg(-\epsilon+\frac{E^{2}g_{\phi\phi}+2ELg_{t\phi}+L^{2}g_{tt}}{g_{t\phi}^2-g_{tt}g_{\phi\phi}}\bigg),
\end{equation}
which also defines the effective potential $V_{\rm eff}$. The existence of a circular orbit requires $\dot r = 0$ and $\ddot r = 0$, which at constant orbital radius reduce to $V_{\rm eff}=0$ and $\partial_r V_{\rm eff}=0$. These requirements simplify to the following algebraic relations:
\begin{eqnarray}
  && E^{2}g_{\phi\phi}+2EL\,g_{t\phi}+L^{2}g_{tt}-\epsilon\,(g_{t\phi}^2-g_{tt}g_{\phi\phi})=0, \label{LR_Condition_Veff_1} \\ 
  && E^{2}\partial_{r}g_{\phi\phi}+2EL\,\partial_{r}g_{t\phi}+L^{2}\,\partial_{r}g_{tt}-\epsilon\,\partial_{r}(g_{t\phi}^2-g_{tt}g_{\phi\phi})=0. \label{LR_Condition_Veff_2}
\end{eqnarray}

For a massive particle, the marginally stable circular orbit (MSCO) is defined as the stable orbit of smallest radius that remains continuously connected to spatial infinity through a sequence of stable timelike orbits. It corresponds to an inflection point of $V_{\rm eff}$, characterized by $\partial_r^2 V_{\rm eff}=0$ and $\partial_r^3 V_{\rm eff}<0$. At $r=r_{\rm MSCO}$, the marginal stability requirement $\partial_r^2 V_{\rm eff}=0$ leads, in addition to Eqs.~(\ref{LR_Condition_Veff_1}) and (\ref{LR_Condition_Veff_2}), to the algebraic relation
\begin{equation}
E^{2}\,\partial^{2}_{r} g_{\phi\phi}
+ 2 E L\,\partial^{2}_{r} g_{t\phi}
+ L^{2}\,\partial^{2}_{r} g_{tt}
- \epsilon\,\partial^{2}_{r}\!\left(g_{t\phi}^{2} - g_{tt} g_{\phi\phi}\right)
= 0 .
\end{equation}

In Kerr-like spacetimes this equation admits a unique solution, and the MSCO coincides with the innermost stable circular orbit (ISCO). In more general spacetimes, however, the structure of timelike circular motion may be richer, allowing for disconnected regions of stability, in which case the ISCO does not necessarily coincide with the MSCO~\cite{Delgado2022}.

For a particle on a circular orbit of radius $r$ with angular frequency $\Omega := u^\phi/u^t$, the conserved energy and angular momentum associated with the spacetime stationarity and axial symmetry can be written in terms of $\Omega$ using the normalization of the four-velocity, and take the form
\begin{equation}\label{Enegry}
    E=-\frac{g_{tt}+g_{t\phi}\Omega}{\sqrt{-g_{tt}-2g_{t\phi}\Omega-g_{\phi\phi}\Omega^{2}}},
\end{equation}
\begin{equation}\label{Momentum}
    L=\frac{g_{t\phi}+g_{\phi\phi}\Omega}{\sqrt{-g_{tt}-2g_{t\phi}\Omega-g_{\phi\phi}\Omega^{2}}},
\end{equation}
In addition the orbital angular frequency of particles reads
\begin{equation}\label{Omega_pm}
    \Omega_{\pm}=\frac{-\partial_{r}g_{t\phi}\pm\sqrt{\partial_{r}g_{t\phi}^2-\partial_{r}g_{tt}\partial_{r}g_{\phi\phi}}}{\partial_{r}g_{\phi\phi}},
\end{equation}
where the plus sign corresponds to the angular frequency of co-rotating particles, while the minus sign corresponds to the angular frequency of counter-rotating particles.

\subsection{Stability of TCOs}

The existence of stable TCOs at a given coordinate radius $r_{\mathrm{TCO}}$ is subject to a set of fundamental physical requirements. First, the motion must be physically admissible, which implies that the conserved quantities, specifically the energy $E$ and the angular momentum $L$ associated with the geodesic, must be real-valued. This requirement is fulfilled if and only if the following inequality holds: 
\begin{equation}
\rho_{\pm} \equiv -g_{tt} - 2g_{t\varphi} \Omega_\pm - g_{\varphi\varphi} \Omega_\pm^2 > 0.
\end{equation}
This constraint ensures that the normalization of the four-velocity of the particle remains time-like, and thus excludes nonphysical space-like orbits.

In addition to the reality of energy and angular momentum, the angular velocities $\Omega_{\pm}$, as observed from spatial infinity, must also be real. This translates into the requirement that the expression under the square root in Eq. (\ref{Omega_pm}) be nonnegative:
\begin{equation}\label{Discriminant}
D=\partial^{2}_{r}g_{t\phi}-\partial_{r}g_{tt}\partial_{r}g_{\phi\phi}\geq 0.
\end{equation}

Lastly, an orbit is dynamically \textit{stable} when the effective potential given by Eq. (\ref{Veff}) exhibits a stationary point at $r = r_{\mathrm{TCO}}$, at which the second radial derivative is \textit{negative}, namely:
\begin{equation}\label{V2eff}
\partial^{2}_{r}V_{\rm eff} < 0.
\end{equation}

Together, these three conditions define the radial domain in which stable circular motion of massive particles is allowed. In ultracompact geometries admitting stable and unstable light rings, the domain of stable circular motion may be disconnected, and the ISCO can occur at the threshold $D=0$, while the stability condition (\ref{V2eff}) remains satisfied.

\subsection{Existence and stability of Light Rings}

Light rings (LRs) are defined as circular null geodesics confined to the equatorial plane of symmetry, $\theta = \pi/2$.  
For lightlike particles, $\epsilon = 0$, these orbits satisfy the vanishing of the effective potential and its radial derivative:
\begin{equation}
    V_{\rm eff}(r_{\mathrm{LR}}) = 0, 
    \qquad 
    \partial_r V_{\rm eff}(r_{\mathrm{LR}}) = 0.
    \label{LR_Condition_Veff}
\end{equation}

Introducing the impact parameter function $\eta \equiv L/E$, the first condition~\eqref{LR_Condition_Veff} becomes an algebraic equation in $\eta$:
\begin{equation}
    g_{\phi\phi} + 2\,g_{t\phi}\,\eta + g_{tt}\,\eta^2 = 0,
    \label{QuadEq_eta}
\end{equation}
whose solution yields two branches,
\begin{equation}
    \eta_{\pm} = \frac{-g_{t\phi} \pm \sqrt{g_{t\phi}^{2} - g_{tt}\,g_{\phi\phi}}}{g_{tt}},
    \label{eta_branches}
\end{equation}
corresponding to photon orbits that are co-rotating and counter-rotating, respectively.

Evaluating Eq.~\eqref{QuadEq_eta} at the critical radius \(r = r_{\mathrm{LR}}\) and combining it with the second requirement in~\eqref{LR_Condition_Veff} leads to the light-ring condition:
\begin{equation}
    \partial_r g_{\phi\phi} + 2\,\partial_r g_{t\phi} \,\eta_{\pm}
    + \partial_r g_{tt} \, \eta_{\pm}^{2} = 0.
    \label{LR_eq_eta}
\end{equation}

The coordinate angular velocity of the photon along the ring is related to the impact parameter via
\begin{equation}
    \Omega_{\pm} \equiv \frac{d\phi}{dt} = \frac{1}{\eta_{\pm}}.
    \label{Omega_Eta}
\end{equation}
Substituting this into Eq.~\eqref{LR_eq_eta} yields an equivalent expression for the light-ring condition:
\begin{equation}
    \partial_r g_{tt} + 2\,\partial_r g_{t\phi} \,\Omega_{\pm}
    + \partial_r g_{\phi\phi}\,\Omega_{\pm}^{2} = 0,
    \label{LR_eq_Omega}
\end{equation}
where $\Omega_{+} > 0$ corresponds to co-rotating photon orbits and $\Omega_{-} < 0$ to counter-rotating ones.

Moreover, since Eq.~\eqref{Omega_Eta} holds, the null condition also implies the identity
\begin{equation}
    \rho_{\pm} \equiv -g_{tt} - 2\,g_{t\phi}\,\Omega_{\pm} - g_{\phi\phi}\,\Omega_{\pm}^{2} = 0,
    \label{NullIdentity}
\end{equation}
which is satisfied identically on the photon ring.

The stability of a photon ring is determined by the sign of the second radial derivative of the effective potential:
\begin{equation}
    \partial_{r}^{2} V_{\rm eff}  \propto \partial_{r}^{2} g_{\phi\phi} + 2\,\partial_{r}^{2} g_{t\phi} \,\eta_{\pm} + \partial_{r}^{2} g_{tt} \, \eta_{\pm}^{2}.
    \label{LR_stability}
\end{equation}
The orbit is \textit{unstable }if the right-hand side of Eq.~\eqref{LR_stability} is \textit{positive}, and \textit{stable} otherwise.

\subsection{TCOs structure for the Kerr black hole with synchronized scalar hair}

The black hole solutions with synchronized scalar fields \textbf{I}$^{\,0}_{\,0.01}$, \textbf{II}$^{\,0}_{\,0.01}$, \textbf{III}$^{\,0}_{\,0.05}$, \textbf{IV}$^{\,0}_{\,0.1}$, \textbf{V}$^{\,0}_{\,0.2}$, and \textbf{VI}$^{\,0}_{\,0.3}$, as illustrated in Fig. \ref{fig:M-Omega Space}, exhibit a non-trivial discrete structure of TCOs in all cases considered. The spatial distribution of these orbits above the event horizon depends strongly on the value of the normalized charge $q$ and differs significantly between prograde and retrograde motion with respect to the black hole's rotation. Although the Keplerian angular velocity admits two mathematical solution branches, $\Omega_{+}$ and $\Omega_{-}$, the distinction between prograde and retrograde motion is defined by the sign of the angular velocity as measured by an asymptotic observer. Consequently, any spacetime region in which $\Omega>0$, including portions of the $\Omega_{-}$ branch, corresponds to locally prograde motion and contributes to the prograde disk. Among all configurations, \textbf{I}$^{\,0}_{\,0.01}$ stands out by presenting four distinct spatial regions with stable prograde TCOs, while allowing only a single region of stability for retrograde motion.

In contrast, configurations \textbf{II}$^{\,0}_{\,0.01}$, \textbf{III}$^{\,0}_{\,0.05}$, and \textbf{IV}$^{\,0}_{\,0.1}$ share similar characteristics in terms of the distribution and stability of TCOs. In these cases, a noticeable simplification occurs, with the prograde structure reducing to a single continuous stability domain starting at $r_{+}^{\,\text{ISCO}_{1}}$ and extending outward to infinity. For retrograde motion, the stability zones are typically larger in radial extent, but the number of distinct stable regions is smaller, and in most of these configurations they appear as a single continuous domain beyond $r_{-}^{\,\text{ISCO}_{2}}$.

The first four configurations (\textbf{I}$^{\,0}_{\,0.01}$, \textbf{II}$^{\,0}_{\,0.01}$, \textbf{III}$^{\,0}_{\,0.05}$, and \textbf{IV}$^{\,0}_{\,0.1}$) are characterized by the presence of a system of four equatorial light rings -- arranged from the inside out as two unstable rings, followed by one stable ring, and the outermost unstable ring -- while the last two solutions (\textbf{V}$^{\,0}_{\,0.2}$, and \textbf{VI}$^{\,0}_{\,0.3}$) are distinguished by having only two unstable light rings, one prograde and one retrograde. The radial stability of these LRs provides a direct diagnostic of the zones of stability or instability for TCOs in each rotational direction. For a comprehensive discussion of the behavior of equatorial time-like geodesics and their stability, we refer the reader to \cite{Collodel_2021} and \cite{Delgado2022}.

The detailed orbital properties for each configuration are summarized in the tables in this section, listing the radial positions of the event horizon, light rings, and the boundaries of stable and unstable TCO regions. The tables first present the prograde TCO structure (Tables \ref{tab:ProgradeTCOs1} and \ref{tab:ProgradeTCOs2}) for all configurations, and then the retrograde TCO structure (Tables \ref{tab:RetrogradeTCOs1}--\ref{tab:RetrogradeTCOs5}), which corresponds to globally retrograde TCOs containing radially embedded regions of locally prograde motion where the angular velocity $\Omega_{-}$ becomes positive.

\subsection*{\textit{1. Prograde TCOs Structure}}

In solution \textbf{I}$^{\,0}_{\,0.01}$, the structure of the prograde TCOs is markedly intricate, as indicated by the behavior of the circular frequencies $\Omega_{\pm}$ and the corresponding stability analysis. The profile of $\Omega_{+}$ reveals two disconnected stability zones, separated by alternating unstable and forbidden regions, as shown in Table \ref{tab:ProgradeTCOs1} and Fig. \ref{fig:Omega_I}. The innermost prograde stable zone is relatively narrow and emerges immediately above the first prograde unstable light ring at $r_{+}^{\,\text{LR}_{1}} \simeq 0.0141$, extending over the interval $(r_{1}^{\,\text{TCO}}, r_{2}^{\,\text{TCO}}) \simeq [0.0292, 0.0488]$ before encountering the next outer unstable region. A subsequent prograde stable zone appears at larger radii, with the outermost one extending well beyond $r^{\,\Omega_{+}}_{\max} \simeq 0.2708$ towards infinity. Within the radial interval $(r_{1}^{\,D}, r_{2}^{\,D}) \simeq [0.0974, 0.1769]$, the discriminant $D$, defined in Eq.~(\ref{Discriminant}), becomes negative, indicating that circular orbits cease to exist in this region. 

\begin{table}[htbp]
\centering
\footnotesize
\setlength{\tabcolsep}{9.0pt}
\renewcommand{\arraystretch}{1.3}
\caption{\footnotesize Prograde TCOs properties for configuration \textbf{I}$^{\,0}_{\,0.01}$. Physical parameters are listed in Table~\ref{tab_7} in Appendix~A.}
\label{tab:ProgradeTCOs1}
\begin{tabular}{c*{10}{c}}
\hline
\hline
Configuration & Orientation & $r_{H}$ & $r_{+}^{\,\text{LR}_{1}}$  &  $r_{1}^{\,\text{TCO}}$ & $r_{2}^{\,\text{TCO}}$ & $r_{1}^{\,D}$ & $r_{2}^{\,D}$ & $r_{\max}^{\,\Omega_{+}}$ & $r^{\,\text{OUT}}$ \\
\hline
\multirow{3}{*}{\textbf{I}$^{\,0}_{\,0.01}$}  & \text{Prograde}   & $0.01$ & 0.0141 & 0.0293 & 0.0488 & 0.0974 & 0.1769 & 0.2708 & $+\infty$ \\
 & \textit{Regions} & \multicolumn{8}{l}{\hspace{6pt}$\bigg{|}\textcolor[rgb]{0.89,0.00,0.00}{\begin{array}{c} {\scriptsize\text{TCOs}} \\ \scriptsize{\text{forbidden}}\end{array}}\bigg{|}\textcolor[rgb]{1.00,0.36,0.06}{\begin{array}{c} {\scriptsize\text{Unstable}} \\ {\scriptsize\text{TCOs}} \end{array}}\bigg{|}\;\;\textcolor[rgb]{0.35,0.71,0.00}{\begin{array}{c} {\scriptsize\text{Stable}} \\ {\scriptsize\text{TCOs}} \end{array}}\;\;\bigg{|}\,\textcolor[rgb]{1.00,0.36,0.06}{\begin{array}{c} {\scriptsize\text{Unstable}} \\ {\scriptsize\text{TCOs}} \end{array}}\,\bigg{|}\textcolor[rgb]{0.89,0.00,0.00}{\begin{array}{c} {\scriptsize\text{TCOs}} \\ \scriptsize{\text{forbidden}} \end{array}}\bigg{|}\,\textcolor[rgb]{1.00,0.36,0.06}{\begin{array}{c} {\scriptsize\text{Unstable}} \\ {\scriptsize\text{TCOs}} \end{array}}\,\bigg{|}\;\,\textcolor[rgb]{0.35,0.71,0.00}{\begin{array}{c} {\scriptsize\text{Stable}} \\ {\scriptsize\text{TCOs}} \end{array}}\;\,\bigg{|}$} \\
\hline
\end{tabular}
\end{table}

In contrast to the highly fragmented structure of the high-normalized-charge configuration \textbf{I}$^{\,0}_{\,0.01}$, the prograde TCOs for configurations \textbf{II}$^{\,0}_{\,0.01}$ through \textbf{VI}$^{\,0}_{\,0.3}$ display a gradual simplification of the orbital morphology. As summarized in Table~\ref{tab:ProgradeTCOs2}, each of these configurations supports a single continuous stability region extending from the prograde ISCO radius $r_{+}^{\,\text{ISCO}_{1}}$ to spatial infinity. The ISCO position increases systematically with $r_{\text{H}}$, shifting from $r_{+}^{\,\text{ISCO}_{1}} \simeq 0.0539$ in configuration \textbf{II}$^{\,0}_{\,0.01}$ to $0.2454$ in \textbf{III}$^{\,0}_{\,0.05}$, $0.2387$ in \textbf{IV}$^{\,0}_{\,0.1}$, and $0.6328$ in both \textbf{V}$^{\,0}_{\,0.2}$, and \textbf{VI}$^{\,0}_{\,0.3}$. This outward shift of the ISCO enlarges the inner unstable region without producing additional stability regions. Similarly, the radius of the prograde light ring $r_{+}^{\text{LR}_{1}}$ increases with $r_{\text{H}}$, pushing the onset of the stability zone farther away from the event horizon. In all cases from \textbf{II}$^{\,0}_{\,0.01}$ to \textbf{VI}$^{\,0}_{\,0.3}$, the outer boundary of the prograde TCO region lies at spatial infinity, and the radial stability profile remains monotonic, lacking the multi-zone fragmentation of configuration \textbf{I}$^{\,0}_{\,0.01}$. The corresponding profiles of the circular orbital frequencies $\Omega_{+}$ for each configuration are presented in Figs.~\ref{fig:Omega_II}\,--\,\ref{fig:Omega_VI}, providing a detailed view of the radial behavior of prograde motion across the stability domains.

\begin{table}[htbp]
\centering
\footnotesize
\setlength{\tabcolsep}{23.4pt}
\renewcommand{\arraystretch}{1.3}
\caption{\footnotesize Prograde TCOs properties for configurations \textbf{II}$^{\,0}_{\,0.01}$\,--\,\textbf{VI}$^{\,0}_{\,0.3}$. See Table~\ref{tab_7} in Appendix~A for details.}
\label{tab:ProgradeTCOs2}
\begin{tabular}{c*{5}{c}}
\hline\hline
Configuration & Orientation & $r_{H}$ & $r_{+}^{\text{LR}_{1}}$ &  $r_{+}^{\text{ISCO}_{1}}$ & $r^{\text{OUT}}$ \\
\hline
\textbf{II}$^{\,0}_{\,0.01}$  & \text{Prograde}   & $0.01$ & 0.0130   & 0.0539 & $+\infty$ \\
\textbf{III}$^{\,0}_{\,0.05}$  & \text{Prograde}   & $0.05$ & 0.0634   & 0.2454 & $+\infty$ \\
\textbf{IV}$^{\,0}_{\,0.1}$  & \text{Prograde}   & $0.10$ & 0.1169   & 0.2387 & $+\infty$ \\
\textbf{V}$^{\,0}_{\,0.2}$  & \text{Prograde}   & $0.20$ & 0.2252   & 0.6328 & $+\infty$ \\
\textbf{VI}$^{\,0}_{\,0.3}$  & \text{Prograde}   & $0.30$ & 0.3552   & 0.6328 & $+\infty$ \\
                                              & \textit{Regions} & \multicolumn{4}{l}{\hspace{6pt}$\bigg{|}\hspace{12.5pt}\textcolor[rgb]{0.89,0.00,0.00}{\begin{array}{c} \text{TCOs} \\ {\scriptsize\text{forbidden}} \end{array}}\hspace{12.5pt}\bigg{|}\hspace{16.5pt}\textcolor[rgb]{1.00,0.36,0.06}{\begin{array}{c} {\scriptsize\text{Unstable}} \\ \text{TCOs} \end{array}}\hspace{16.5pt}\bigg{|}\hspace{18.5pt}\textcolor[rgb]{0.35,0.71,0.00}{\begin{array}{c} {\scriptsize\text{Stable}} \\ \text{TCOs} \end{array}}\hspace{18.5pt}\bigg{|}$} \\

\hline		
\end{tabular}
\end{table}

\subsection*{\textit{2. Retrograde TCOs Structure}}

The retrograde TCO structure, summarized in Tables~\ref{tab:RetrogradeTCOs1}\,--\,\ref{tab:RetrogradeTCOs5}, exhibits several distinct morphological patterns across the considered configurations \textbf{I}$^{\,0}_{\,0.01}$ through \textbf{VI}$^{\,0}_{\,0.3}$.

\begin{figure}[t!]
    \centering
    \includegraphics[width=0.55\textwidth]{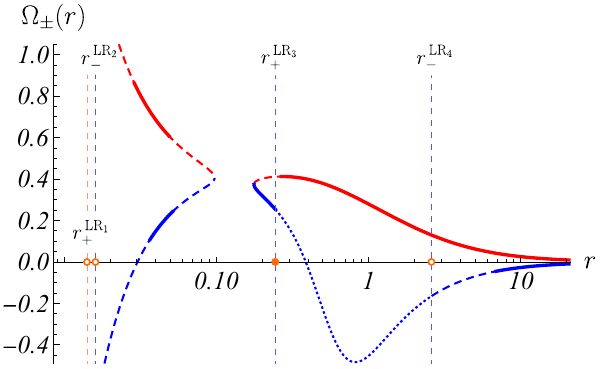}\\
    \caption{\small The Keplerian angular velocities $\Omega_{\pm}$ for massive particles are shown for configuration $\mathbf{I}^{\,0}_{\,0.01}$ with parameters $\kappa = 0$, $\omega_s/\mu = 0.679241$, $M\mu = 0.881991$, and $q = 0.999875$. Red curves correspond to co-rotating orbits with $\Omega_{+} > 0$. Blue curves represent counter-rotating orbits when $\Omega_{-} < 0$ and co-rotating orbits when $\Omega_{-} > 0$. Long-dashed segments indicate regions where circular massive-particle orbits are dynamically unstable. The short-dashed curve delineates the domain in which timelike circular orbits are forbidden. Circular orbits terminate at the boundary where $(\partial_{r} g_{t\phi})^{2} < \partial_{r} g_{tt}\, \partial_{r} g_{\phi\phi}$. Vertical dashed lines together with the orange markers identify a sequence of light rings: a co-rotating \textit{unstable} light ring (red), a counter-rotating \textit{unstable} light ring (first blue), a co-rotating \textit{stable} light ring (second blue), and a counter-rotating unstable light ring (third blue).
} 
	\label{fig:Omega_I}
\end{figure}

The first configuration examined, \textbf{I}$^{\,0}_{\,0.01}$, lying close to the solitonic limit and characterized by a high normalized charge, supports a single retrograde stability domain extending from the retrograde ISCO radius $r_{-}^{\,\text{ISCO}_{2}} \simeq 6.8973$ to spatial infinity, as can be seen from Table~\ref{tab:RetrogradeTCOs1} and Fig.~\ref{fig:Omega_I}. No fragmentation of the stability region is observed. Below this stable domain lies an unstable region, followed by a wide region in which no TCOs exist, extending down to the prograde stable light ring $r_{+}^{\,\text{LR}_{3}}$.

A distinctive feature of this class of scalarized Kerr black holes is the sign reversal of the retrograde orbital frequency $\Omega_{-}$ near $\text{LR}_{3}$, producing a narrow prograde stability band $(r_{2}^{\,D}, r_{+}^{\,\text{LR}_{3}}) \simeq [0.1769, 0.2437)$ immediately below the light ring. This creates an embedded prograde TCO region within the outer retrograde domain, aligned with the light ring's rotation.

Moreover, moving further inward across the TCO-forbidden region, one encounters an additional unstable zone, beneath which lies a second, significantly narrower stability domain capable of sustaining prograde TCO orbits around the black hole. No further stable TCO structures are present below this inner region.

\begin{table}[!b]
\centering
\footnotesize
\setlength{\tabcolsep}{5.65pt}
\renewcommand{\arraystretch}{1.3}
\caption{\footnotesize Retrograde TCOs properties for configuration \textbf{I}$^{\,0}_{\,0.01}$ with embedded locally stable prograde subregions characterized by $\Omega_{-} > 0$ within the retrograde branch. See Table \ref{tab_7} in Appendix~A for physical details.}
\label{tab:RetrogradeTCOs1}
\begin{tabular}{c@{\hspace{7pt}}c*{11}{c}}
\hline
\hline
Configuration & Orientation & $r_{H}$ & $r_{-}^{\,\text{LR}_{2}}$  &  $r_{1}^{\,\text{TCO}}$ &  &  &  & $r_{+}^{\,\text{LR}_{3}}$ & $r_{-}^{\text{LR}_{4}}$ & $r_{-}^{\,\text{ISCO}_{2}}$ & $r^{\,\text{OUT}}$ \\
\hline
\multirow{6}{*}{\textbf{I}$^{\,0}_{\,0.01}$}  & \text{Retrograde}   & $0.01$ & 0.0161 & 0.0363 &  &  &  & 0.2437 & 2.5949 & 6.8973 & $+\infty$ \\
                                              & \multirow{3}{*}{\textit{Regions}} & \multicolumn{10}{l}{\hspace{4.3pt}$\bigg{|}\hspace{-3pt}\textcolor[rgb]{0.89,0.00,0.00}{\begin{array}{c} {\scriptsize\text{TCOs}} \\ \scriptsize{\text{forbidden}}\end{array}}\hspace{-3pt}\bigg{|}\hspace{-3pt}\textcolor[rgb]{1.00,0.36,0.06}{\begin{array}{c} {\scriptsize\text{Unstable}} \\ {\scriptsize\text{TCOs}} \end{array}}\hspace{-3pt}\bigg{|}\hspace{136.3pt}\bigg{|}\hspace{-3pt}\textcolor[rgb]{0.89,0.00,0.00}{\begin{array}{c} {\scriptsize\text{TCOs}} \\ \scriptsize{\text{forbidden}} \end{array}}\hspace{-3pt}\bigg{|}\hspace{-3pt}\textcolor[rgb]{1.00,0.36,0.06}{\begin{array}{c} {\scriptsize\text{Unstable}} \\ {\scriptsize\text{TCOs}} \end{array}}\hspace{-3pt}\bigg{|}\,\textcolor[rgb]{0.35,0.71,0.00}{\begin{array}{c} {\scriptsize\text{Stable}} \\ {\scriptsize\text{TCOs}} \end{array}}\,\bigg{|}$} \\

\cline{3-12}		
&  & \multicolumn{10}{l}{\hspace{77.3pt}$\bigg{|}\textcolor[rgb]{0.35,0.71,0.00}{\begin{array}{c} {\scriptsize\text{Stable}} \\ {\scriptsize\text{TCOs}} \end{array}}\bigg{|}\hspace{-2pt}\textcolor[rgb]{1.00,0.36,0.06}{\begin{array}{c} {\scriptsize\text{Unstable}} \\ {\scriptsize\text{TCOs}} \end{array}}\hspace{-2pt}\bigg{|}\hspace{-3pt}\textcolor[rgb]{0.89,0.00,0.00}{\begin{array}{c} {\scriptsize\text{TCOs}} \\ \scriptsize{\text{forbidden}} \end{array}}\hspace{-3pt}\bigg{|}\,\textcolor[rgb]{0.35,0.71,0.00}{\begin{array}{c} {\scriptsize\text{Stable}} \\ {\scriptsize\text{TCOs}} \end{array}}\,\bigg{|}$} \\
   & \text{Prograde}   &  &  & 0.0363 & 0.0517 & 0.0974 & 0.1769 & 0.2437 &  &  &  \\
\cline{2-12}
   &  &  &   &  $r_{1}^{\,\text{TCO}}$ & $r_{2}^{\,\text{TCO}}$ & $r_{1}^{\,D}$ & $r_{2}^{\,D}$ & $r_{+}^{\,\text{LR}_{3}}$ &  &  &  \\
\hline
\end{tabular}
\end{table}

\begin{figure}[h!]
    \centering
    \includegraphics[width=0.55\textwidth]{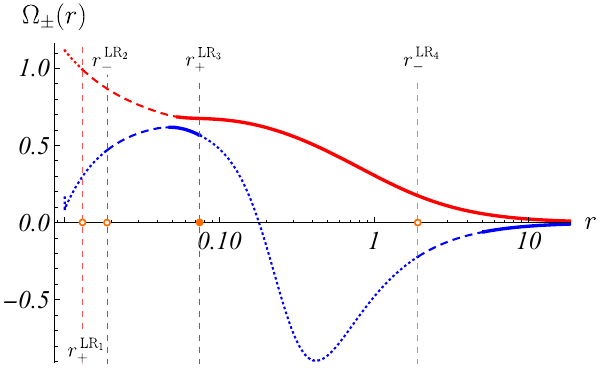}\\
    \caption{\small The Keplerian angular velocities $\Omega_{\pm}$ for massive particles are shown for configuration $\mathbf{II}^{\,0}_{\,0.01}$ with parameters $\kappa = 0$, $\omega_s/\mu = 0.835272$, $M\mu = 0.648229$, and $q = 0.997293$. Red curves correspond to co-rotating orbits with $\Omega_{+} > 0$. Blue curves represent counter-rotating orbits when $\Omega_{-} < 0$ and co-rotating orbits when $\Omega_{-} > 0$. Long-dashed segments indicate regions where circular orbits of massive particles are dynamically unstable. The short-dashed curve delineates the domain in which timelike circular orbits are forbidden. Vertical dashed lines together with the orange markers identify a sequence of light rings: a co-rotating \textit{unstable} light ring (red), a counter-rotating \textit{unstable} light ring (first blue), a co-rotating \textit{stable} light ring (second blue), and a counter-rotating \textit{unstable} light ring (third blue).} 
	\label{fig:Omega_II}
\end{figure}

\begin{figure}[t!]
    \centering
    \includegraphics[width=0.55\textwidth]{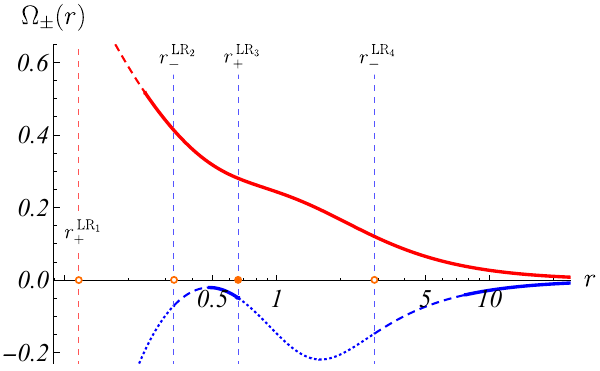}\\
    \caption{\small The plots present the Keplerian angular velocities $\Omega_{\pm}$ associated with massive particles in configuration \textbf{IV}$^{\,0}_{\,0.1}$ with $\kappa = 0$, $\omega_s/\mu = 0.738499$, $M\mu = 1.00043$, and $q = 0.9644770.96447$. The red and blue curves correspond to co-rotating and counter-rotating orbits, respectively. Long-dashed lines indicate regions of dynamical instability, whereas the short-dashed line delineates the boundary beyond which timelike circular orbits cease to exist. Dashed vertical lines indicate the positions of light rings, both stable and unstable, as illustrated in Fig. \ref{fig:Omega_I}.} 
	\label{fig:Omega_IV}
\end{figure}

By contrast, configurations \textbf{II}$^{\,0}_{\,0.01}$, \textbf{III}$^{\,0}_{\,0.05}$, and \textbf{IV}$^{\,0}_{\,0.1}$ share a qualitatively similar retrograde structure, yet exhibit distinct features in hosting internal prograde or retrograde stability regions, as shown in Tables~\ref{tab:RetrogradeTCOs2}\,--\,\ref{tab:RetrogradeTCOs4} and Figs.~\ref{fig:Omega_II}\,--\,\ref{fig:Omega_IV}. In all three cases, the retrograde TCO region extends from the retrograde ISCO to spatial infinity, with the ISCO radius increasing as the normalized charge $q$ decreases: from $r_{-}^{\,\mathrm{ISCO}_{2}} \simeq 5.1287$ in \textbf{II}$^{\,0}_{\,0.01}$, to $7.4839$ in \textbf{III}$^{\,0}_{\,0.05}$, and to $7.7256$ in \textbf{IV}$^{\,0}_{\,0.1}$.

\begin{table}[!t]
\centering
\footnotesize
\setlength{\tabcolsep}{11.4pt}
\renewcommand{\arraystretch}{1.3}
\caption{\footnotesize Retrograde TCOs properties for configuration \textbf{II}$^{\,0}_{\,0.01}$ with embedded locally stable prograde subregions characterized by $\Omega_{-} > 0$ within the retrograde branch. See Table~\ref{tab_7} in Appendix~A for details.}
\label{tab:RetrogradeTCOs2}
\begin{tabular}{c*{9}{c}}
\hline
\hline
Configuration & Orientation &   & &  & $r_{+}^{\,\text{LR}_{3}}$ & $r_{-}^{\,\text{LR}_{4}}$ & $r_{-}^{\,\text{ISCO}_{2}}$ & $r^{\,\text{OUT}}$ \\
\hline
\multirow{6}{*}{\textbf{II}$^{\,0}_{\,0.01}$}  & \text{Retrograde}  &   & &  & 0.0747 & 1.9186 & 5.1287 & $+\infty$ \\
                                              & \multirow{3}{*}{\textit{Regions}} & \multicolumn{7}{l}{\hspace{144.3pt}$\bigg{|}\hspace{2pt}\textcolor[rgb]{0.89,0.00,0.00}{\begin{array}{c} {\scriptsize\text{TCOs}} \\ \scriptsize{\text{forbidden}} \end{array}}\hspace{2pt}\bigg{|}\hspace{3pt}\textcolor[rgb]{1.00,0.36,0.06}{\begin{array}{c} {\scriptsize\text{Unstable}} \\ {\scriptsize\text{TCOs}} \end{array}}\hspace{3pt}\bigg{|}\hspace{7.5pt}\textcolor[rgb]{0.35,0.71,0.00}{\begin{array}{c} {\scriptsize\text{Stable}} \\ {\scriptsize\text{TCOs}} \end{array}}\hspace{7.5pt}\bigg{|}$} \\
\cline{3-9}		
   &       & \multicolumn{7}{l}{\hspace{7pt}$\bigg{|}\hspace{-0.5pt}\textcolor[rgb]{0.89,0.00,0.00}{\begin{array}{c} {\scriptsize\text{TCOs}} \\ \scriptsize{\text{forbidden}} \end{array}}\hspace{-0.5pt}\bigg{|}\hspace{3.5pt}\textcolor[rgb]{1.00,0.36,0.06}{\begin{array}{c} {\scriptsize\text{Unstable}} \\ {\scriptsize\text{TCOs}} \end{array}}\hspace{3.5pt}\bigg{|}\hspace{7.5pt}\textcolor[rgb]{0.35,0.71,0.00}{\begin{array}{c} {\scriptsize\text{Stable}} \\ {\scriptsize\text{TCOs}} \end{array}}\hspace{7.5pt}\bigg{|}$} \\
   & \text{Prograde}   & $0.01$ & 0.0187 & 0.0479 & 0.0747 &  &  &  \\
\cline{2-9}
   & Disk & $r_{H}$ & $r_{-}^{\,\text{LR}_{2}}$ & $r_{\max}^{\,\Omega_{-}}$ & $r_{+}^{\,\text{LR}_{3}}$ &  & &   \\
\hline
\end{tabular}
\end{table}

Inside the outer retrograde domain lies a forbidden zone and, in certain cases, embedded stability regions. Configurations \textbf{II}$^{\,0}_{\,0.01}$ and \textbf{III}$^{\,0}_{\,0.05}$ develop an internal \textit{prograde} stability region, bounded above by the prograde stable light ring $r_{+}^{\,\mathrm{LR}_{3}}$, which acts as the upper boundary of the internally located prograde stability zone. In contrast, \textbf{IV}$^{\,0}_{\,0.1}$ features an internal \textit{retrograde} stability region, also bounded above by $r_{+}^{\,\text{LR}_{3}}$, which acts as the upper boundary of the internally located retrograde stability zone. In all cases, the lower boundary of the embedded zone coincides with the local maximum of $\Omega_{-}$ at $r_{\min}^{\,\Omega_{-}}$, whose radial position indicates that the width of the zone increases as $q$ decreases. Beyond this internal region, the standard sequence of an unstable zone followed by a TCO--forbidden domain is recovered, within which no stable circular orbits exist.

\begin{table}[!t]
\centering
\footnotesize
\setlength{\tabcolsep}{11pt}
\renewcommand{\arraystretch}{1.3}
\caption{\footnotesize Retrograde TCOs properties for configuration \textbf{III}$^{\,0}_{\,0.05}$ that include embedded locally stable prograde subregions with $\Omega_{-} > 0$ within the retrograde branch. See Table~\ref{tab_7} in Appendix~A for details.}
\label{tab:RetrogradeTCOs3}
\begin{tabular}{c*{9}{c}}
\hline
\hline
Configuration & Orientation & $r_{H}$ & \hspace{4pt} $r_{-}^{\,\text{LR}_{2}}$  &   &  $r_{+}^{\,\text{LR}_{3}}$  & $r_{-}^{\,\text{LR}_{4}}$ & $r_{-}^{\,\text{ISCO}_{2}}$ & $r^{\,\text{OUT}}$ \\
\hline
\multirow{6}{*}{\textbf{III}$^{\,0}_{\,0.05}$}  & \text{Retrograde}   & $0.05$ & \hspace{4pt} 0.1004 &  & 0.4260 & 2.8289 & 7.4839 & $+\infty$ \\
                                              & \multirow{3}{*}{\textit{Regions}} & \multicolumn{7}{l}{\hspace{6.5pt}$\bigg{|}\hspace{2.5pt}\textcolor[rgb]{0.89,0.00,0.00}{\begin{array}{c} {\scriptsize\text{TCOs}} \\ \scriptsize{\text{forbidden}}\end{array}}\hspace{2.5pt}\bigg{|}\hspace{90pt}\bigg{|}\hspace{1.25pt}\textcolor[rgb]{0.89,0.00,0.00}{\begin{array}{c} {\scriptsize\text{TCOs}} \\ \scriptsize{\text{forbidden}} \end{array}}\hspace{1.25pt}\bigg{|}\hspace{3pt}\textcolor[rgb]{1.00,0.36,0.06}{\begin{array}{c} {\scriptsize\text{Unstable}} \\ {\scriptsize\text{TCOs}} \end{array}}\hspace{3pt}\bigg{|}\hspace{7pt}\textcolor[rgb]{0.35,0.71,0.00}{\begin{array}{c} {\scriptsize\text{Stable}} \\ {\scriptsize\text{TCOs}} \end{array}}\hspace{7pt}\bigg{|}$} \\
\cline{3-9}		
&  & \multicolumn{7}{l}{\hspace{56pt}$\bigg{|}\hspace{3pt}\textcolor[rgb]{1.00,0.36,0.06}{\begin{array}{c} {\scriptsize\text{Unstable}} \\ {\scriptsize\text{TCOs}} \end{array}}\hspace{3pt}\bigg{|}\hspace{7.865pt}\textcolor[rgb]{0.35,0.71,0.00}{\begin{array}{c} {\scriptsize\text{Stable}} \\ {\scriptsize\text{TCOs}} \end{array}}\hspace{7.865pt}\bigg{|}$} \\
   & \text{Prograde}   & & \hspace{4pt} 0.1004 & 0.2581 & 0.4260 &  &  &    \\
\cline{2-9}
   & Disk &  & \hspace{4pt} $r_{-}^{\,\text{LR}_{2}}$  & $r_{\max}^{\,\Omega_{-}}$  & $r_{+}^{\,\text{LR}_{3}}$ &  &  &   \\
\hline
\end{tabular}
\end{table}

\begin{table}[!t]
\centering
\footnotesize
\setlength{\tabcolsep}{13.5pt}
\renewcommand{\arraystretch}{1.2}
\caption{\footnotesize Retrograde TCOs properties for configuration \textbf{IV}$^{\,0}_{\,0.1}$ with embedded locally stable retrograde subregions characterized by $\Omega_{-} < 0$ in the retrograde branch. See Table \ref{tab_7} in Appendix A for details.}
\label{tab:RetrogradeTCOs4}
\begin{tabular}{c@{\hspace{9pt}}c@{\hspace{14pt}}c*{6}{c}}
\hline
\hline
Configuration & Orientation & $r_{H}$ & $r_{-}^{\text{LR}_{2}}$ & $r^{\,\Omega_{-}}_{\max}$  &  $r_{+}^{\text{LR}_{3}}$ & $r_{-}^{\text{LR}_{4}}$ & $r_{-}^{\text{ISCO}_{2}}$ & $r^{\text{OUT}}$ \\
\hline
\multirow{3}{*}{\textbf{IV}$^{\,0}_{\,0.1}$}  & \text{Retrograde}   & $0.1$ & 0.3284 & 0.4827 & 0.6602 & 2.8773 & 7.7256 & $+\infty$ \\
                                              & \textit{Regions} & \multicolumn{7}{l}{\hspace{-9.5pt}$\bigg{|}\hspace{1pt}\textcolor[rgb]{0.89,0.00,0.00}{\begin{array}{c} {\scriptsize\text{TCOs}} \\ \scriptsize{\text{forbidden}} \end{array}}\hspace{1pt}\bigg{|}\hspace{5.5pt}\textcolor[rgb]{1.00,0.36,0.06}{\begin{array}{c} {\scriptsize\text{Unstable}} \\ {\scriptsize\text{TCOs}} \end{array}}\hspace{5.5pt}\bigg{|}\hspace{10.5pt}\textcolor[rgb]{0.35,0.71,0.00}{\begin{array}{c} {\scriptsize\text{Stable}} \\ {\scriptsize\text{TCOs}} \end{array}}\hspace{10.5pt}\bigg{|}\hspace{3.5pt}\textcolor[rgb]{0.89,0.00,0.00}{\begin{array}{c} {\scriptsize\text{TCOs}} \\ \scriptsize{\text{forbidden}} \end{array}}\hspace{3.5pt}\bigg{|}\hspace{5.5pt}\textcolor[rgb]{1.00,0.36,0.06}{\begin{array}{c} {\scriptsize\text{Unstable}} \\ {\scriptsize\text{TCOs}} \end{array}}\hspace{5.5pt}\bigg{|}\hspace{9.3pt}\textcolor[rgb]{0.35,0.71,0.00}{\begin{array}{c} {\scriptsize\text{Stable}} \\ {\scriptsize\text{TCOs}} \end{array}}\hspace{9.3pt}\bigg{|}$} \\

\hline		
\end{tabular}
\end{table}

\begin{figure}[t!]
    \centering
    \includegraphics[width=0.55\textwidth]{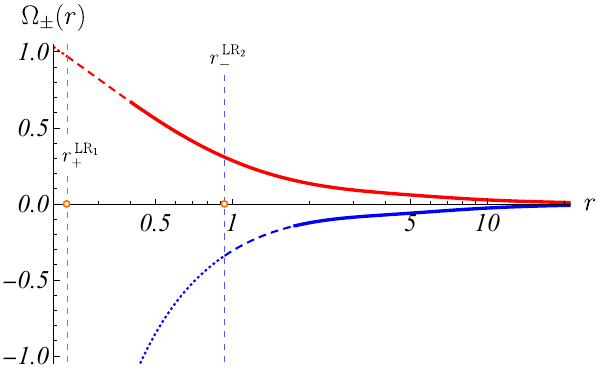}\\
    \caption{\small The Keplerian angular velocities $\Omega_{\pm}$ for massive particles are shown for configuration $\mathbf{V}^{\,0}_{\,0.2}$ with parameters $\kappa = 0$, $\omega_s/\mu = 0.895538$, $M\mu = 0.878726$, and $q = 0.850778$. Red and blue curves correspond to co-rotating and counter-rotating orbits, respectively. Timelike circular motion is forbidden in the regions marked by short-dashed segments, while the long-dotted segments indicate domains of dynamical instability. Two \textit{unstable} light rings are identified by vertical dashed lines together with the orange markers: the inner ring corresponds to photons on co-rotating trajectories, whereas the outer one is associated with counter-rotating trajectories, following the convention of Fig. \ref{fig:Omega_I}.} 
	\label{fig:Omega_V}
\end{figure}

\begin{figure}[t!]
    \centering
    \includegraphics[width=0.55\textwidth]{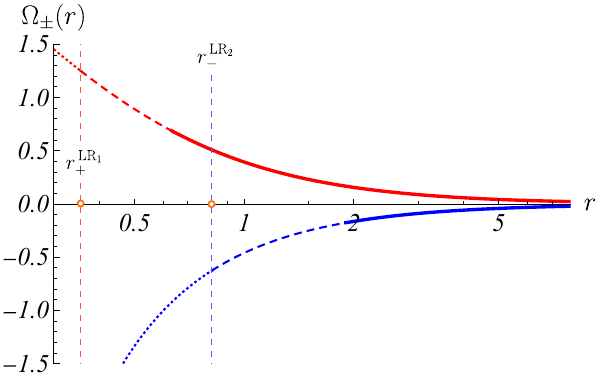}\\
    \caption{\small The Keplerian angular velocities $\Omega_{\pm}$ for massive particles are shown for configuration $\mathbf{VI}^{\,0}_{\,0.3}$ with parameters $\kappa = 0$, $\omega_s/\mu = 0.988000$, $M\mu = 0.320010$, and $q = 0.647791$. Red and blue curves correspond to co-rotating and counter-rotating orbits, respectively. Short-dashed segments indicate regions in which no timelike circular orbits exist, whereas long-dashed segments mark domains of dynamical instability. Two \textit{unstable} light rings are identified by vertical dashed lines together with the orange markers: the inner ring corresponds to co-rotating photon trajectories, while the outer ring is associated with counter-rotating trajectories.} 
	\label{fig:Omega_VI}
\end{figure}

Finally, for the last two configurations, \textbf{V}$^{\,0}_{\,0.2}$ and \textbf{VI}$^{\,0}_{\,0.3}$, the retrograde topology simplifies markedly. Only a single continuous stability domain remains, extending from $r_{-}^{\text{ISCO}_{2}}$ to spatial infinity, closely paralleling the prograde structure of the same solutions, as summarized in Table~\ref{tab:RetrogradeTCOs5}. The ISCO radius increases from $r_{-}^{\text{ISCO}} \simeq 1.7651$ in \textbf{V}$^{\,0}_{\,0.2}$ to $1.8999$ in \textbf{VI}$^{\,0}_{\,0.3}$, while the retrograde light ring $r_{-}^{\mathrm{LR}_{2}}$ shifts steadily outward from the event horizon, as illustrated in Figs.~\ref{fig:Omega_V} and \ref{fig:Omega_VI}. Within this stability region, no sign reversal of $\Omega_{-}$ occurs, and thus no embedded prograde zones are formed.  

In summary, the occurrence of nested orbital structures in the retrograde sector is directly tied to the sign behavior of $\Omega_{-}$. Configurations \textbf{II}$^{\,0}_{\,0.01}$\,--\,\textbf{IV}$^{\,0}_{\,0.1}$ uniquely develop a positive $\Omega_{-}$ interval, giving rise to embedded prograde stability islands, whereas \textbf{I}$^{\,0}_{\,0.01}$, \textbf{V}$^{\,0}_{\,0.2}$, and \textbf{VI}$^{\,0}_{\,0.3}$ preserve purely retrograde domains free of internal fragmentation.

\begin{table}[!t]
\centering
\footnotesize
\setlength{\tabcolsep}{23.6pt}
\renewcommand{\arraystretch}{1.3}
\caption{\footnotesize Retrograde TCOs properties for configurations \textbf{V}$^{\,0}_{\,0.2}$, and \textbf{VI}$^{\,0}_{\,0.3}$. See Table~\ref{tab_7} in Appendix~A for details.}
\label{tab:RetrogradeTCOs5}
\begin{tabular}{c*{5}{c}}
\hline
\hline
Configuration & Orientation & $r_{H}$ & $r_{-}^{\text{LR}_{2}}$ &  $r_{-}^{\text{ISCO}_{2}}$ & $r^{\text{OUT}}$ \\
\hline
\textbf{V}$^{\,0}_{\,0.2}$  & \text{Retrograde} & $0.2$ & 0.9357   & 1.7651 & $+\infty$ \\
\textbf{VI}$^{\,0}_{\,0.3}$ & \text{Retrograde} & $0.3$ & 0.8146   & 1.8999 & $+\infty$ \\
                                              & \textit{Regions} & \multicolumn{4}{l}{\;\,$\bigg{|}\hspace{10.5pt}\textcolor[rgb]{0.89,0.00,0.00}{\begin{array}{c} \scriptsize{\text{TCOs}} \\ \scriptsize{\text{forbidden}} \end{array}}\hspace{10.5pt}\bigg{|}\hspace{16pt}\textcolor[rgb]{1.00,0.36,0.06}{\begin{array}{c} \scriptsize{\text{Unstable}} \\ \scriptsize{\text{TCOs}} \end{array}}\hspace{15pt}\bigg{|}\hspace{19.5pt}\textcolor[rgb]{0.35,0.71,0.00}{\begin{array}{c} \scriptsize{\text{Stable}} \\ \scriptsize{\text{TCOs}} \end{array}}\hspace{19.5pt}\bigg{|}$} \\ 
\hline
\end{tabular}
\end{table}

\subsection{Modeling the Radial Structure and Orbital Placement of Thin Disks}

The intricate distribution of stable and unstable TCOs, as discussed in the previous section, has direct implications for the formation and radial extension of thin accretion disks around Kerr black holes with synchronized scalar hair. In particular, the location of the ISCO, or its effective counterpart whenever fragmentation or discontinuities occur, provides a natural criterion for the inner edge of a thin disk. The radial placement of the disk therefore reflects the orbital morphology encoded in the $\Omega_{\pm}$ profiles and the associated stability analysis.

In configurations where the prograde TCO structure consists of a single continuous stability domain extending from $r_{+}^{\,\mathrm{ISCO}_{1}}$ to infinity, the inner edge of the disk 
is naturally identified with the ISCO radius. Such cases include solutions \textbf{II}$^{\,0}_{\,0.01}$ through \textbf{VI}$^{\,0}_{\,0.3}$. The monotonic stability profile in these geometries suggests that thin disks would form in a manner closely resembling their Kerr counterparts, with the dynamics governed primarily by the outward shift of the ISCO as the normalized charge increases. This displacement enlarges the inner unstable zone, thereby reducing the efficiency of accretion but without introducing additional fragmented substructures.

By contrast, configuration \textbf{I}$^{\,0}_{\,0.01}$ exhibits a much more complex orbital morphology. Multiple prograde stability zones are separated by forbidden and unstable regions. In such a scenario, the accretion flow could in principle be interrupted, giving rise to segmented or multi-ring thin disks. The physical viability of these fragmented disk configurations critically depends on the mechanism of angular momentum transport across the forbidden regions. Under standard magneto-rotational instability (MRI) conditions, which are typical of well-ionized and geometrically thin disks, such transport is inefficient, since MRI vanishes in zones without differential rotation or with reversed shear, as demonstrated in boundary-layer simulations 
\cite{Pessah2012}. Unless supplemented by alternative drivers, such as magnetically induced instabilities in magnetically arrested disks (MADs), these inner fragments are unlikely to contribute significantly to persistent disk formation \cite{Marshall2018}. Additional non-MRI mechanisms for angular momentum transport may operate in strongly magnetized or radiation-dominated environments, but a detailed analysis of such cases lies beyond the scope 
of the present work.

\begin{figure}[t!]
    \centering
    \includegraphics[width=0.55\textwidth]{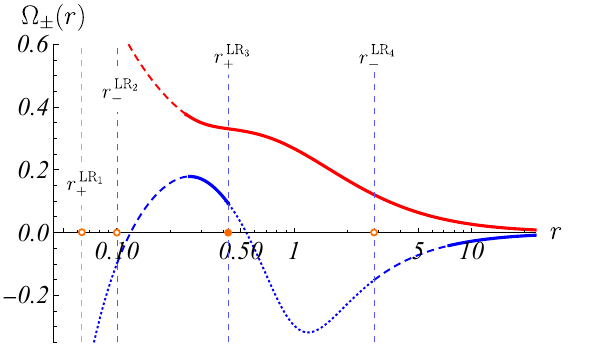}\\
    \caption{\small The Keplerian angular velocities $\Omega_{\pm}$ for massive particles are shown for configuration $\mathbf{III}^{\,0}_{\,0.05}$ with parameters $\kappa = 0$, $\omega_s/\mu = 0.690049$, $M\mu = 0.966595$, and $q = 0.994490$. Red and blue curves correspond to co-rotating and counter-rotating orbits, respectively. Short-dashed segments indicate regions where circular circular timelike orbits are dynamically unstable, while the long-dashed curve delineates the boundary beyond which timelike circular motion is forbidden. Vertical dashed lines together with the orange markers identify a sequence of light rings, including both stable and unstable ones, in accordance with the convention used in Fig.~\ref{fig:Omega_I}.} 
	\label{fig:Omega_III}
\end{figure} 

An astrophysically motivated mechanism for the emergence of such fragmented disks may arise from the tidal disruption of a main-sequence star that approaches the central compact object too closely \cite{Sengo_2024}. During this process, a fraction of the stellar matter is disrupted and can acquire either positive or negative angular momentum with respect to the black hole, thus giving rise to a co-rotating or counter-rotating accretion disk. If the inflowing material possesses the appropriate energy and angular momentum to fall into the intricate sequence of stability zones described above, the outcome may be a segmented accretion structure in which individual disk fragments are sustained within separated prograde or retrograde domains of 
stability. The long-term persistence of such multi-ring disks depends critically on the efficiency of angular momentum transport across the intervening unstable regions. Nevertheless, the tidal disruption process provides a natural astrophysical setting that could populate these zones and transiently sustain fragmented accretion flows.

In the retrograde sector, the situation is similarly diverse. Configurations \textbf{II}$^{\,0}_{\,0.01}$\;--\;\textbf{IV}$^{\,0}_{\,0.1}$ develop embedded stability zones associated with a local sign reversal of $\Omega_{-}$. These internal domains can, in principle, host localized rings of prograde matter, yet their astrophysical relevance remains uncertain. Unless supported by external fueling mechanisms, they are unlikely to form persistent disk structures. In contrast, configurations \textbf{V}$^{\,0}_{\,0.2}$ and \textbf{VI}$^{\,0}_{\,0.3}$ feature a single continuous retrograde stability domain from $r_{-}^{\,\mathrm{ISCO}_{2}}$ outward, closely paralleling the Kerr case.

To model the radiation output of the thin accretion disks in these geometries, we adopt the Novikov--Thorne formalism \cite{Novikov1973, Page1974}, which provides a physically motivated description of emission from a geometrically thin, optically thick disk in hydrodynamic equilibrium. Within this framework, the energy flux radiated from the disk surface at a radius $r$ is obtained from the conservation of rest mass, energy, and angular momentum of the orbiting matter. The resulting expression reads
\begin{equation}
    F(r) = -\frac{\dot{M}}{4\pi \sqrt{-g^{(3)}}}\,
    \frac{\partial_{r}\Omega}{(E-\Omega L)^{2}}
    \int_{r_{\mathrm{ISCO}}}^{r} (E-\Omega L)\,\partial_{r}L\,dr ,
\end{equation}
where $\dot{M}$ is the mass accretion rate, $E$ and $L$ are the specific energy and angular momentum of the circular geodesics, $\Omega$ denotes the orbital angular frequency, and $g^{(3)}$ is the determinant of the induced metric on the equatorial plane. For stationary accretion, $\dot{M}$ is constant, while the kinematic quantities $E$, $L$, and $\Omega$ are determined uniquely by the geodesic structure, according to Eqs.~(\ref{Enegry}) and (\ref{Momentum}). 

For convenience, one may further introduce a normalized flux $\tilde{F}(r)$, scaled by the ADM mass $M$ of the black hole solution, in order to enable direct comparison between spacetimes with different horizon radii and spin parameters:
\begin{equation}
    F(r) = \frac{1}{M^2}\, \tilde{F}(r).
\end{equation}
This normalization highlights the influence of scalar hair on the emission profile and allows one to contrast the accretion disk fluxes obtained in the hairy Kerr family with the Kerr case ($q=0$). The resulting flux distributions $\tilde{F}(r)$ reflect the orbital morphology described above, providing a consistent and observable diagnostic of the impact of synchronized scalar fields on thin-disk accretion, as discussed in detail in \cite{Collodel_2021}.

\section{A Numerical Approach to Calculating the Image of a Thin Accretion Disk}

To construct the apparent images of a thin accretion disk surrounding rotating black holes, we apply the backward ray-tracing technique. In this framework, photon trajectories are numerically integrated until they either cross the event horizon or intersect the equatorial plane hosting the disk. The evolution of these null geodesics is determined by the equations of motion derived from Hamilton's formalism,  
\vspace{-2mm}
\begin{equation}
\dot{x}^\mu = \frac{\partial \mathcal{H}}{\partial p_\mu}, 
\qquad 
\dot{p}_\mu = -\frac{\partial \mathcal{H}}{\partial x^\mu},
\label{eq_Hamilton}
\end{equation}
where differentiation is taken with respect to an affine parameter. The Hamiltonian is constructed from the contravariant components of the spacetime metric $g^{\mu\nu}$, and under the assumption of minimal coupling between null particles with four-momentum $p_\mu$ and the gravitational background, it takes the form
\vspace{-1mm}
\begin{equation}
\mathcal{H} \equiv \tfrac{1}{2} g^{\mu\nu} p_\mu p_\nu = 0 .
\label{Hamiltonian}
\end{equation}

Because the geometry under consideration is stationary and axisymmetric, the Hamiltonian exhibits no explicit dependence on the coordinates $t$ and $\varphi$. Here, stationarity and axisymmetry ensure the conservation of the photon energy and azimuthal angular momentum, and the corresponding equations of motion for $t$ and $\varphi$ follow directly from the metric coefficients, as shown in Eqs.~(\ref{four_velocity}). 


To initialize the photon trajectories, we adopt the zero-angular-momentum observer (ZAMO) frame~\cite{cunha2016shadows, lora2022osiris, Gyulchev2024}, defined by the orthonormal tetrad $\{\hat{e}_{(t)}, \hat{e}_{(r)}, \hat{e}_{(\theta)}, \hat{e}_{(\varphi)}\}$. 
This tetrad can be expanded in the coordinate basis $\{\partial_t, \partial_r, \partial_\theta, \partial_\varphi\}$ as
\begin{equation}
\begin{aligned}
\hat{e}_{(\theta)} &= A^{\theta}\,\partial_\theta, \qquad 
\hat{e}_{(r)} = A^{r}\,\partial_r, \\
\hat{e}_{(\varphi)} &= A^{\varphi}\,\partial_\varphi, \qquad 
\hat{e}_{(t)} = \zeta\,\partial_t + \gamma\,\partial_\varphi ,
\end{aligned}
\end{equation}
with the normalization condition $\hat{e}_{(\mu)} \cdot \hat{e}_{(\nu)} = \eta_{\mu\nu}$. 
This choice guarantees that the tetrad reduces to the standard static frame at spatial infinity. 
The coefficients are given by
\begin{equation}
A^{\theta} = \frac{1}{\sqrt{g_{\theta\theta}}}, \qquad 
A^{r} = \frac{1}{\sqrt{g_{rr}}}, \qquad 
A^{\varphi} = \frac{1}{\sqrt{g_{\varphi\varphi}}},
\end{equation}
\begin{equation}\label{zeta_and_gamma}
\zeta = \sqrt{\frac{g_{\varphi\varphi}}{g_{t\varphi}^2 - g_{tt}g_{\varphi\varphi}}}, 
\qquad 
\gamma = -\frac{g_{t\varphi}}{g_{\varphi\varphi}}\,\zeta . 
\end{equation}

In this tetrad basis, the constants of motion $\{E, p_r, p_\theta, L\}$ of a photon can be expressed in terms of the observation angles $(\alpha,\beta)$ as
\begin{equation}
E = |\vec{P}|\left(\frac{1 + \gamma \sqrt{g_{\varphi\varphi}} \sin\beta \cos\alpha}{\zeta}\right), 
\qquad 
p_r = |\vec{P}|\sqrt{g_{rr}} \cos\beta \cos\alpha ,
\label{Epr}
\end{equation}
\begin{equation}
p_\theta = |\vec{P}|\sqrt{g_{\theta\theta}} \sin\alpha, 
\qquad 
L = |\vec{P}|\sqrt{g_{\varphi\varphi}} \sin\beta \cos\alpha .
\end{equation}

Here, the modulus $|\vec{P}|$ determines only the photon frequency and does not influence the geodesic trajectory. 
Within the backward ray-tracing approach, the apparent image of the accretion disk -- including the characteristic deformations produced by the central compact object -- is reconstructed by numerically integrating the null geodesic equations for each pixel of the observer's image plane. 
The initial conditions are set by the angles $(\alpha,\beta)$, while the outcome of each photon trajectory determines the pixel's color according to the normalized observed flux.

To assign a geometrical position to each pixel of the observer's screen, we introduce the circumferential distance at the observer's location, defined through the azimuthal metric component as
\vspace{-1mm}
\begin{equation}
\tilde r_{O} = \frac{1}{2\pi}\int_{0}^{2\pi}d\phi\sqrt{g_{\phi\phi}(r_{O},\theta_{O})}=r_{O}\,e^{F_{2}(r_{O})},
\end{equation}
which represents the proper radius of a circle centered on the symmetry axis. This definition is consistent with the shadow formalism adopted in Ref. \cite{Gyulchev2024}.

The observer is positioned at $(r_{O},\theta_{O})$ and views a two--dimensional flat screen described by Cartesian coordinates $(x,y)$. 
These coordinates are related to the local angles $(\alpha,\beta)$ through
\begin{equation}
x = - \tilde r_{O}\tan\beta, 
\qquad 
y = \tilde r_{O}\sin\alpha ,
\end{equation}
thus establishing a one-to-one mapping between the angular parametrization of the photon momentum and points on the observer's image plane.

\section{Redshift and Observed Flux}

As noted by Luminet~\cite{Luminet_1979}, the bolometric flux detected by a distant observer, $F_{O}$, is reduced relative to the intrinsic source flux, $F_{S}$, by a factor proportional to $(1+z)^{-4}$. This quartic dependence arises from three relativistic effects: the redshift-induced loss of photon energy, the reduced photon arrival rate due to time dilation, and the solid-angle correction from relativistic aberration. In terms of frequency, the emitted value $\nu_{\text{S}}$ is shifted during propagation so that the observer measures
\vspace{-2mm}
\begin{equation}
\nu_{O} = \frac{\nu_{S}}{1+z},
\end{equation}
and, since the photon energy is given by $E=h\nu$ with $h$ denoting Planck's constant, the same factor also governs the ratio of emitted and observed energies, leading to the observed flux suppression.

To express this relation in a metric framework, consider the photon kinematics in the disk rest frame. The photon energy is obtained by projecting its four-momentum $p_{\mu}$ onto the four-velocity $u^{\mu}$ of the emitter,
\vspace{-1mm}
\begin{equation}
E_{S} = -\left(p_{t}u^{t} + p_{\phi}u^{\phi}\right).
\end{equation}
Taking advantage of the photon's impact parameter $\eta \equiv -p_{\phi}/p_{t}$ and using the orbital angular velocity of prograde or retrograde motion in the disk, $\Omega_{\pm} = u^{\phi}/u^{t}$, this reduces to
\begin{equation}
E_{S} = -p_{t}u^{t}(1-\Omega_{\pm}\eta).
\end{equation}
Here, $p_{t}$ and $p_{\phi}$ are conserved along null geodesics, corresponding to the photon's energy and angular momentum at infinity. The temporal component of the emitter's four-velocity follows from the normalization $u^{\mu}u_{\mu}=-1$, giving
\begin{equation}
u^{t} = \left(-g_{tt}-2\Omega_{\pm} g_{t\phi}-\Omega_{\pm}^{2} g_{\phi\phi}\right)^{-1/2}.
\end{equation}

For a locally nonrotating ZAMO observer at finite radius, the photon energy measured in the local frame is
\begin{equation}
E_{O} = -\left(\hat{e}_{(t)}^{\,\mu}p_{\mu}\right) = -(\zeta p_{t} + \gamma p_{\phi}) = -p_{t}(\zeta - \gamma\eta),
\end{equation}
where the coefficients $\zeta$ and $\gamma$ are given by Eq. \eqref{zeta_and_gamma}. Combining these relations yields the general expression for the redshift factor between the emitted and observed photon frequencies \cite{Sengo_2024}:
\begin{equation}\label{1_plus_z}
1+z = \frac{\nu_{S}}{\nu_{O}}=\frac{E_{S}}{E_{O}} =
\left.\left(\frac{1}{\zeta - \gamma \eta}\right)\right|_{O}
\frac{(1-\Omega_{\pm} \eta)}{\sqrt{-g_{tt}-2\Omega_{\pm} g_{t\phi}-\Omega_{\pm}^{2}g_{\phi\phi}}} \, .
\end{equation}
This formula extends the standard redshift relation to arbitrary observer locations, consistently linking local disk emission with the frequency detected at infinity. Hence, the apparent radiation flux is modified by the gravitational redshift, and the intensity $F_{O}$ at each point of the observer's sky is
\vspace{-2mm}
\begin{equation}
    F_{O}(r) = \frac{\tilde{F}(r)}{(1+z)^{4}}.
\end{equation}

\section{Results}

In this section, we present the observational signatures of thin accretion disks surrounding Kerr black holes endowed with synchronized scalar hair. Using the ray-tracing procedure described above, we compute the apparent bolometric flux $F_{O}$ and the corresponding gravitational red-shift and blue-shift distributions $z$ for both prograde and retrograde disk configurations. The analysis focuses on the dependence of the observable quantities on the radius of the horizon $r_{\text{H}}$ and the normalized scalar charge $q$, highlighting the deviations from the standard Kerr spacetime. The results summarized in Tables \ref{tab_KerrBH_SH}$-$\ref{Comparison_Kerr_BH_SH} demonstrate that the presence of a scalar hair significantly alters the near-horizon emission geometry, producing notable amplification or suppression of the apparent luminosity and characteristic shifts in the frequency distribution across the image plane.

\begin{table}[b!]
\centering
\footnotesize
\setlength{\tabcolsep}{5.4pt}
\renewcommand{\arraystretch}{1.4}
\caption{\small Numerical estimates of the maximal apparent energy flux $F_{O}^{\,\max}$ and the corresponding values of the redshift parameter $z$, evaluated at the locations where the observed flux attains its maximum, at orbital radius $r$ for prograde and retrograde thin accretion disks corresponding to different rotating Kerr black hole solutions with synchronized scalar fields. The observer is located at $\tilde{r}_{O} = 200\,M$ and at an inclination angle $\theta_{O} = 4\pi/9$. The grid resolution in the $\alpha$--$\beta$ plane is of order $10^{-4}$. Fluxes are given in units of $\dot{M} \times 10^{-5}$.}
\label{tab_KerrBH_SH}
\begin{tabular}{lcccccccccccc}
\hline\hline
\multicolumn{13}{c}{\textbf{Kerr BH with synchronized scalar hair}} \\ 
\hline
                              & \multicolumn{3}{c}{Outer \emph{prograde} disk} & \multicolumn{3}{c}{Inner \emph{prograde} disk} & \multicolumn{3}{c}{Outer \emph{retrograde} disk} & \multicolumn{3}{c}{Inner \emph{retrograde} disk} \\
\cmidrule(lr){2-4} \cmidrule(lr){5-7} \cmidrule(lr){8-10} \cmidrule(lr){11-13}\\[-20pt]
Solution                 &   $r$   &    $z$    &  $F_{\,O}^{\,\max}$  & $r$   &    $z$   &  $F_{\,O}^{\,\max}$ & $r$   &    $z$   &  $F_{\,O}^{\,\max}$ & $r$   &    $z$   &  $F_{\,O}^{\,\max}$\\
\hline
\textbf{I}$^{\,0}_{\,0.01}$   & 0.494 & \textcolor[rgb]{0.00,0.07,1.00}{$-0.308$} & \textcolor[rgb]{0.00,0.59,0.00}{25148.5}  & 0.199    & \textcolor[rgb]{1.00,0.00,0.00}{2.483}    & \textcolor[rgb]{0.00,0.59,0.00}{432.224}   & 11.04  & \textcolor[rgb]{0.00,0.07,1.00}{$-$0.206} & \textcolor[rgb]{0.00,0.59,0.00}{0.99806} & $\ldots$ & $\ldots$ & $\ldots$    \\
\textbf{II}$^{\,0}_{\,0.01}$  & 0.157 & \textcolor[rgb]{0.00,0.07,1.00}{$-$0.291} & \textcolor[rgb]{0.00,0.59,0.00}{106\,085}  & 0.066    & \textcolor[rgb]{1.00,0.00,0.00}{4.827}    & \textcolor[rgb]{0.00,0.59,0.00}{328.228}    & 8.221  & \textcolor[rgb]{0.00,0.07,1.00}{$-$0.205} & \textcolor[rgb]{0.00,0.59,0.00}{0.96144} & $\ldots$ & $\ldots$ & $\ldots$    \\
\textbf{III}$^{\,0}_{\,0.05}$ & 0.854 & \textcolor[rgb]{0.00,0.07,1.00}{$-$0.227} & \textcolor[rgb]{0.00,0.59,0.00}{7873.90} & 0.375    & \textcolor[rgb]{1.00,0.00,0.00}{4.371}    & \textcolor[rgb]{0.00,0.59,0.00}{65.9684}   & 11.96  & \textcolor[rgb]{0.00,0.07,1.00}{$-$0.206} & \textcolor[rgb]{0.00,0.59,0.00}{1.03941} & $\ldots$ & $\ldots$ & $\ldots$    \\
\textbf{IV}$^{\,0}_{\,0.1}$   & 1.522 & \textcolor[rgb]{0.00,0.07,1.00}{$-$0.257} & \textcolor[rgb]{0.00,0.59,0.00}{2579.35} & $\ldots$ & $\ldots$ & $\ldots$ & 12.34  & \textcolor[rgb]{0.00,0.07,1.00}{$-$0.207} & \textcolor[rgb]{0.00,0.59,0.00}{1.06231} & 0.631    & \textcolor[rgb]{1.00,0.00,0.00}{7.482}    & \textcolor[rgb]{0.00,0.59,0.00}{18.9843}  \\
\textbf{V}$^{\,0}_{\,0.2}$    & 0.669 & \textcolor[rgb]{0.00,0.07,1.00}{$-$0.170} & \textcolor[rgb]{0.00,0.59,0.00}{881.542} & $\ldots$ & $\ldots$ & $\ldots$ & 4.842  & \textcolor[rgb]{0.00,0.07,1.00}{$-$0.235} & \textcolor[rgb]{0.00,0.59,0.00}{56.9128} & $\ldots$ & $\ldots$ & $\ldots$ \\
\textbf{VI}$^{\,0}_{\,0.3}$   & 0.958 & \textcolor[rgb]{0.00,0.07,1.00}{$-$0.280} & \textcolor[rgb]{0.00,0.59,0.00}{87.2932} & $\ldots$ & $\ldots$ & $\ldots$ & 2.955  & \textcolor[rgb]{0.00,0.07,1.00}{$-$0.204} & \textcolor[rgb]{0.00,0.59,0.00}{2.10851} & $\ldots$ & $\ldots$ & $\ldots$ \\
\hline\hline
\end{tabular}
\end{table}

\begin{table}[h!]
\centering
\footnotesize
\setlength{\tabcolsep}{19.9pt}
\renewcommand{\arraystretch}{1.2}
\caption{\small Numerical estimates of the maximal apparent energy flux $F_{O}^{\,\max}$ and the corresponding value of the redshift parameter $z$, evaluated at the location where the observed flux attains its maximum, at orbital radius $r$ for prograde and retrograde thin accretion disks, for different values of the dimensionless spin parameter $a/M$ of a Kerr black hole. The spin parameter $a$ and the ADM mass $M$ correspond to the configurations listed in Table~\ref{tab_KerrBH_SH}. The observer is located at $\tilde{r}_{O} = 200\,M$ and at an inclination angle $\theta_{O} = 4\pi/9$. The grid resolution in the $\alpha$--$\beta$ plane is $3.9 \times 10^{-5}$. The maximal observed flux $F_{O}^{\,\max}$ is expressed in units of $\dot{M} \times 10^{-5}$.}
\label{tab_KerrBH}
\begin{tabular}{ccccccc}
\hline\hline
\multicolumn{7}{c}{\textbf{Kerr BH}} \\ 
\hline
                              & \multicolumn{3}{c}{\emph{Prograde} disk} & \multicolumn{3}{c}{\emph{Retrograde} disk} \\
\cmidrule(lr){2-4} \cmidrule(lr){5-7}\\[-20pt]
$|a/M|$                 &   $r$   &    $z$    &  $F_{\,O}^{\,\max}$  & $r$   &    $z$   &  $F_{\,O}^{\,\max}$\\
\hline
0.999984  & 1.136 & \textcolor[rgb]{0.00,0.07,1.00}{$-0.365$} & \textcolor[rgb]{0.00,0.59,0.00}{9035.86}  & 13.82  & \textcolor[rgb]{0.00,0.07,1.00}{$-0.204$} & \textcolor[rgb]{0.00,0.59,0.00}{0.92779}  \\
0.999970  & 1.162 & \textcolor[rgb]{0.00,0.07,1.00}{$-0.360$} & \textcolor[rgb]{0.00,0.59,0.00}{8040.10}  & 13.82  & \textcolor[rgb]{0.00,0.07,1.00}{$-0.204$} & \textcolor[rgb]{0.00,0.59,0.00}{0.92781}  \\
0.999665  & 1.268 & \textcolor[rgb]{0.00,0.07,1.00}{$-0.331$} & \textcolor[rgb]{0.00,0.59,0.00}{3937.02} & 13.82  & \textcolor[rgb]{0.00,0.07,1.00}{$-0.204$} & \textcolor[rgb]{0.00,0.59,0.00}{0.92812}  \\
0.998750  & 1.412 & \textcolor[rgb]{0.00,0.07,1.00}{$-0.290$} & \textcolor[rgb]{0.00,0.59,0.00}{2123.94} & 13.82  & \textcolor[rgb]{0.00,0.07,1.00}{$-0.204$} & \textcolor[rgb]{0.00,0.59,0.00}{0.92905}  \\
0.993504  & 1.982 & \textcolor[rgb]{0.00,0.07,1.00}{$-0.289$} & \textcolor[rgb]{0.00,0.59,0.00}{1052.12} & 13.79  & \textcolor[rgb]{0.00,0.07,1.00}{$-0.204$} & \textcolor[rgb]{0.00,0.59,0.00}{0.93441}  \\
0.883338  & 3.488 & \textcolor[rgb]{0.00,0.07,1.00}{$-0.298$} & \textcolor[rgb]{0.00,0.59,0.00}{117.455} & 13.27  & \textcolor[rgb]{0.00,0.07,1.00}{$-0.207$} & \textcolor[rgb]{0.00,0.59,0.00}{1.05776}  \\
\hline\hline
\end{tabular}
\end{table}

\begin{table}[h!]
\centering
\footnotesize
\setlength{\tabcolsep}{19.7pt}
\renewcommand{\arraystretch}{1.2}
\caption{\small Relative difference, $|F^{\,\text{max}}_{\,O(\text{Kerr SH})}-F^{\,\text{max}}_{\,O(\text{Kerr})}|/F^{\,\text{max}}_{\,O(\text{Kerr})}$ in percentages, between the maximal values of the apparent energy flux of the Kerr BH with synchronized scalar hair and Kerr BH, shown in Tables \ref{tab_KerrBH_SH} and \ref{tab_KerrBH} respectively. The $\uparrow$ indicates $F^{\,\text{max}}_{\,O(\text{Kerr SH})}>F^{\,\text{max}}_{\,O(\text{Kerr})}$, while $\downarrow$ indicates $F^{\,\text{max}}_{\,O(\text{Kerr SH})}<F^{\,\text{max}}_{\,O(\text{Kerr})}$. }
\label{Comparison_Kerr_BH_SH}
\begin{tabular}{ccccc}
\hline\hline
\multicolumn{5}{c}{\textbf{Relative flux difference between Kerr BH with synchronized scalar hair and Kerr BH}} \\
\hline
Solution & Outer \textit{prograde} & Inner \textit{prograde} & Outer \textit{retrograde} & Inner \textit{retrograde} \\
\hline
\textbf{I}$^{\,0}_{\,0.01}$   & $\uparrow\,178.3\;\%$ & $\downarrow\,95.22\;\%$ & $\uparrow\,7.574\;\%$ & $\ldots$ \\
\textbf{II}$^{\,0}_{\,0.01}$  & $\uparrow\,1219\;\%$ & $\downarrow\,95.92\;\%$ & $\uparrow\,3.625\;\%$ & $\ldots$ \\
\textbf{III}$^{\,0}_{\,0.05}$ & $\uparrow\,10.00\;\%$ & $\downarrow\,98.32\;\%$ & $\uparrow\,11.99\;\%$ & $\ldots$ \\
\textbf{IV}$^{\,0}_{\,0.1}$   & $\uparrow\,21.44\;\%$ & $\ldots$ & $\uparrow\,14.34\;\%$ & $\uparrow\,1943\;\%$ \\
\textbf{V}$^{\,0}_{\,0.2}$    & $\downarrow\,16.21\;\%$ & $\ldots$ & $\uparrow\,5991\;\%$ & $\ldots$ \\
\textbf{VI}$^{\,0}_{\,0.3}$   & $\downarrow\,25.68\;\%$ & $\ldots$ & $\uparrow\,99.34\;\%$ & $\ldots$ \\
\hline\hline
\end{tabular}
\end{table}

\begin{figure}[h!]
    \centering
    \includegraphics[width=0.985\textwidth]{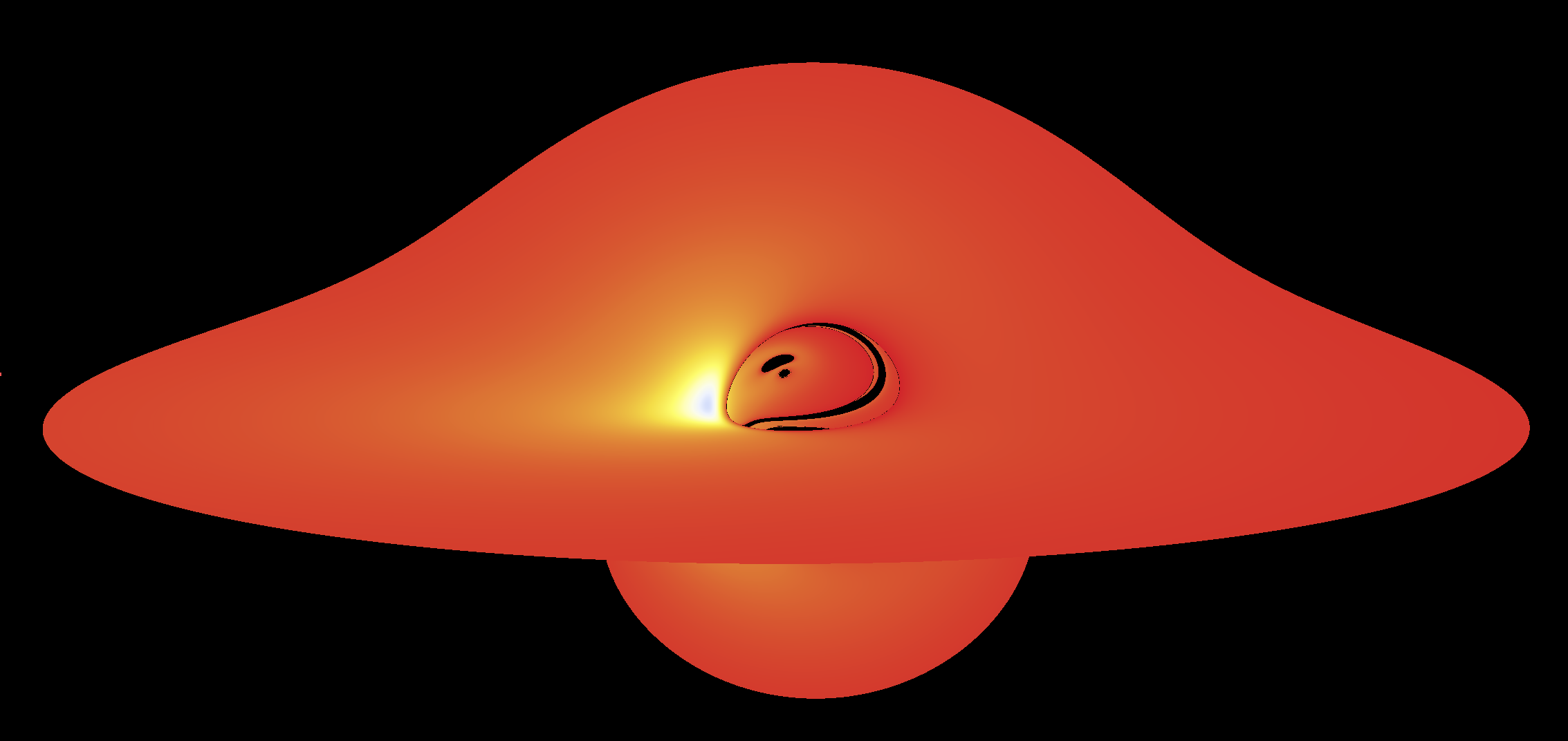}\\ \vspace{0.8mm}
    \includegraphics[width=0.985\textwidth]{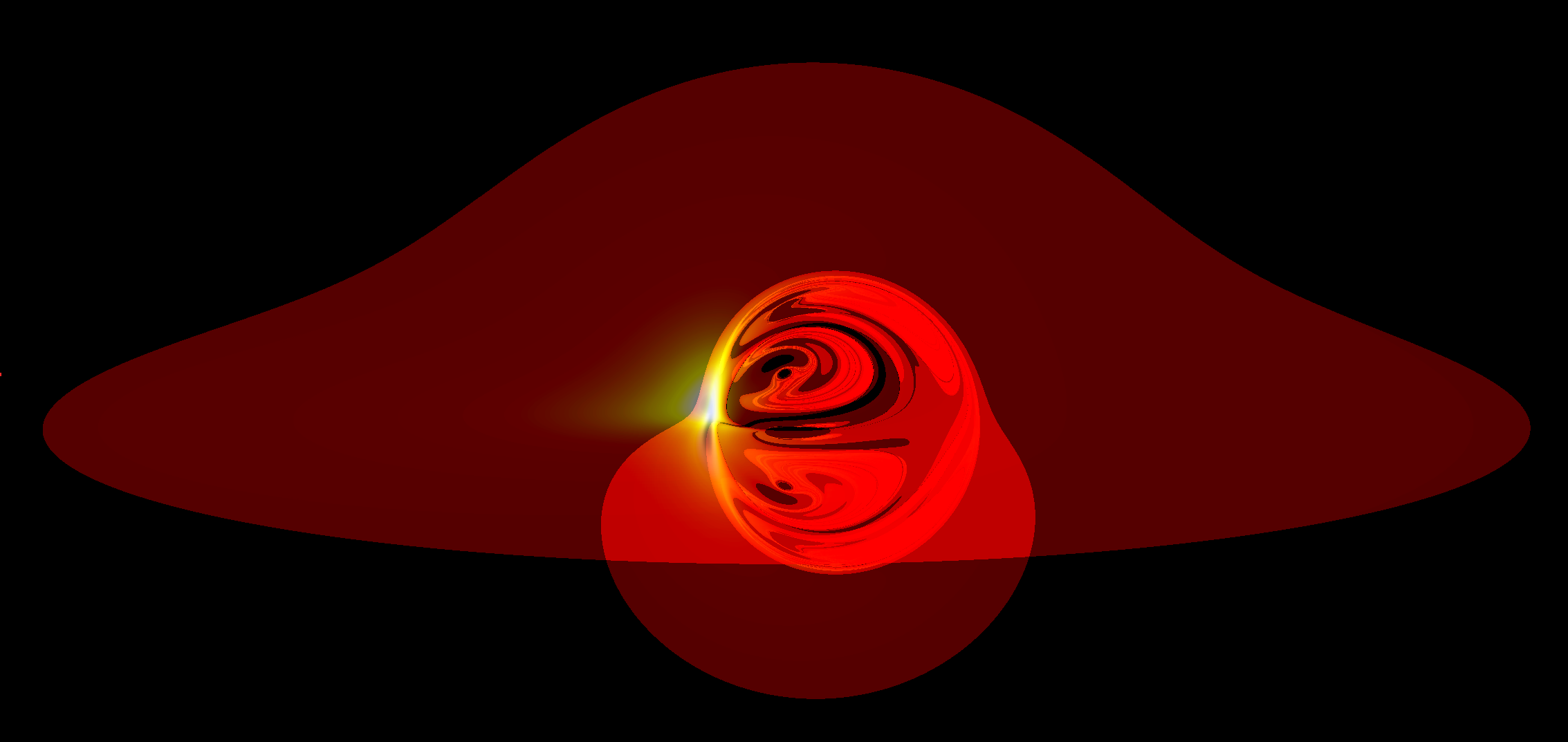}
    \caption{\small Illustrations of the image produced by the outermost region of a prograde-rotating geometrically thin accretion disk around a black hole embedded in a target space with vanishing Gaussian curvature ($\kappa = 0$) and horizon radius $r_H = 0.01$, as viewed by an observer located at a circumferential distance $\tilde r_{O} = 200\,M$ and inclination angle $\theta_{\text{obs}} = 80^\circ$. The images correspond to configuration \textbf{I}$^{\,0}_{\,0.01}$ with parameters $\omega_s/\mu = 0.679241$, $M\mu = 0.881991$, and $q = 0.999875$. The color palette represents the apparent radiation flux $F_{\,O}$ normalized by its maximal value $F_{\,O}^{\,\max}$ listed in Table~\ref{tab_KerrBH_SH}, with the highest flux shown in blue and the lowest in red. A cube-root scaling is used to enhance the visibility of both bright and faint emission regions. The upper panel shows the disk as optically opaque and dominated by direct emission, while the lower panel uses a semi-transparent ($50\%$) rendering that reveals both direct and lensed relativistic images of the disk together with the black hole shadow. Further parameters are given in Table~\ref{tab_7} (Appendix~A). } 
	\label{fig:Disk_P_OUT_I}
\end{figure} 

\subsection{Emission Signatures of the Most Highly Scalarized Solution \texorpdfstring{$\textbf{I}^{\,0}_{\,0.01}$}{Lg}}

Let us consider the first of the most highly scalarized solution, $\textbf{I}^{\,0}_{\,0.01}$, with normalized charge $q=0.999874$, and with almost all the mass ($99.60\%$) and angular momentum ($99.99\%$) stored in the scalar field (Table \ref{tab_7}). In this case, the apparent distribution of the bolometric flux and the associated redshift pattern reveal a strong imprint on the synchronized scalar field, as well as on the geometry of the emission near the event horizon. The accretion disk possesses two distinct prograde rotating components with high emission, one inner and one outer, corresponding to radii $r \simeq 0.199$ and $r \simeq 0.494$, respectively. The maximum observed fluxes of the emitting regions reach $F_{O}^{\,\mathrm{max}} \simeq 432.22 \,\dot{M}\times 10^{-5}$ and $25148.5 \,\dot{M}\times 10^{-5}$, significantly exceeding the expected luminosity of the accretion disk for the Kerr black hole of the corresponding spin. The inner prograde ring, located closer to the black hole, produces the strongest redshift in this configuration, reaching up to $z \simeq 2.483$. 
This enhancement is indicative of strong Doppler boosting combined with pronounced gravitational time dilation in the vicinity of the stable prograde light ring at $r_{+}^{\,\text{LR}_{3}} \simeq 0.244$. In contrast, the outer prograde emission region produces a softer blueshift ($z \simeq -0.308$), consistent with the gravitational blueshift in the receding sector of the disk. 

\begin{figure}[t!]
    \centering
    \includegraphics[width=0.49\textwidth]{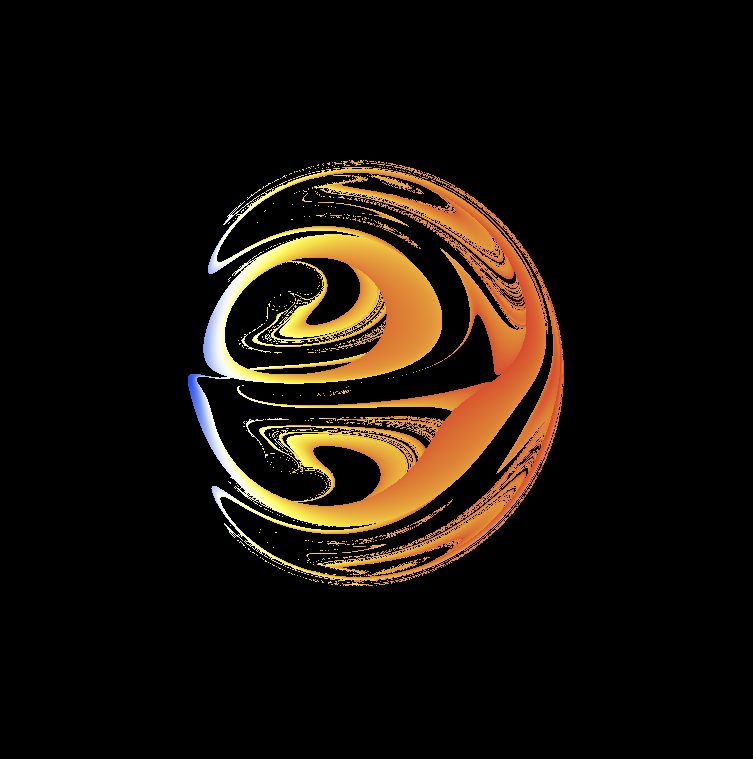}
    \includegraphics[width=0.49\textwidth]{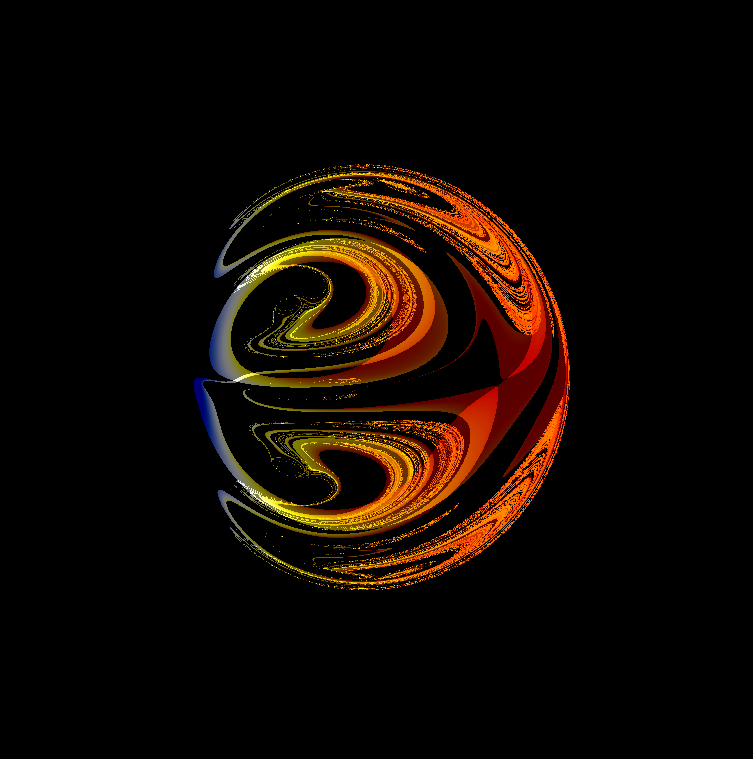}
    \caption{\small Close-up views of the image produced by the innermost region of a prograde-rotating geometrically thin accretion disk around a black hole embedded in a target space with vanishing Gaussian curvature ($\kappa = 0$) and horizon radius $r_H = 0.01$, as viewed by an observer located at a circumferential distance $\tilde r_{\text{obs}} = 200\,M$ and inclination angle $\theta_{\text{obs}} = 80^\circ$. The images correspond to configuration \textbf{I}$^{\,0}_{\,0.01}$, defined by the parameters $\omega_s/\mu = 0.679241$, $M\mu = 0.881991$, and $q = 0.999875$. The color palette and panel rendering follow the same conventions as in Fig.~\ref{fig:Disk_P_OUT_I}. The zoomed-in view highlights fine relativistic features in the vicinity of the black hole shadow. Further parameters are given in Table~\ref{tab_7} (Appendix~A).} 
	\label{fig:Disk_P_IN_I}
\end{figure} 

This pronounced asymmetry between the red-shifted and blue-shifted sides of the image reflects the high degree of frame dragging in the strongly scalarized spacetime and the compact orbital structure characteristic of this configuration. The retrograde component contributes only marginally, with a significantly lower apparent flux ($F_{O}^{\,\mathrm{max}} \simeq 0.99806 \,\dot{M}\times10^{-5}$) and a moderate blueshift ($z \simeq -0.206$). Consequently, the emission morphology deviates substantially from the Kerr black hole case, producing sharper brightness contrasts and a wider dynamical range in frequency shifts-features directly tied to the scalar field distribution and the modified innermost circular orbits.

To assess these features quantitatively, we compare $\mathbf{I}^{\,0}_{\,0.01}$ with a Kerr black hole of identical dimensionless spin, $|a/M| = 0.999984$ (Table \ref{tab_KerrBH}). Relative to the Kerr black hole with the same spin (Table \ref{Comparison_Kerr_BH_SH}), the outer prograde disk of $\mathbf{I}^{\,0}_{\,0.01}$ attains a luminosity that exceeds the Kerr value by $178.3\%$, despite the markedly inward displacement of its emission maximum. The innermost prograde feature, which has no Kerr analogue, is instead considerably less luminous, with its peak flux reduced by about $95.2\%$ compared to the Kerr value. The corresponding frequency shift follows the opposite behavior, with the inner ring exhibiting a substantially enhanced redshift relative to the Kerr prediction, as a direct consequence of the strong gravitational and kinematic effects near the prograde light ring. 

\begin{figure}[t!]
    \centering
    \includegraphics[width=0.985\textwidth]{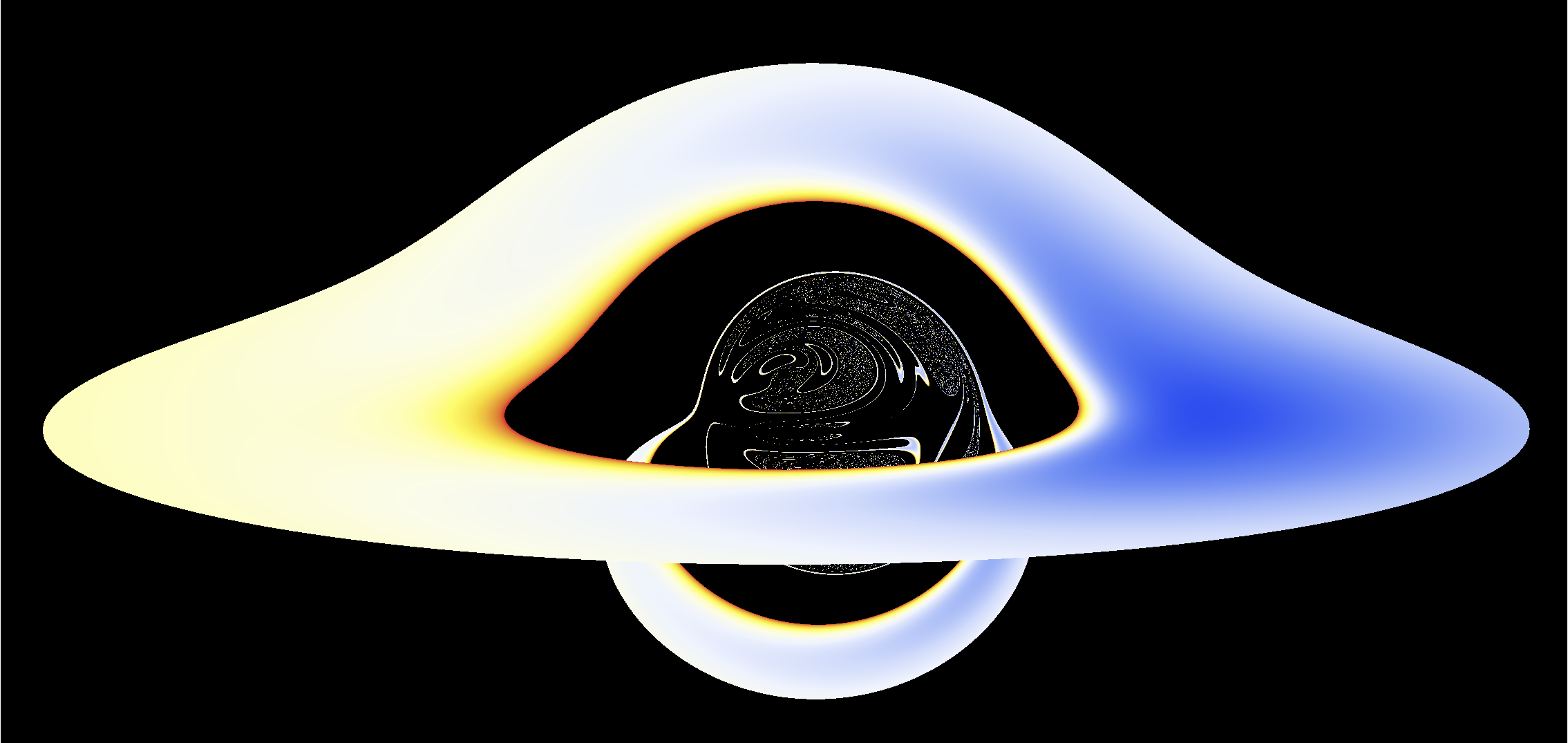}\\ \vspace{0.8mm}
    \includegraphics[width=0.985\textwidth]{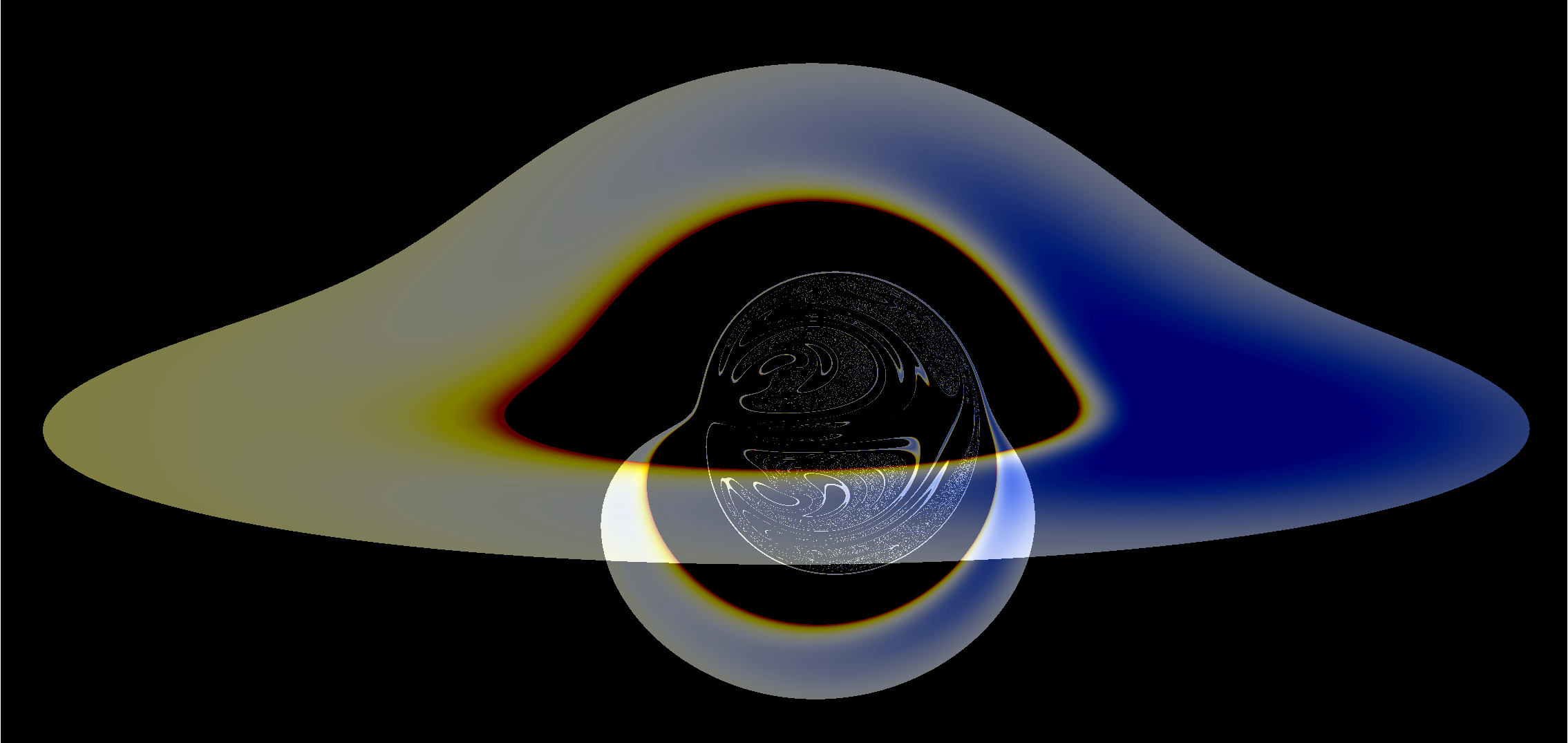}
    \caption{\small Images of a retrograde geometrically thin accretion disk around a black hole for a spacetime with vanishing Gaussian curvature ($\kappa = 0$) and horizon radius $r_H = 0.01$. The system is observed from a circumferential distance $\tilde r_{\text{obs}} = 200\,M$ at an inclination angle $\theta_{\text{obs}} = 80^\circ$. The images correspond to configuration \textbf{I}$^{\,0}_{\,0.01}$ with parameters $\omega_s/\mu = 0.679241$, $M\mu = 0.881991$, and $q = 0.999875$. The color palette and panel rendering follow the same conventions as in Fig.~\ref{fig:Disk_P_OUT_I}, with the black hole shadow fully visible. Further parameters are given in Table~\ref{tab_7} (Appendix~A).} 
	\label{fig:Disk_R_I}
\end{figure} 

For the retrograde disk, the deviations remain modest. The maximal flux, $F_{\mathrm{O}}^{\,\max} \simeq 0.99806\,\dot{M}\times10^{-5}$ at $r \simeq 11.04$, exceeds the Kerr value by only $7.57\%$, while the associated blueshift ($z \simeq -0.21$) remains nearly identical in both spacetimes. Since retrograde orbits lie far from the strongly deformed near-horizon region, their emission retains only weak imprints of the scalar hair. Consequently, the overall radiative appearance of the retrograde disk remains close to that predicted for the Kerr spacetime.

Figures \ref{fig:Disk_P_OUT_I}, \ref{fig:Disk_P_IN_I}, and \ref{fig:Disk_R_I} illustrate the corresponding ray-traced images of the outer prograde, inner prograde, and retrograde thin accretion disks for configuration $\mathbf{I}^{\,0}_{\,0.01}$, constructed by backward integration of the photon Hamiltonian equations in the highly scalarized spacetime. At the center of each image, one clearly observes the shadow of the rotating black hole, whose detailed structure in this class of scalarized geometries was recently investigated in Ref.~\cite{Gyulchev2024}.

\subsection{Emission Signatures of Very Highly Scalarized Solution \texorpdfstring{\textbf{II}$^{\,0}_{\,0.01}$}{Lg}}

\vspace{-2mm}
We now turn to the second configuration in the sequence $\mathbf{II}^{\,0}_{\,0.01}$, characterized by a very high level of scalarization with almost all the mass ($99.23\%$) and angular momentum ($99.73\%$) stored in the scalar field, corresponding to a normalized charge of $q\simeq 0.997292$. In this case, the accretion structure simplifies significantly compared to the highly scalarized solution $\mathbf{I}_{\,0}^{\,0.01}$: the fragmented multi-ring morphology disappears, and the prograde thin disk occupies a single continuous region extending from the prograde ISCO at $r_{+}^{\mathrm{ISCO}_{1}}\simeq 0.0539$ outward to spatial infinity. 

\vspace{-2mm}
\begin{figure}[b!]
    \centering
    \includegraphics[width=0.985\textwidth]{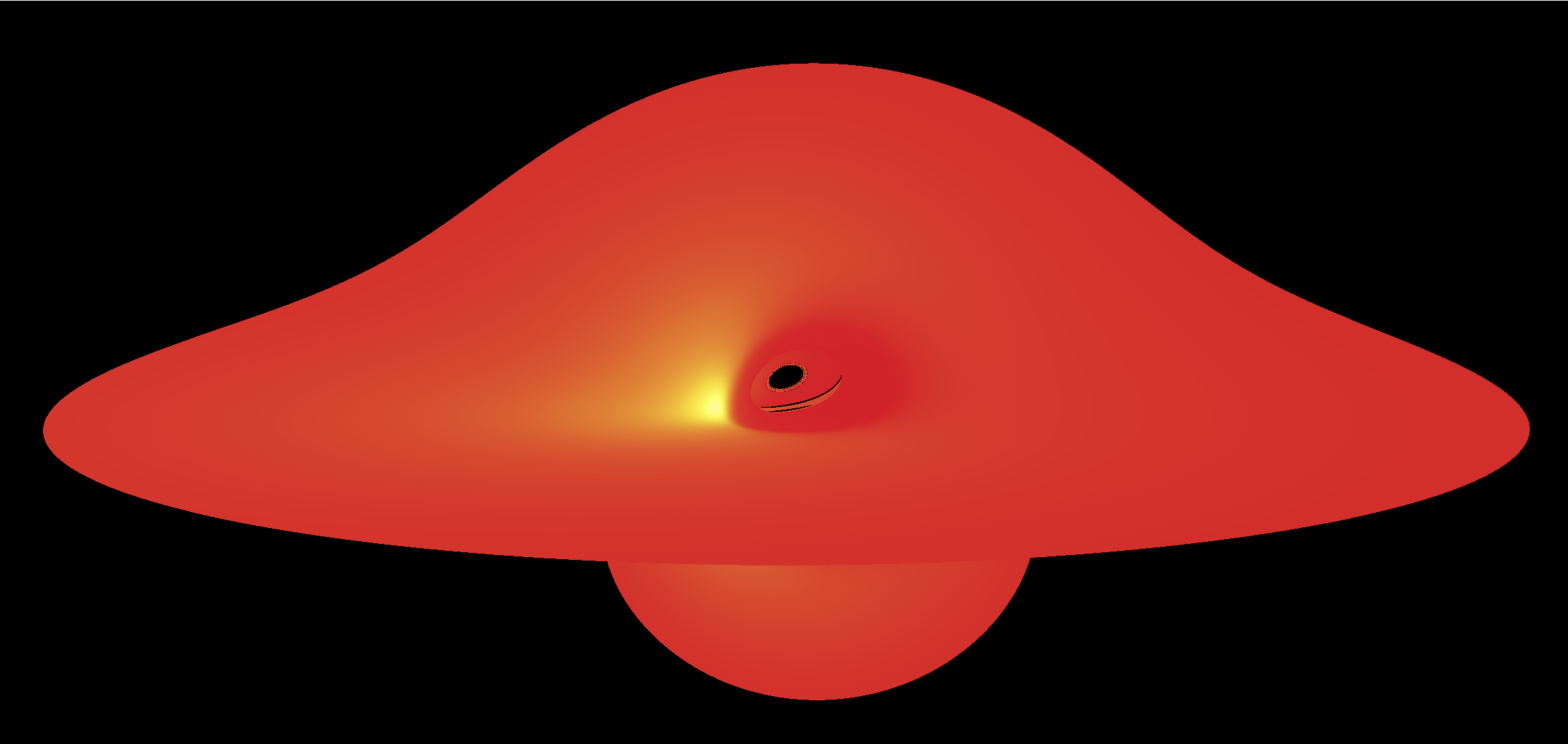}\\ \vspace{0.8mm}
    \includegraphics[width=0.985\textwidth]{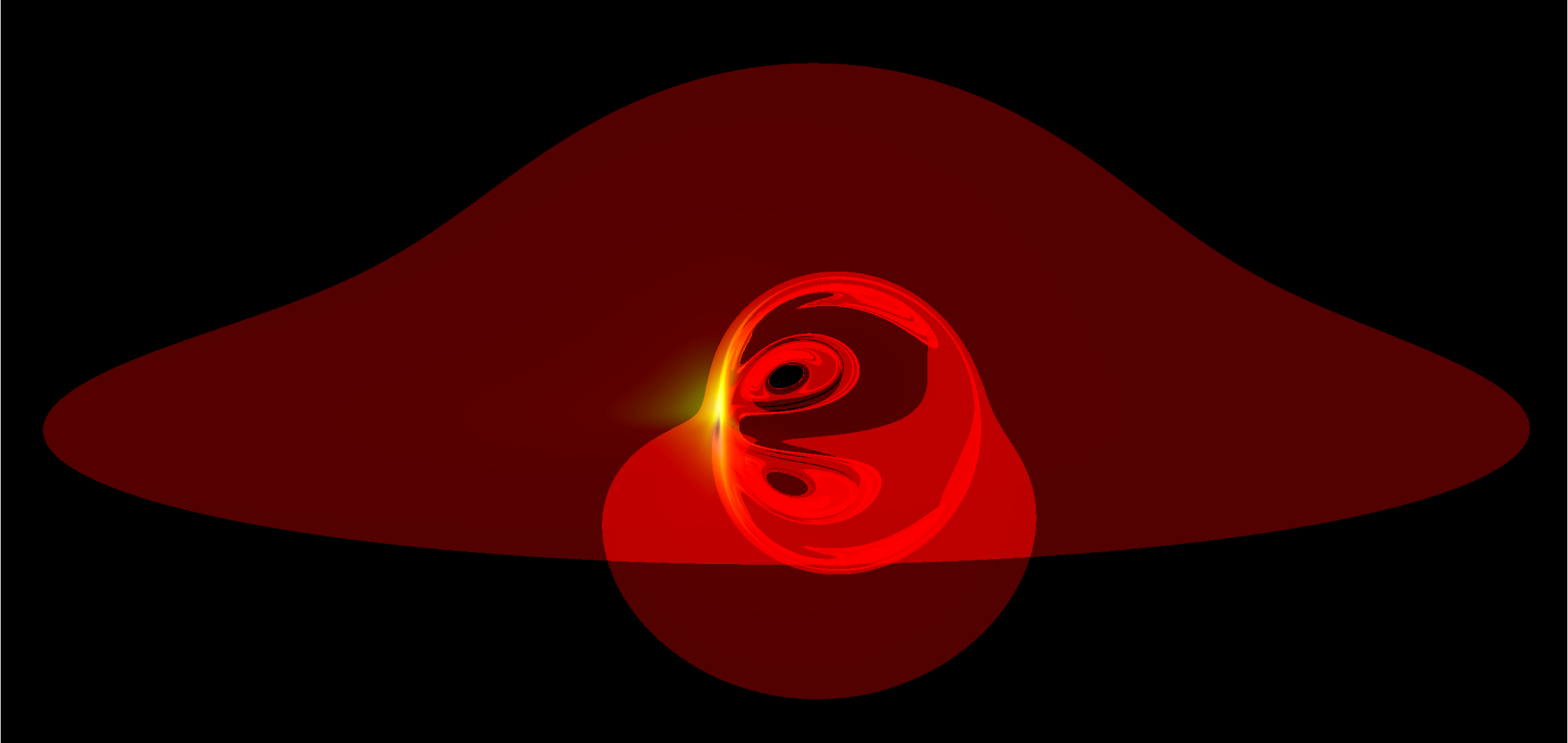}
    \caption{\small Images of the outermost region of a prograde-rotating geometrically thin accretion disk around a black hole are shown for a spacetime with vanishing Gaussian curvature ($\kappa = 0$) and horizon radius $r_H = 0.01$. The system is observed from a circumferential distance $\tilde r_{\text{obs}} = 200\,M$ at an inclination angle $\theta_{\text{obs}} = 80^\circ$. The images correspond to configuration \textbf{II}$^{\,0}_{\,0.01}$ with parameters $\omega_s/\mu = 0.835272$, $M\mu = 0.648229$, and $q = 0.997293$. The color palette and panel rendering follow the same conventions as in Fig.~\ref{fig:Disk_P_OUT_I}, with the black hole shadow fully visible. Additional parameters are listed in Table~\ref{tab_7} (Appendix~A). } 
	\label{fig:Disk_P_OUT_II}
\end{figure} 

The bolometric flux distribution reveals two distinct maxima. The outer prograde emission peak $F_{O}^{\,\max} \simeq 106085 \,\dot{M}\times 10^{-5}$ occurs at $r \simeq 0.157$ and is associated with a modest blueshift of $z\simeq -0.291$, as reported in Table VIII. This flux exceeds the corresponding Kerr black hole value by approximately $1219\%$ (Table X), corresponding to an enhancement by more than one order of magnitude, which results from the inward shift of the emitting region relative to the Kerr reference radius (Table IX).

A second, inner emitting region develops at $r \simeq 0.066$, where the bolometric flux reaches $F_{O}^{\,\max} \simeq 328.228 \,\dot{M}\times 10^{-5}$ and features a large redshift of $z \simeq 4.827$, nearly twice the value obtained for $\mathbf{I}^{\,0}_{\,0.01}$. Such amplification originates from the steep gravitational potential and the rapid variation of the orbital frequency $\Omega_{-}$ near the stable prograde light ring at $r_{+}^{\mathrm{LR}_{3}} \simeq 0.0747$ (Table IV), where the denominator of the frequency-shift expression, given by Eq.~(\ref{1_plus_z}), approaches zero. 

The retrograde disk shows only weak deviations from the Kerr black hole case. Its maximal flux, $F_{O}^{\,\max} \simeq 0.96144 \,\dot{M}\times10^{-5}$ at $r \simeq 8.221$, exceeds the Kerr value by only $\sim 3.63\%$, while the associated blueshift ($z \simeq -0.205$) remains nearly identical to that of the Kerr black hole (Tables VIII--X). Since retrograde orbits reside far from the deformed near-horizon region, their observational properties carry only weak imprints of the synchronized scalar field. 

\begin{figure}[t!]
    \centering
    \includegraphics[width=0.49\textwidth]{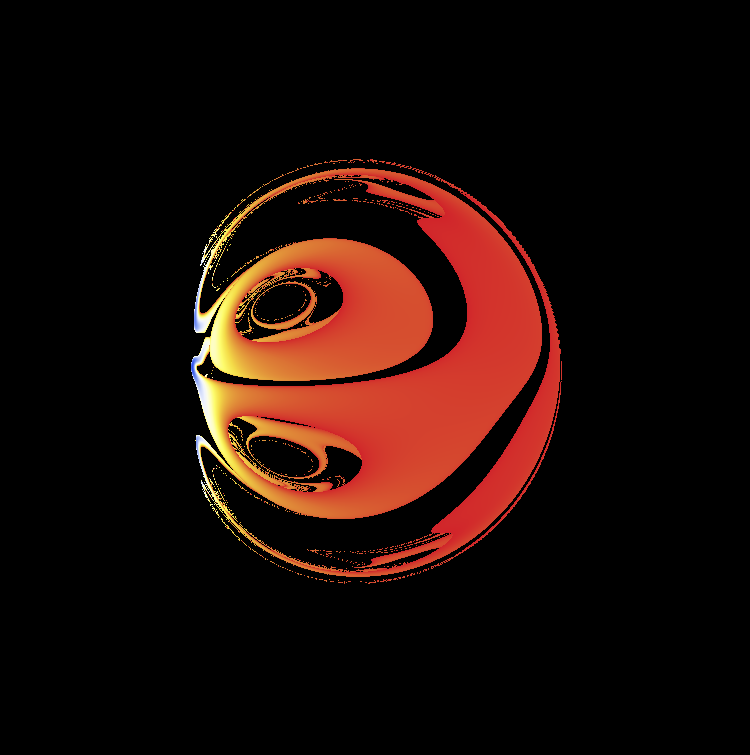}
    \includegraphics[width=0.49\textwidth]{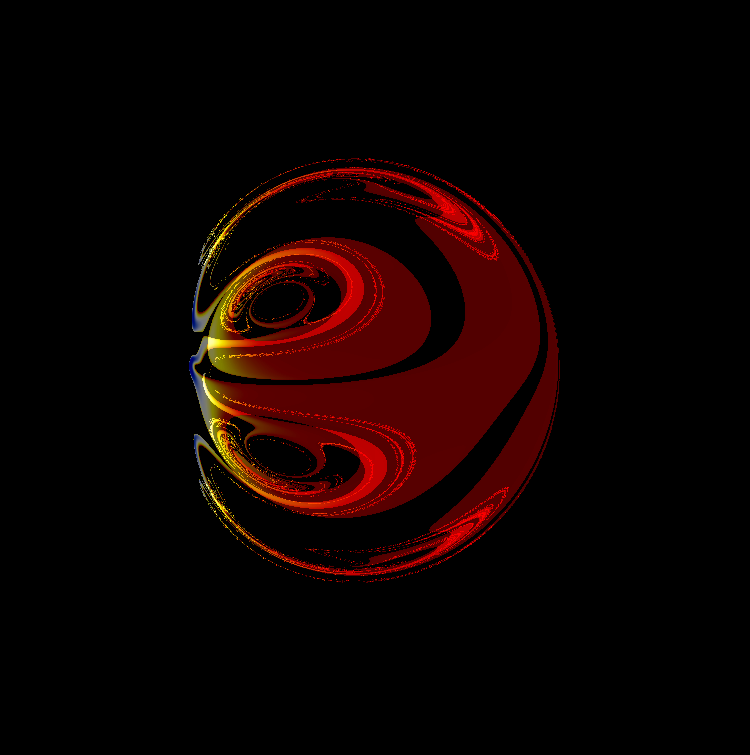}
    \caption{\small Close-up views of the innermost region of a prograde-rotating geometrically thin accretion disk around a black hole with vanishing Gaussian curvature ($\kappa = 0$) and horizon radius $r_H = 0.01$, as observed from a circumferential distance $\tilde r_{\text{obs}} = 200\,M$ at an inclination angle $\theta_{\text{obs}} = 80^\circ$. The images correspond to configuration \textbf{II}$^{\,0}_{\,0.01}$, defined by the parameters $\omega_s/\mu = 0.835272$, $M\mu = 0.648229$, and $q = 0.997293$. The color palette and panel rendering follow the same conventions as in Fig. \ref{fig:Disk_P_OUT_I}. The zoomed-in view highlights fine relativistic features in the vicinity of the black hole shadow. Further physical parameters of this configuration are provided in Table \ref{tab_7} of Appendix A.} 
	\label{fig:Disk_P_IN_II}
\end{figure}

\begin{figure}[t!]
    \centering
    \includegraphics[width=0.985\textwidth]{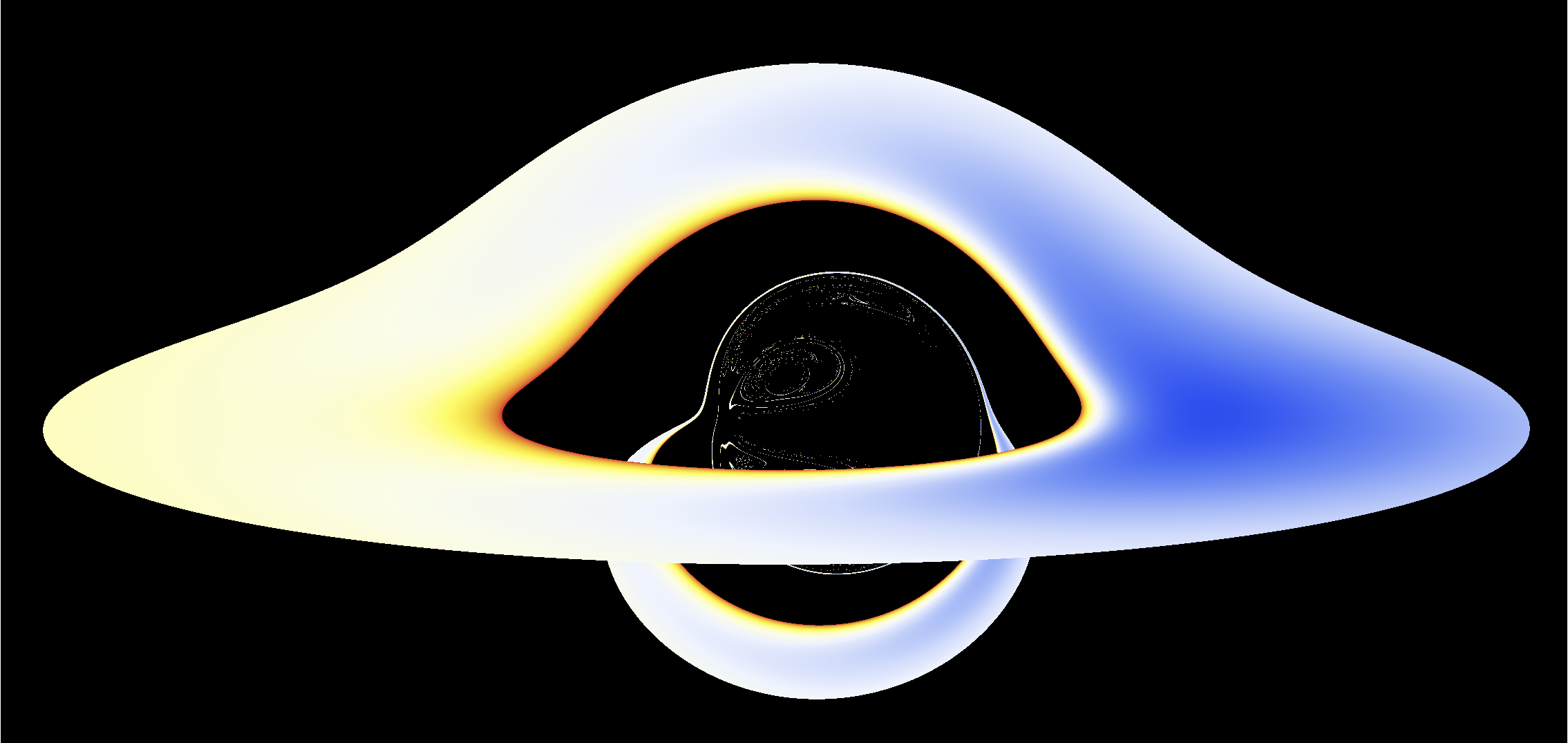}\\ \vspace{0.8mm}
    \includegraphics[width=0.985\textwidth]{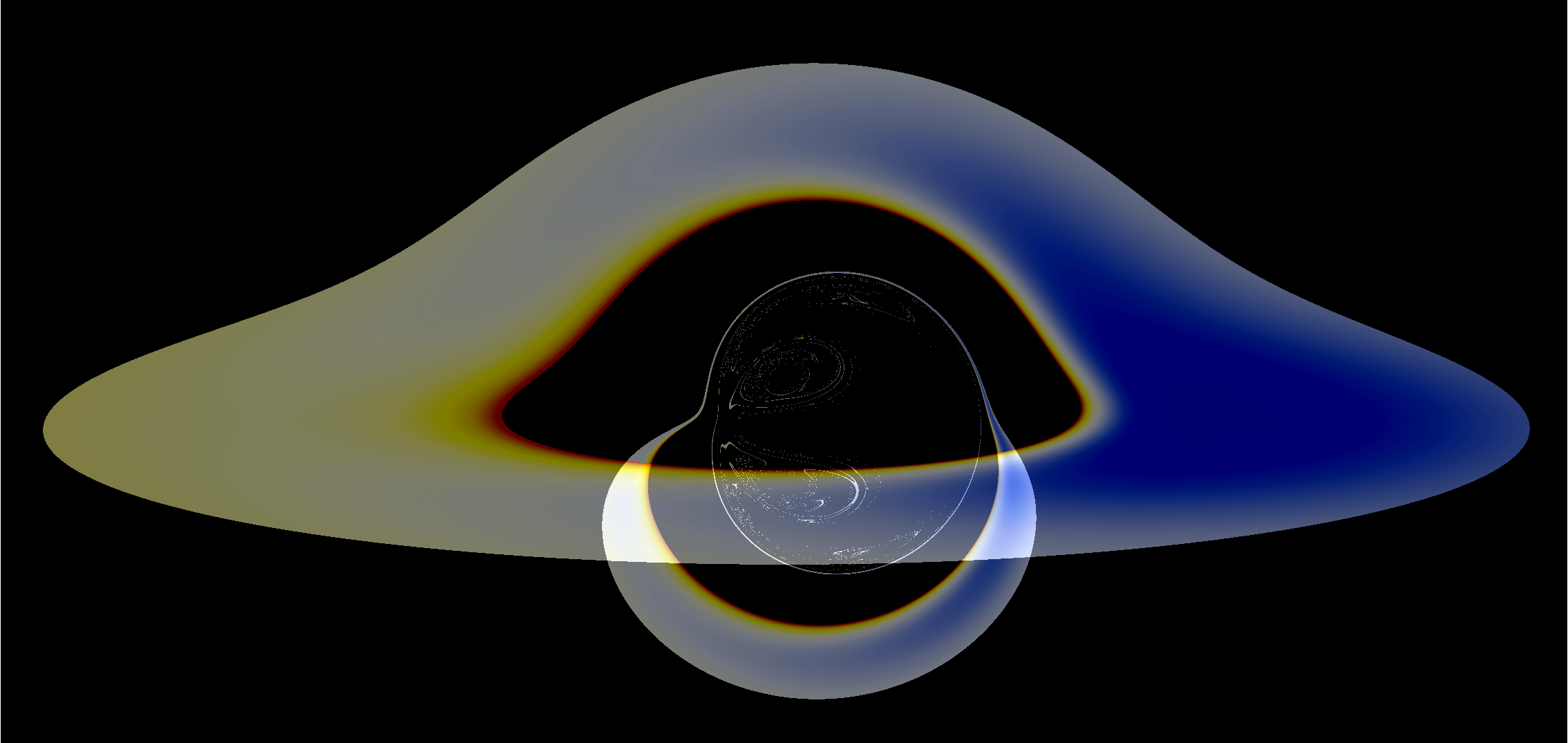}
    \caption{\small Images of a retrograde geometrically thin accretion disk around a black hole with vanishing Gaussian curvature ($\kappa = 0$) and horizon radius $r_H = 0.01$, observed from a circumferential distance $\tilde r_{\text{obs}} = 200\,M$ at an inclination angle $\theta_{\text{obs}} = 80^\circ$. The images correspond to configuration \textbf{II}$^{\,0}_{\,0.01}$ with parameters $\omega_s/\mu = 0.835272$, $M\mu = 0.648229$, and $q = 0.997293$. The color palette and panel rendering follow the conventions of Fig.~\ref{fig:Disk_P_OUT_I}, with the black hole shadow fully visible. Further parameters are given in Table~\ref{tab_7} (Appendix~A).} 
	\label{fig:Disk_R_II}
\end{figure}

Figures \ref{fig:Disk_P_OUT_II}, \ref{fig:Disk_P_IN_II}, and \ref{fig:Disk_R_II} show the optical appearance of the outer and inner prograde components, as well as the retrograde accretion disk, for configuration $\mathbf{II}^{\,0}_{\,0.01}$.

\subsection{Emission Signatures of Highly Scalarized Solution \texorpdfstring{$\mathbf{III}_{\,0.05}^{\,0}$}{Lg}}

We now examine the third configuration in the family, $\mathbf{III}_{\,0.05}^{\,0}$, which represents a highly scalarized state. According to Table \ref{tab_7}, the scalar field still stores a 
dominant fraction of the total mass ($97.40\%$) and angular momentum ($99.45\%$), corresponding to a normalized charge of $q \simeq 0.994490$. Relative to the more strongly scalarized solutions 
$\mathbf{I}^{\,0}_{\,0.01}$ and $\mathbf{II}^{\,0}_{\,0.01}$, the accretion structure becomes increasingly Kerr-like: only a single prograde emitting region remains, which smoothly extends from the ISCO at $r^{\mathrm{ISCO}_{1}}_{+} \simeq 0.2454$ outward to large radii (see Table~\ref{tab:ProgradeTCOs2}).

\begin{figure}[t!]
    \centering
    \includegraphics[width=0.985\textwidth]{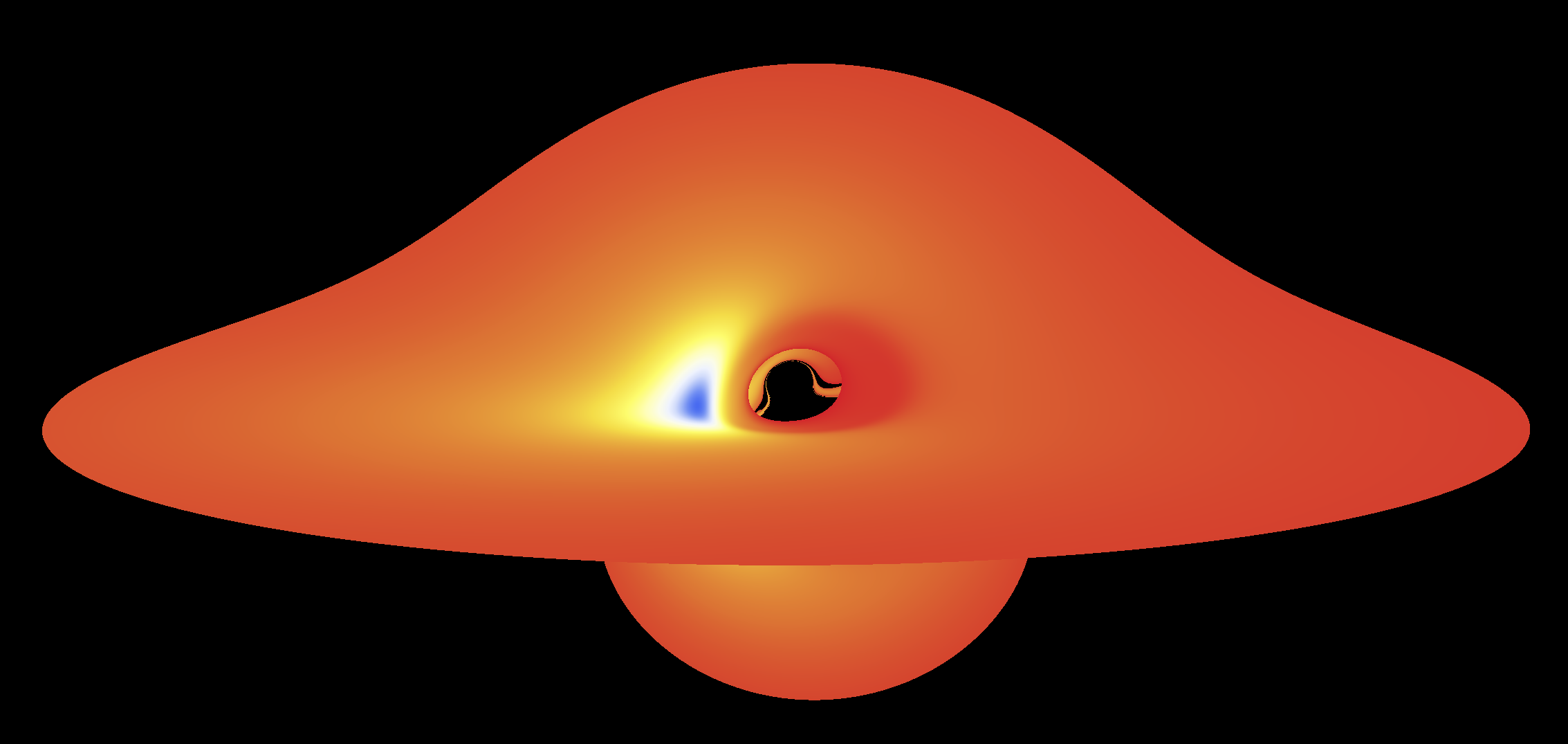}\\ \vspace{0.8mm}
    \includegraphics[width=0.985\textwidth]{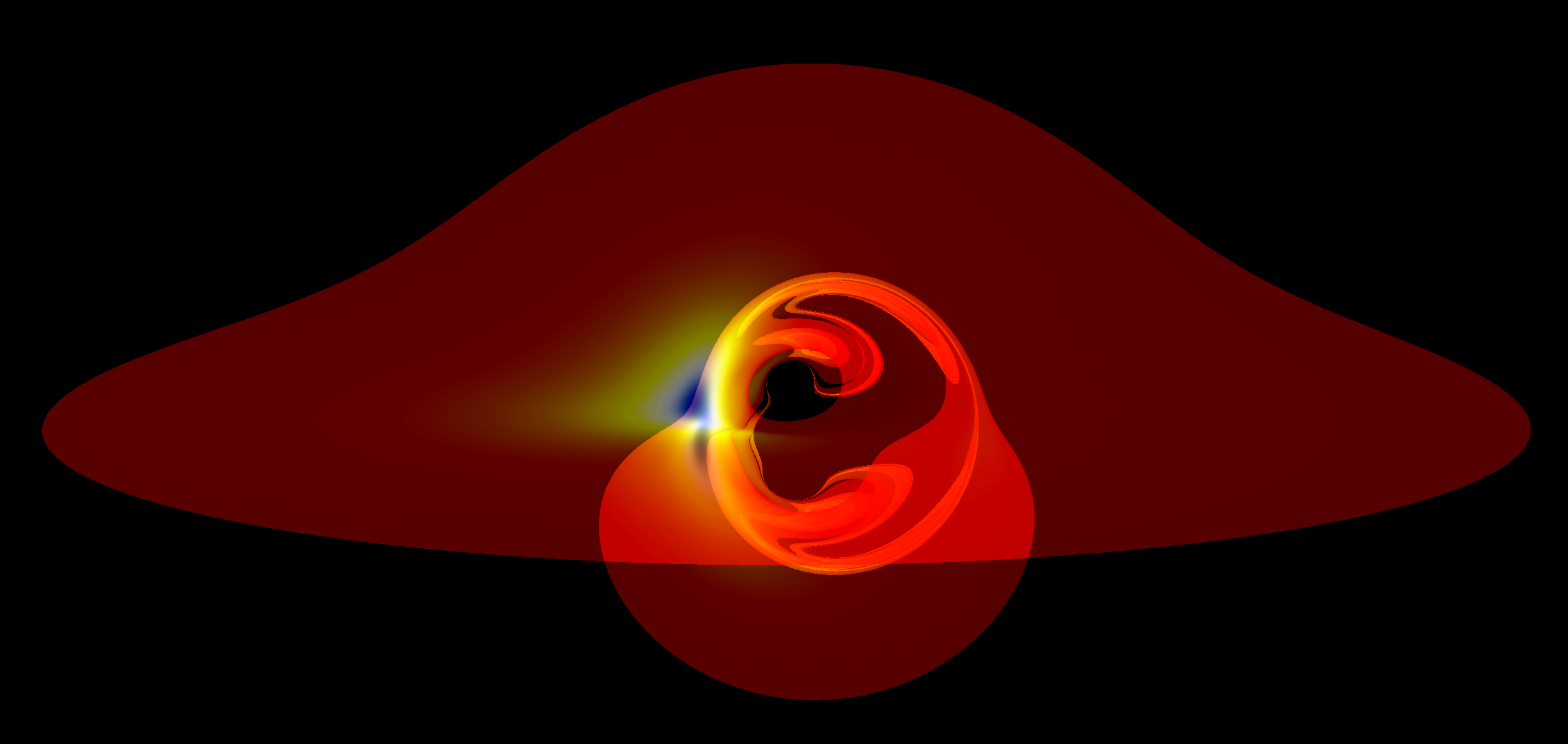}
    \caption{\small Images of the outermost region of a prograde-rotating geometrically thin accretion disk around a black hole with vanishing Gaussian curvature ($\kappa = 0$) and horizon radius $r_H = 0.05$, observed from a circumferential distance $\tilde r_{\text{obs}} = 200\,M$ at an inclination angle $\theta_{\text{obs}} = 80^\circ$. The images correspond to configuration \textbf{III}$^{\,0}_{\,0.05}$ with parameters $\omega_s/\mu = 0.690049$, $M\mu = 0.966595$, and $q = 0.994490$. The color palette and panel rendering follow the conventions of Fig.~\ref{fig:Disk_P_OUT_I}, with the black hole shadow fully visible. Further parameters are given in Table~\ref{tab_7} (Appendix~A).} 
	\label{fig:Disk_P_OUT_III}
\end{figure}

\begin{figure}[t!]
    \centering
    \includegraphics[width=0.49\textwidth]{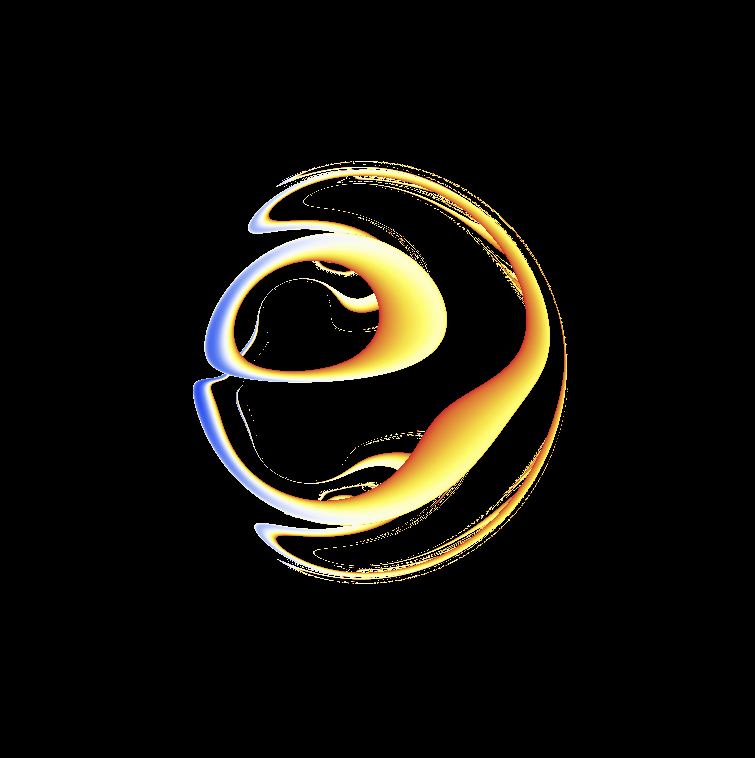}
    \includegraphics[width=0.49\textwidth]{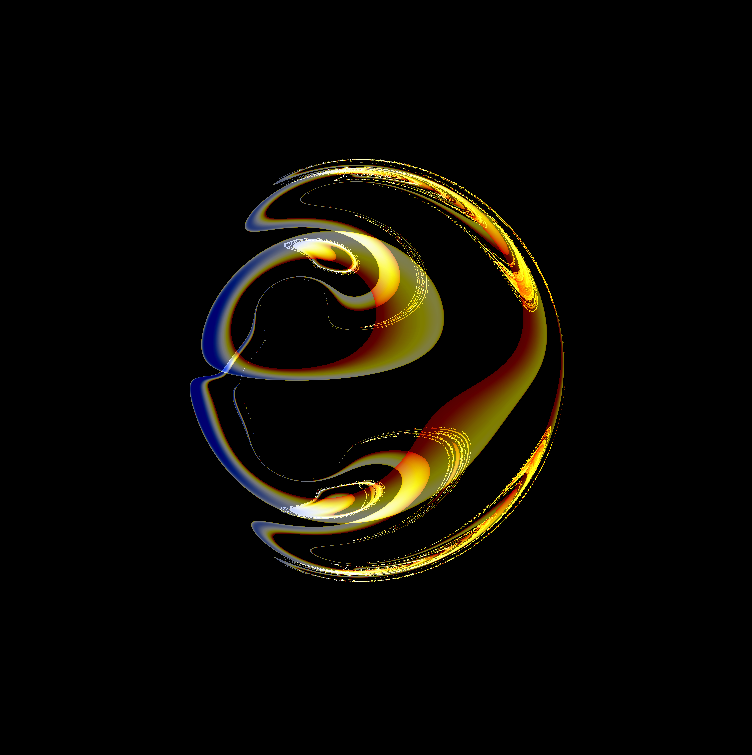}
    \caption{\small Close-up views of the innermost sector of a prograde, geometrically thin accretion disk around a black hole with $\kappa = 0$, horizon radius $r_H = 0.05$, and observer inclination $\theta_{\mathrm{obs}} = 80^\circ$. The panels correspond to configuration \textbf{III}$^{\,0}_{\,0.05}$, defined by the parameters $\omega_s/\mu = 0.690049$, $M\mu = 0.966595$, and $q = 0.994490$. The color palette and panel rendering follow the same conventions as in Fig.~\ref{fig:Disk_P_OUT_I}, with the disk shown opaque in the left panel and semi-transparent in the right panel, revealing both direct and higher-order relativistic images. Further physical parameters of this configuration are provided in Table~\ref{tab_7} of Appendix~A.} 
	\label{fig:Disk_P_IN_III}
\end{figure}

\begin{figure}[t!]
    \centering
    \includegraphics[width=0.985\textwidth]{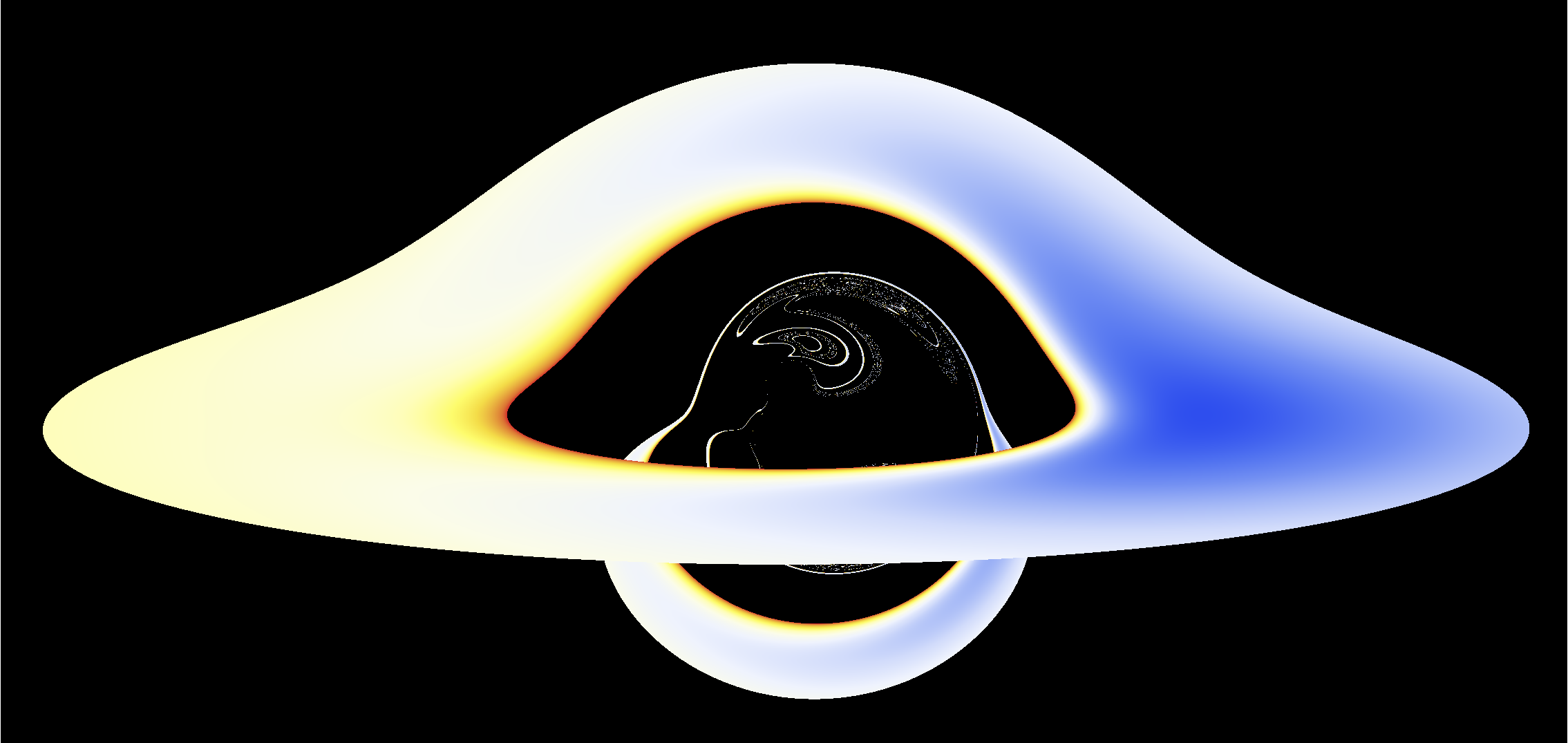}\\ \vspace{0.8mm}
    \includegraphics[width=0.985\textwidth]{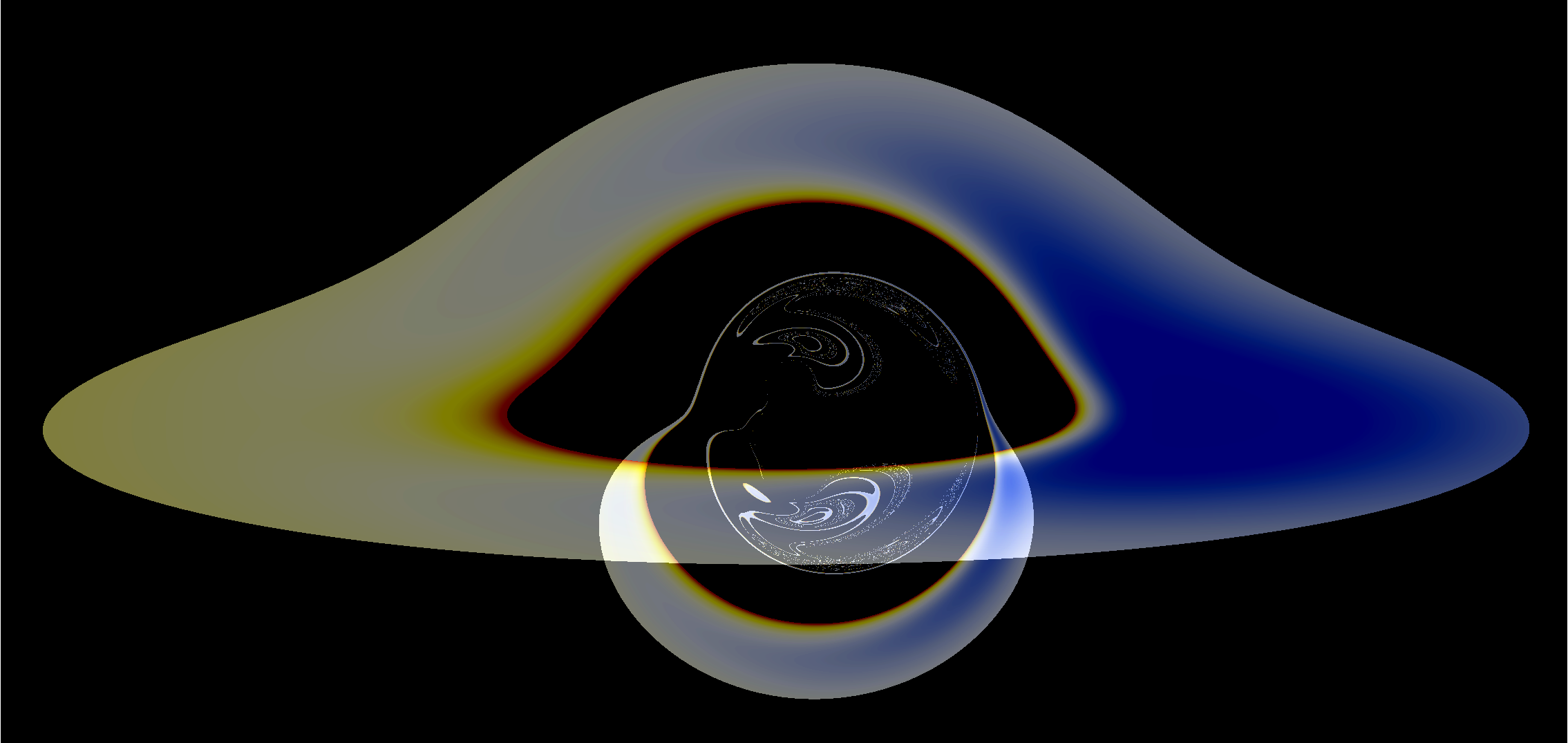}
    \caption{\small Images of a retrograde geometrically thin accretion disk around a black hole with $\kappa = 0$, horizon radius $r_H = 0.05$, and observer inclination $\theta_{\text{obs}} = 80^\circ$. The images correspond to configuration \textbf{III}$^{\,0}_{\,0.05}$ with parameters $\omega_s/\mu = 0.690049$, $M\mu = 0.966595$, and $q = 0.994490$. The color palette and panel rendering follow the conventions of Fig.~\ref{fig:Disk_P_OUT_I}, with the disk shown opaque in the upper panel and semi-transparent in the lower panel, revealing both direct and higher-order relativistic images. Additional parameters are listed in Table~\ref{tab_7} (Appendix~A).} 
	\label{fig:Disk_R_III}
\end{figure} 

The bolometric flux distribution of the prograde disk exhibits a single well--defined maximum, $F_{O}^{\,\max} \simeq 7873.90\,\dot{M}\times10^{-5}$ at $r \simeq 0.854$, accompanied by a moderate blueshift of $z \simeq -0.227$ (Table~\ref{tab_KerrBH_SH}). Compared with a Kerr black hole of the same spin ($|a/M| = 0.999665$, Table~\ref{tab_KerrBH}), the peak flux is enhanced by approximately $10.0\%$ (Table~\ref{Comparison_Kerr_BH_SH}), primarily due to the inward displacement of the emitting radius relative to the Kerr reference configuration.

Analogously to Solutions $\mathbf{I}_{\,0}^{\,0.01}$ and $\mathbf{II}^{\,0}_{\,0.01}$, Solution $\mathbf{III}^{\,0}_{\,0.05}$ also develops a distinct inner prograde radiative component, indicating that scalar-field effects in the strong-field regime remain sufficient to generate a secondary local maximum in the flux profile. This inner region is characterized by a pronounced redshift of $z \simeq 4.371$, which is substantially larger than in the near-extremal Kerr case, reflecting the combined action of gravitational time dilation and relativistic kinematic effects near the prograde light ring at $r_{+}^{\mathrm{LR}_{3}} \simeq 0.4260$.

The strong redshift is accompanied by a substantial suppression of the inner observed flux, leading to a reduction of the peak inner luminosity by approximately $98.3\%$ relative to the Kerr value 
(Table~\ref{Comparison_Kerr_BH_SH}). This suppression reflects the strong gravitational damping of the observed emission in the vicinity of the light ring, where the frequency-shift factor significantly reduces the received flux.

The retrograde disk of $\mathbf{III}_{\,0.05}^{\,0}$ shows only a modest deviation from the Kerr behavior. Its maximal luminosity, $F_{O}^{\,\max} \simeq 1.03941\,\dot{M}\times10^{-5}$ at $r \simeq 11.96$, exceeds the Kerr value by approximately $12.0\%$ (Table~\ref{Comparison_Kerr_BH_SH}), while the associated blueshift, $z \simeq -0.206$, remains nearly identical to that of the Kerr black hole. Since retrograde orbits are located far from the scalar-modified near-horizon region, their observable properties retain only weak sensitivity to the presence of synchronized scalar hair. 

Figures \ref{fig:Disk_P_OUT_III}, \ref{fig:Disk_P_IN_III}, and \ref{fig:Disk_R_III} show the ray-traced appearance of the prograde and retrograde thin accretion disks for configuration $\mathbf{III}_{\,0.05}^{\,0}$.

\subsection{Emission Signatures of the Last Highly Scalarized Solution \texorpdfstring{$\mathbf{IV}_{\,0.1}^{\,0}$}{Lg}}

\vspace{-4mm}
We now turn to the last highly scalarized configuration in the sequence, $\mathbf{IV}^{\,0}_{\,0.1}$. According to Table \ref{tab_7}, the scalar field continues to dominate the energy content of the system, storing $91.30\%$ of the total mass and $96.45\%$ of the total angular momentum, which corresponds to a normalized charge of $q \simeq 0.964477$. As the degree of scalarization decreases relative to $\mathbf{I}^{\,0}_{\,0.01}-\mathbf{III}^{\,0}_{\,0.05}$, the accretion structure becomes increasingly Kerr-like. Only a single continuous prograde emitting region survives, extending outward from the ISCO at $r^{\mathrm{ISCO}_{1}}_{+} \simeq 0.2387$ (Table \ref{tab:ProgradeTCOs2}). 

\begin{figure}[t!]
    \centering
    \includegraphics[width=0.985\textwidth]{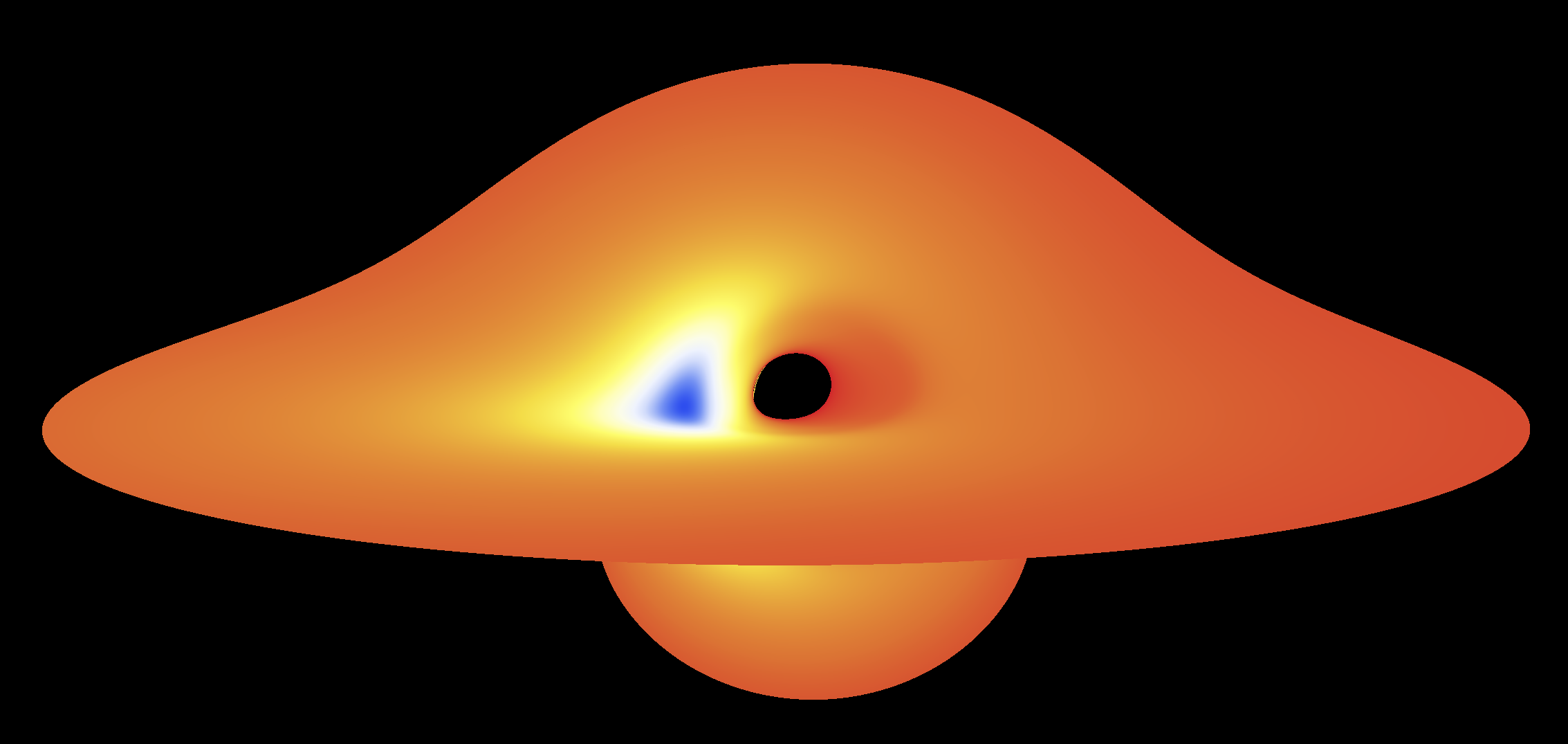}\\ \vspace{0.8mm}
    \includegraphics[width=0.985\textwidth]{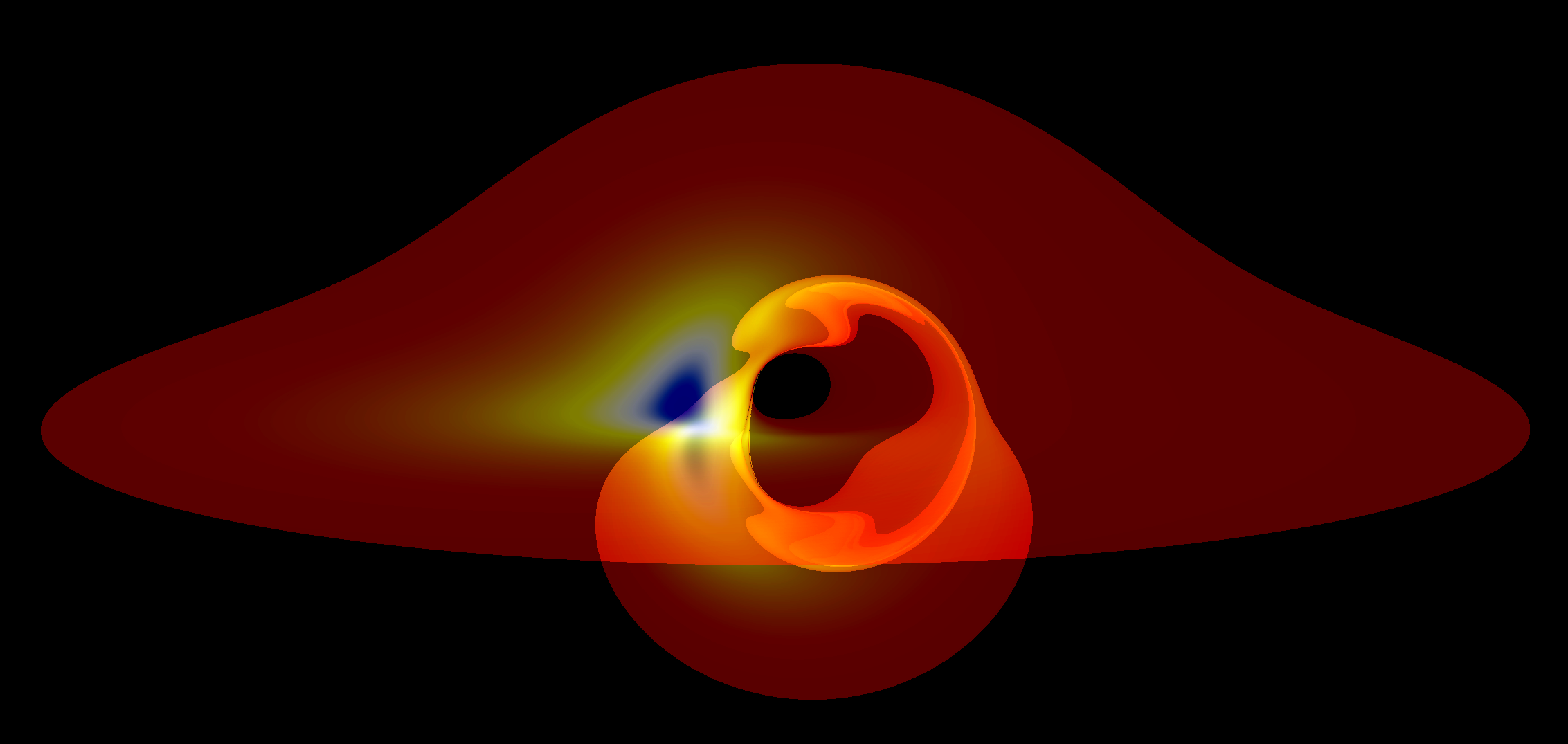}
    \caption{\small Images of a continuous prograde geometrically thin accretion disk around a black hole with vanishing Gaussian curvature ($\kappa = 0$), horizon radius $r_H = 0.1$, and observer inclination $\theta_{\text{obs}} = 80^\circ$. The images correspond to configuration \textbf{IV}$^{\,0}_{\,0.1}$ with parameters $\omega_s/\mu = 0.738499$, $M\mu = 1.00043$, and $q = 0.964477$. The color palette and panel rendering follow the conventions of Fig.~\ref{fig:Disk_P_OUT_I}, with the disk shown opaque in the upper panel and semi-transparent in the lower panel, revealing both direct and higher-order relativistic images. Additional parameters are listed in Table~\ref{tab_7} (Appendix~A).} 
	\label{fig:Disk_P_IV}
\end{figure}  

\begin{figure}[t!]
    \centering
    \includegraphics[width=0.985\textwidth]{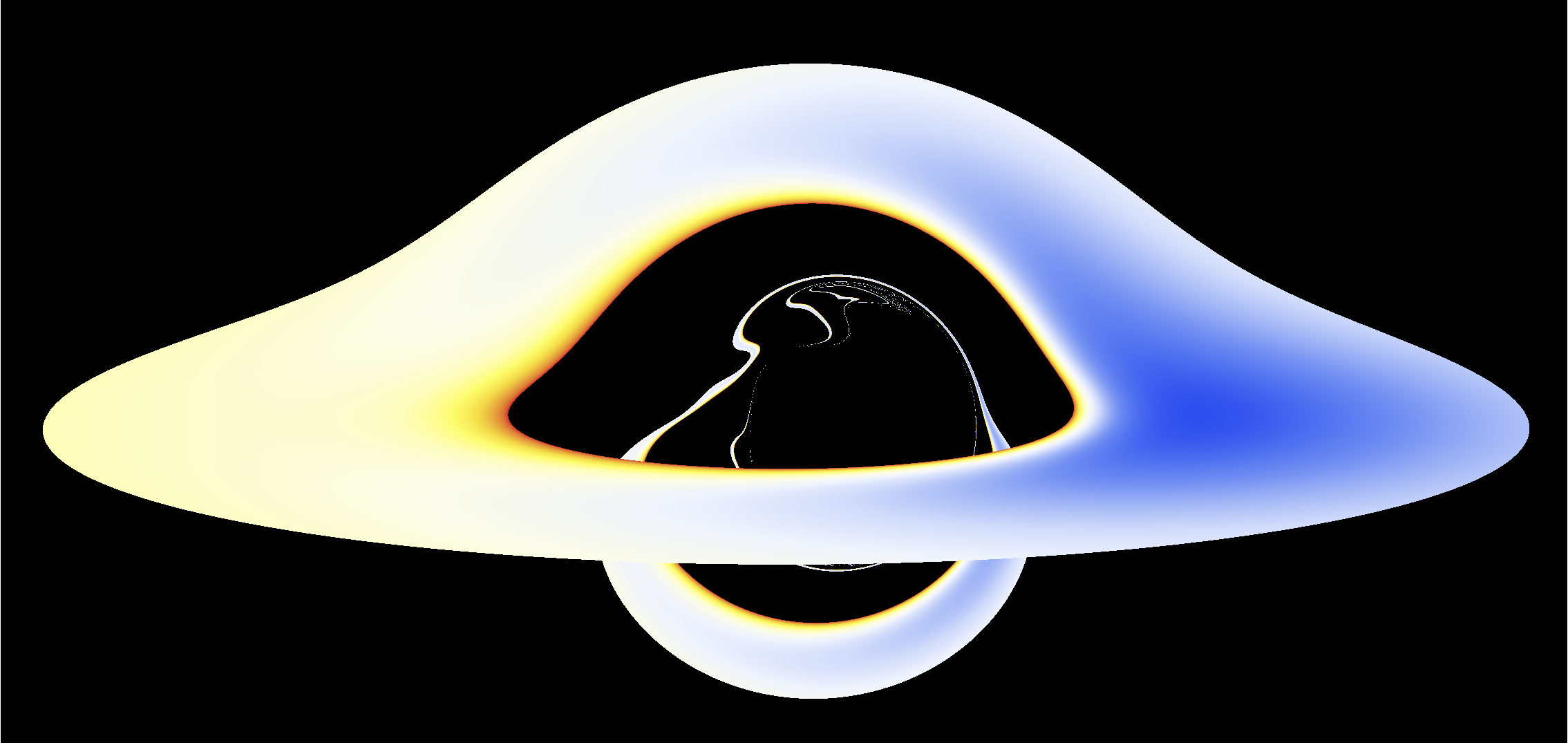}\\ \vspace{0.8mm}
    \includegraphics[width=0.985\textwidth]{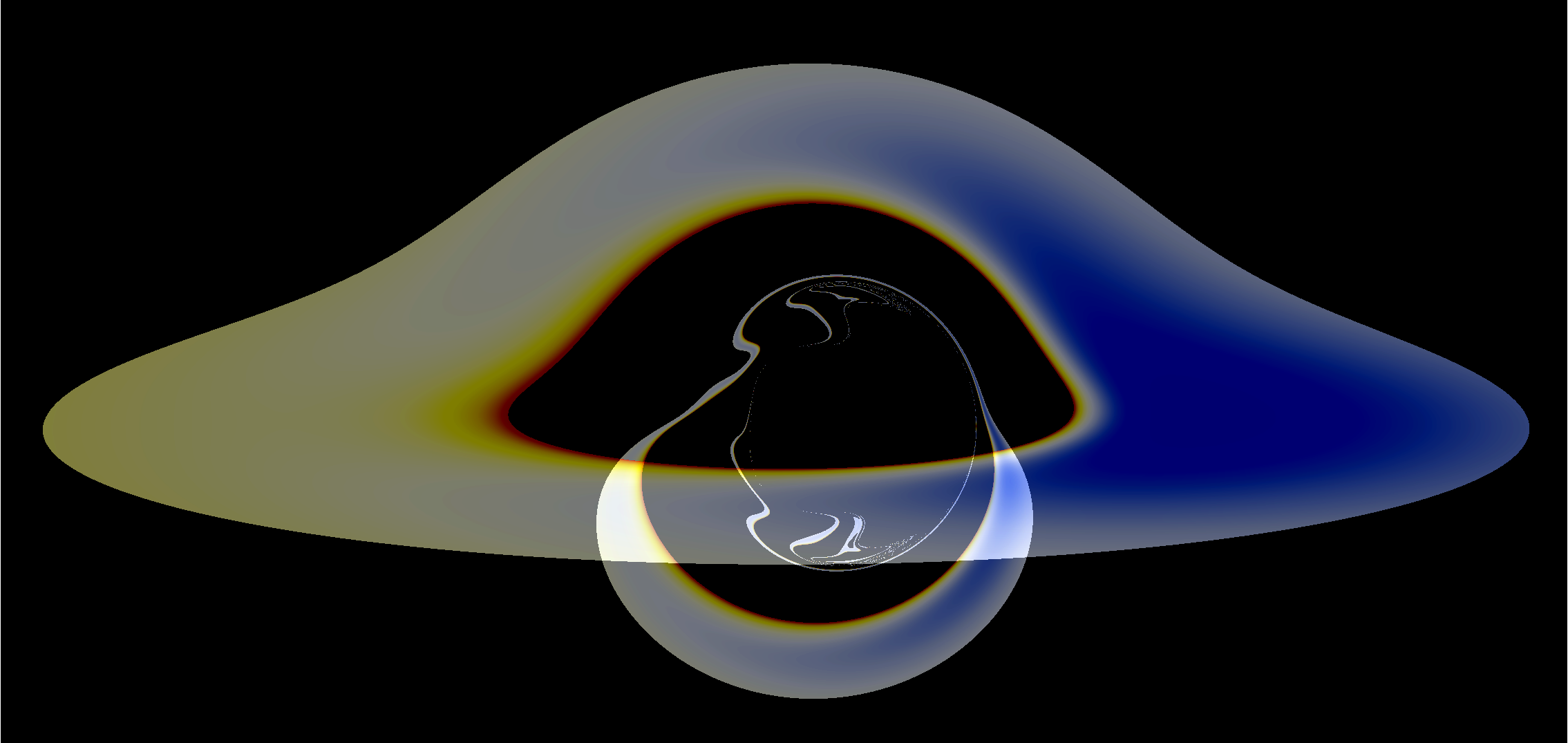}
    \caption{\small Images of the outer region of a retrograde geometrically thin accretion disk around a black hole with vanishing Gaussian curvature ($\kappa = 0$), horizon radius $r_H = 0.1$, and observer inclination $\theta_{\text{obs}} = 80^\circ$. The images correspond to configuration \textbf{IV}$^{\,0}_{\,0.1}$ with parameters $\omega_s/\mu = 0.738499$, $M\mu = 1.00043$, and $q = 0.964477$. The color palette and panel rendering follow the conventions of Fig.~\ref{fig:Disk_P_OUT_I}, with the disk shown opaque in the upper panel and semi-transparent in the lower panel, revealing both direct and higher-order relativistic images. Additional parameters are listed in Table~\ref{tab_7} (Appendix~A).}
    \label{fig:Disk_R_OUT_IV}
\end{figure}

\begin{figure}[t!]
    \centering
    \includegraphics[width=0.49\textwidth]{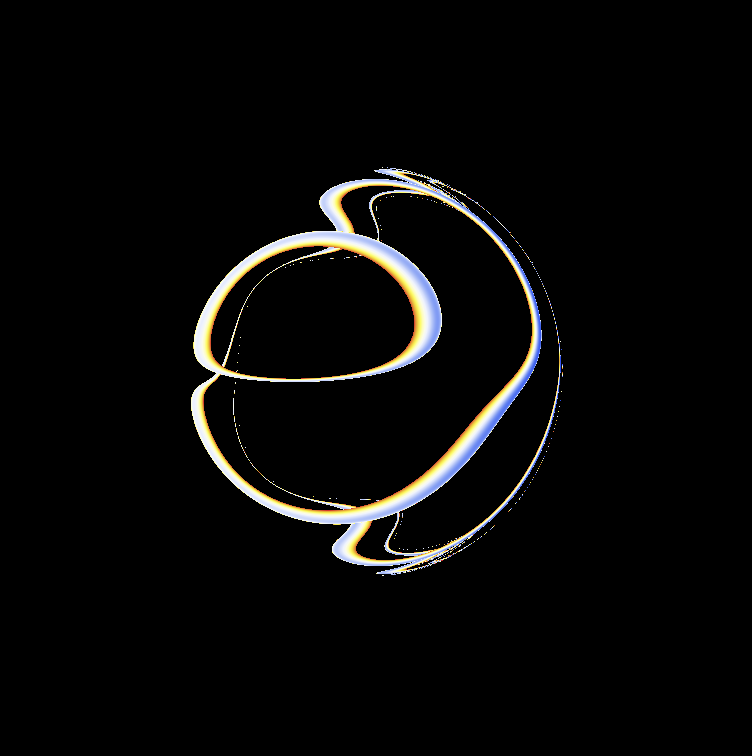}
    \includegraphics[width=0.49\textwidth]{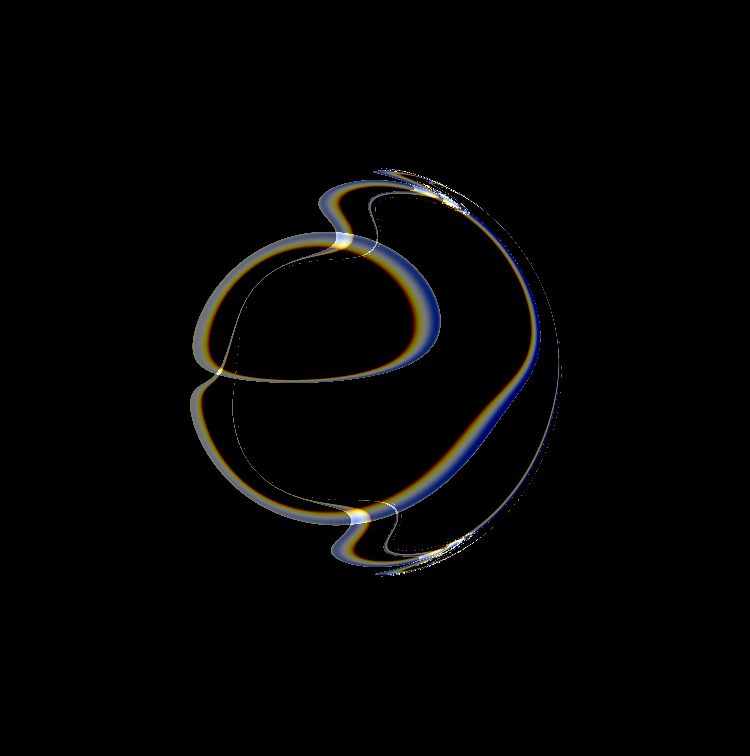}
    \caption{\small Close-up views of the inner region of a retrograde geometrically thin accretion disk around a black hole with vanishing Gaussian curvature ($\kappa = 0$), horizon radius $r_H = 0.1$, and observer inclination $theta_{\text{obs}} = 80^\circ$. The images correspond to configuration \textbf{IV}$^{\,0}_{\,0.1}$, defined by the parameters $\omega_s/\mu = 0.738499$, $M\mu = 1.00043$, and $q = 0.964477$. The color palette and panel rendering follow the same conventions as in Fig.~\ref{fig:Disk_P_OUT_I}, with the disk shown opaque in the left panel and semi-transparent in the right panel, revealing both direct and higher-order relativistic images. Further physical parameters of this configuration are provided in Table~\ref{tab_7} of Appendix~A.} 
	\label{fig:Disk_R_IN_IV}
\end{figure}

The bolometric flux distribution reveals one outer prograde maximum, $F_{O}^{\,\max} \simeq 2579.35 \,\dot{M}\times10^{-5}$ at $r \simeq 1.522$, associated with a mild blueshift of $z \simeq -0.257$ (Table \ref{tab_KerrBH_SH}). When compared with a Kerr black hole of identical spin ($|a/M| = 0.998750$, Table \ref{tab_KerrBH}), the prograde peak flux is reduced by $21.44\%$ (Table \ref{Comparison_Kerr_BH_SH}), primarily reflecting the modified strong-field geometry and the outward displacement of the effective emitting region relative to the Kerr reference configuration. No inner prograde emitting region forms for this configuration, indicating that the strong-field scalar effects responsible for the secondary flux maximum in Solutions $\mathbf{II}^{\,0}_{\,0.01}$ and $\mathbf{III}^{\,0}_{\,0.05}$ have largely diminished at this weaker scalarization level.

A striking new feature of $\mathbf{IV}^{\,0}_{\,0.1}$ emerges in the retrograde sector. Besides the standard outer retrograde peak flux located at $r \simeq 12.34$ with $F_{O}^{\,\max} \simeq 1.06231\,\dot{M}\times10^{-5}$ and a blueshift of $z \simeq -0.207$ (Table \ref{tab_KerrBH_SH}), an additional inner retrograde emitting region appears at $r \simeq 0.631$. This inner retrograde component exhibits a pronounced redshift, $z \simeq 7.482$, and a maximal luminosity of $F_{O}^{\,\max} \simeq 18.9843\,\dot{M}\times10^{-5}$. According to Table \ref{Comparison_Kerr_BH_SH}, this corresponds to an enhancement of approximately $1943\%$, that is, nearly a factor of twenty relative to the Kerr case, which does not support any inner retrograde emission. This feature arises from the intricate structure of retrograde circular orbits revealed in Table~\ref{tab:RetrogradeTCOs3}, where the appearance of a stable retrograde region near $r_{-}^{\mathrm{LR}_{3}}$ enables the formation of this innermost radiative ring.

This feature reflects a qualitative difference between the scalarized spacetime and the Kerr geometry. In the Kerr case retrograde circular orbits remain confined to large radii and therefore cannot support inner retrograde emission. In contrast, the modified structure of retrograde circular orbits in the scalarized solution allows a stable retrograde region to develop in the strong-field regime, giving rise to the innermost retrograde radiative ring identified above. Configuration $\mathbf{IV}^{\,0}_{\,0.1}$ thus provides a clear example of how synchronized scalar hair can introduce additional emitting regions in geometrically thin accretion disks through geometric modifications of the orbital structure. Since no analogous inner retrograde emission exists in the Kerr spacetime, this radiative component represents a potentially distinctive observational feature of scalarized black hole geometries.

The outer retrograde component, by contrast, follows the Kerr prediction more closely. Its luminosity exceeds the Kerr value by $14.34\%$, and the corresponding blueshift is nearly identical to that of the Kerr spacetime (Table \ref{tab_KerrBH}). Since these orbits lie far from the scalar-modified near-horizon region, their observational signatures exhibit only modest deviations from Kerr. 

Figures \ref{fig:Disk_P_IV}, \ref{fig:Disk_R_OUT_IV}, and \ref{fig:Disk_R_IN_IV} present the visual appearance of the prograde and retrograde thin accretion disks for configuration $\mathbf{IV}_{\,0}^{\,0.1}$. The images demonstrate the characteristic light bending in this still highly scalarized geometry and clearly reveal the presence of both the standard outer retrograde disk and the unexpected inner retrograde ring, while the central dark region traces the black hole shadow. The shadow structure for this class of synchronized-hair spacetimes has been discussed in detail in Ref. \cite{Gyulchev2024}.

The behavior observed for configuration $\mathbf{IV}^{\,0}_{\,0.1}$ marks an important transitional stage in the sequence of scalarized solutions. While the prograde disk has already recovered a largely Kerr-like structure, the retrograde sector remains strongly influenced by the modified circular-orbit structure of the scalarized spacetime, as evidenced by the appearance of the inner retrograde radiative ring. This contrast highlights the enhanced sensitivity of counter-rotating orbits to the underlying geodesic structure. It is therefore instructive to examine how these features evolve as the scalarization decreases further and the spacetime approaches the Kerr limit.

\begin{figure}[t!]
    \centering
    \includegraphics[width=0.985\textwidth]{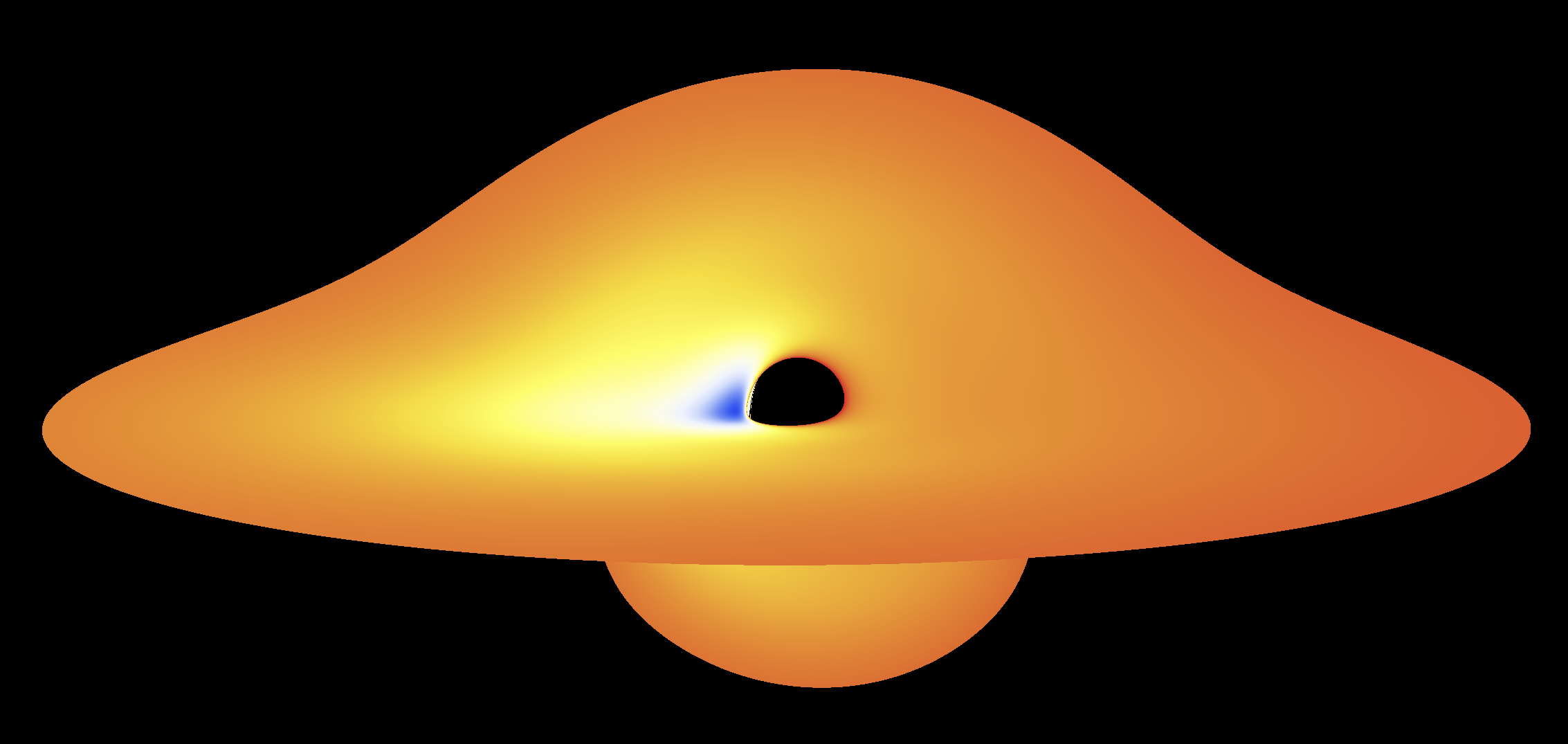}\\ \vspace{0.8mm}
    \includegraphics[width=0.985\textwidth]{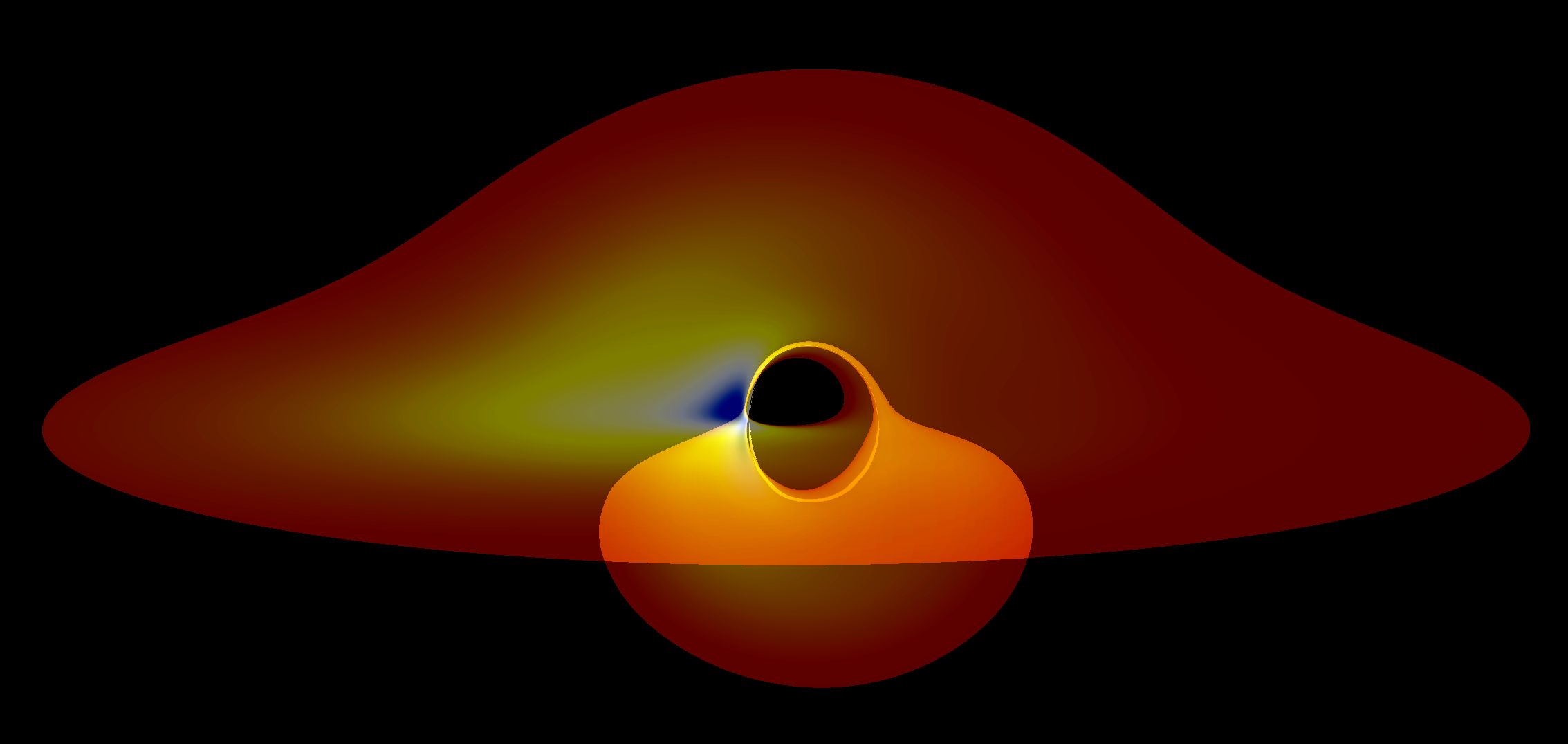}
    \caption{\small Images of a prograde geometrically thin accretion disk around a black hole with vanishing Gaussian curvature ($\kappa = 0$) and horizon radius $r_H = 0.2$, observed from a circumferential distance $\tilde r_{\text{obs}} = 200\,M$ at an inclination angle $\theta_{\text{obs}} = 80^\circ$. The images correspond to configuration \textbf{V}$^{\,0}_{\,0.2}$ with parameters $\omega_s/\mu = 0.895538$, $M\mu = 0.878726$, and $q = 0.850778$. The color palette and panel rendering follow the same conventions as in Fig.~\ref{fig:Disk_P_OUT_I}, with the disk shown opaque in the upper panel and semi-transparent in the lower panel, revealing both direct and higher-order relativistic images. Additional parameters are listed in Table~\ref{tab_7} (Appendix~A). } 
	\label{fig:Disk_P_V}
\end{figure}

\begin{figure}[t!]
    \centering
    \includegraphics[width=0.985\textwidth]{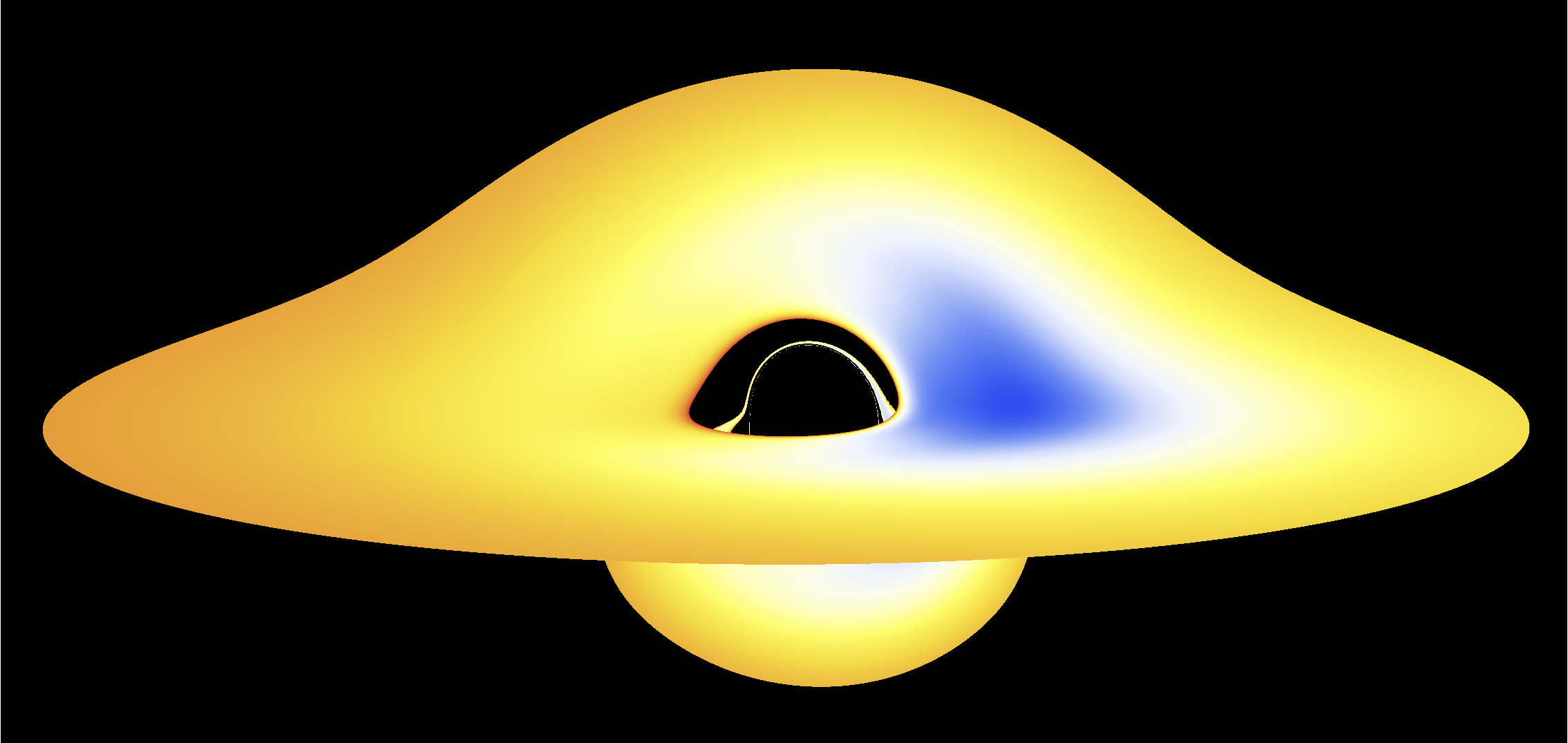}\\ \vspace{0.8mm}
    \includegraphics[width=0.985\textwidth]{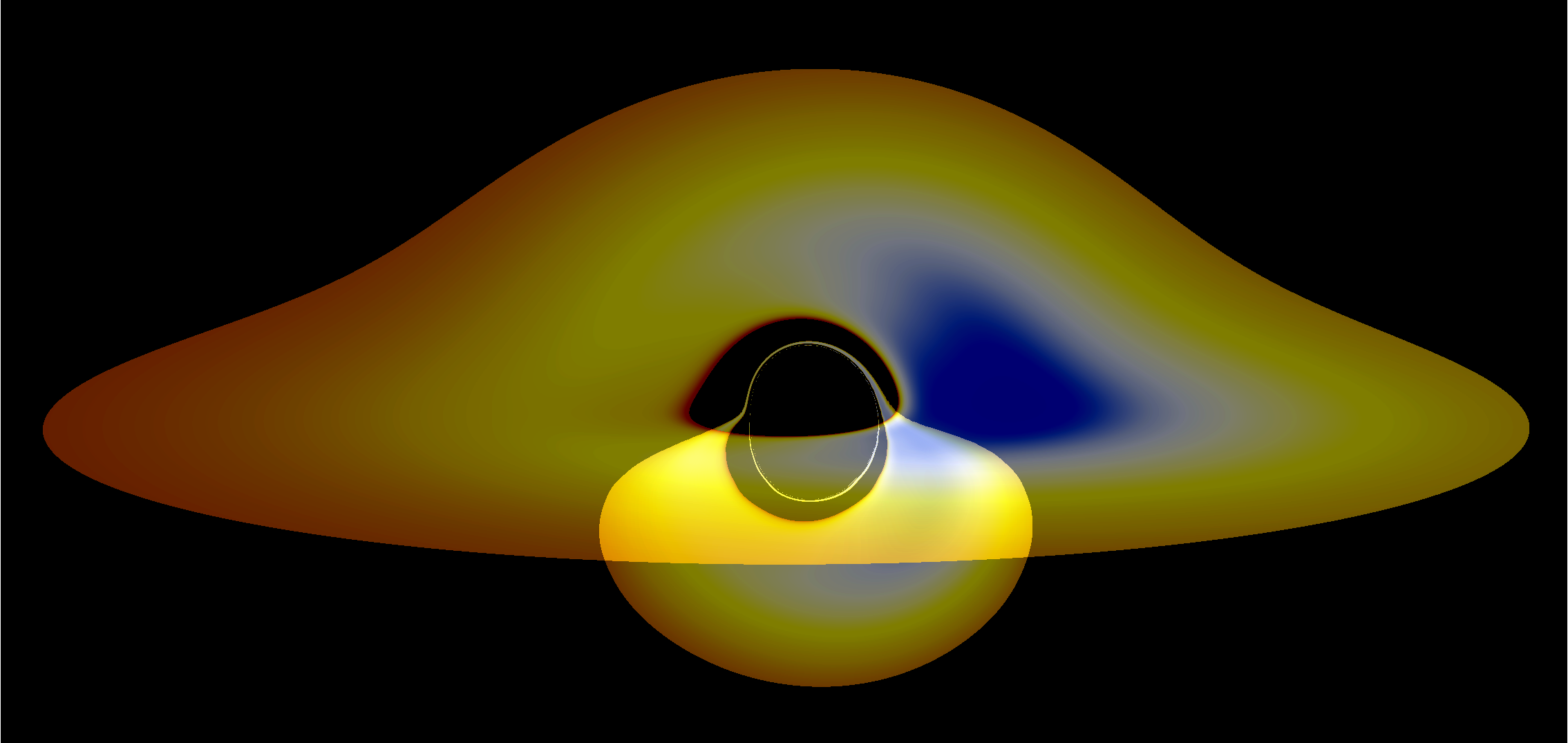}
   \caption{\small Images of a retrograde geometrically thin accretion disk around a black hole with vanishing Gaussian curvature ($\kappa = 0$) and horizon radius $r_H = 0.2$, observed from a circumferential distance $\tilde r_{\text{obs}} = 200\,M$ at an inclination angle $\theta_{\text{obs}} = 80^\circ$. The images correspond to configuration \textbf{V}$^{\,0}_{\,0.2}$ with parameters $\omega_s/\mu = 0.895538$, $M\mu = 0.878726$, and $q = 0.850778$. The color palette and panel rendering follow the same conventions as in Fig.~\ref{fig:Disk_P_OUT_I}, with the disk shown opaque in the upper panel and semi-transparent in the lower panel, revealing both direct and higher-order relativistic images. Additional parameters are listed in Table~\ref{tab_7} (Appendix~A).}
    \label{fig:Disk_R_V}
\end{figure} 

\subsection{Emission Signatures of the Moderately Scalarized Solution \texorpdfstring{$\mathbf{V}^{\,0}_{\,0.2}$}{Lg}}

We now consider the fifth configuration in the sequence, $\mathbf{V}^{\,0}_{\,0.2}$, corresponding to a moderately scalarized configuration. According to Table \ref{tab_7}, the scalar field stores $68.16\%$ of the total mass and $85.08\%$ of the angular momentum, yielding a normalized charge of $q \simeq 0.850778$. At this level of scalarization the disk morphology becomes increasingly Kerr-like, with a single continuous prograde disk extending outward from the ISCO at $r^{\mathrm{ISCO}_{1}}_{+} \simeq 0.6328$ (Table \ref{tab:ProgradeTCOs2}).

The bolometric flux distribution exhibits a single prograde luminosity maximum, $F_{O}^{\,\max} \simeq 881.542\,\dot{M}\times10^{-5}$ at $r \simeq 0.669$, accompanied by a blueshift of $z \simeq -0.170$ (Table~\ref{tab_KerrBH_SH}). When compared with a Kerr black hole of identical spin ($|a/M| = 0.993504$, Table~\ref{tab_KerrBH}), the peak flux is reduced by about $16.21\%$ (Table~\ref{Comparison_Kerr_BH_SH}), indicating that, despite the slightly more compact emitting region, the prograde disk of $\mathbf{V}^{\,0}_{\,0.2}$ is less efficient in converting accretion power into observable radiation at this scalarization level. No inner prograde emitting component appears, confirming that the strong-field scalar effects responsible for the multi-ring morphology in the more highly scalarized solutions have essentially vanished.

The retrograde disk shows far more pronounced deviations from the Kerr black hole case. Its maximal luminosity, $F_{O}^{\,\max} \simeq 56.9128\,\dot{M}\times10^{-5}$ at $r \simeq 4.842$, exceeds the Kerr value by approximately $5991\%$ (Table~\ref{Comparison_Kerr_BH_SH}), corresponding to an enhancement by nearly a factor of sixty. This dramatic increase is primarily driven by the substantial inward displacement of the retrograde emitting region relative to the Kerr reference configuration, which leads to a significantly higher local emissivity and observed flux. This enhancement is directly reflected in the corresponding ray-traced images, where the retrograde component is visibly brighter than in the Kerr spacetime.

Nevertheless, the associated blueshift, $z \simeq -0.235$, remains close to the Kerr prediction, indicating that the large luminosity enhancement is not caused by stronger frequency shifts, but rather by the modified orbital structure of the retrograde circular geodesics in the scalarized spacetime. 

This behavior demonstrates that retrograde disk emission remains highly sensitive to the modified orbital structure of the scalarized spacetime even when the overall disk morphology becomes increasingly similar to that of the Kerr solution. Configuration $\mathbf{V}^{\,0}_{\,0.2}$ therefore illustrates that large luminosity deviations from the Kerr prediction may persist well into the moderately scalarized regime. This highlights the strong diagnostic potential of retrograde thin accretion disks for detecting deviations from the Kerr geometry in scalarized black hole spacetimes.

The ray-traced morphology of the prograde and retrograde thin accretion disks in the spacetime of $\mathbf{V}^{\,0}_{\,0.2}$, along with the shadow of the central black hole, is depicted in Figures \ref{fig:Disk_P_V} and \ref{fig:Disk_R_V}. Consistent with the flux analysis above, the retrograde disk displays a substantially intensified emission region, while the prograde component appears comparatively dimmer. This visual contrast provides a direct illustration of the nearly factor-of-sixty enhancement of the retrograde peak luminosity produced by the scalar-modified orbital structure. 

\subsection{Emission Signatures of the Weakest Scalarized Solution \texorpdfstring{$\mathbf{VI}^{\,0}_{\,0.3}$}{Lg}}

We now analyze the last and weakest scalarized configuration in the sequence, $\mathbf{VI}^{\,0}_{\,0.3}$. As shown in Table \ref{tab_7}, the scalar field contributes $25.73\%$ of the total mass and $64.78\%$ of the total angular momentum, resulting in a normalized charge of $q \simeq 0.647791$. At this minimal level of scalarization, the thin accretion disk structure becomes nearly indistinguishable from the Kerr case: a single prograde and a single retrograde emitting region remain, and no inner radiative components are present.

\begin{figure}[b!]
    \centering
    \includegraphics[width=0.985\textwidth]{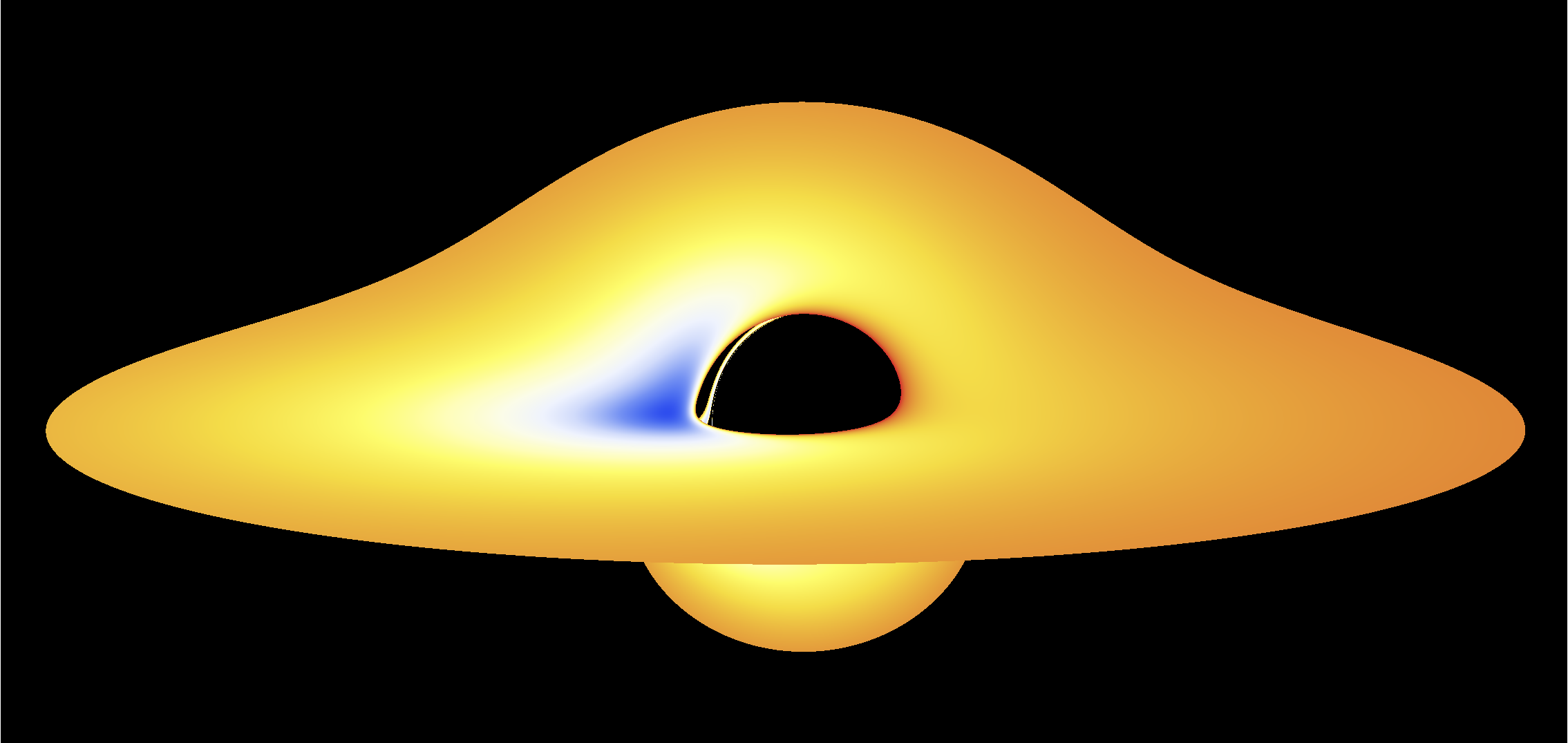}\\ \vspace{0.8mm}
    \includegraphics[width=0.985\textwidth]{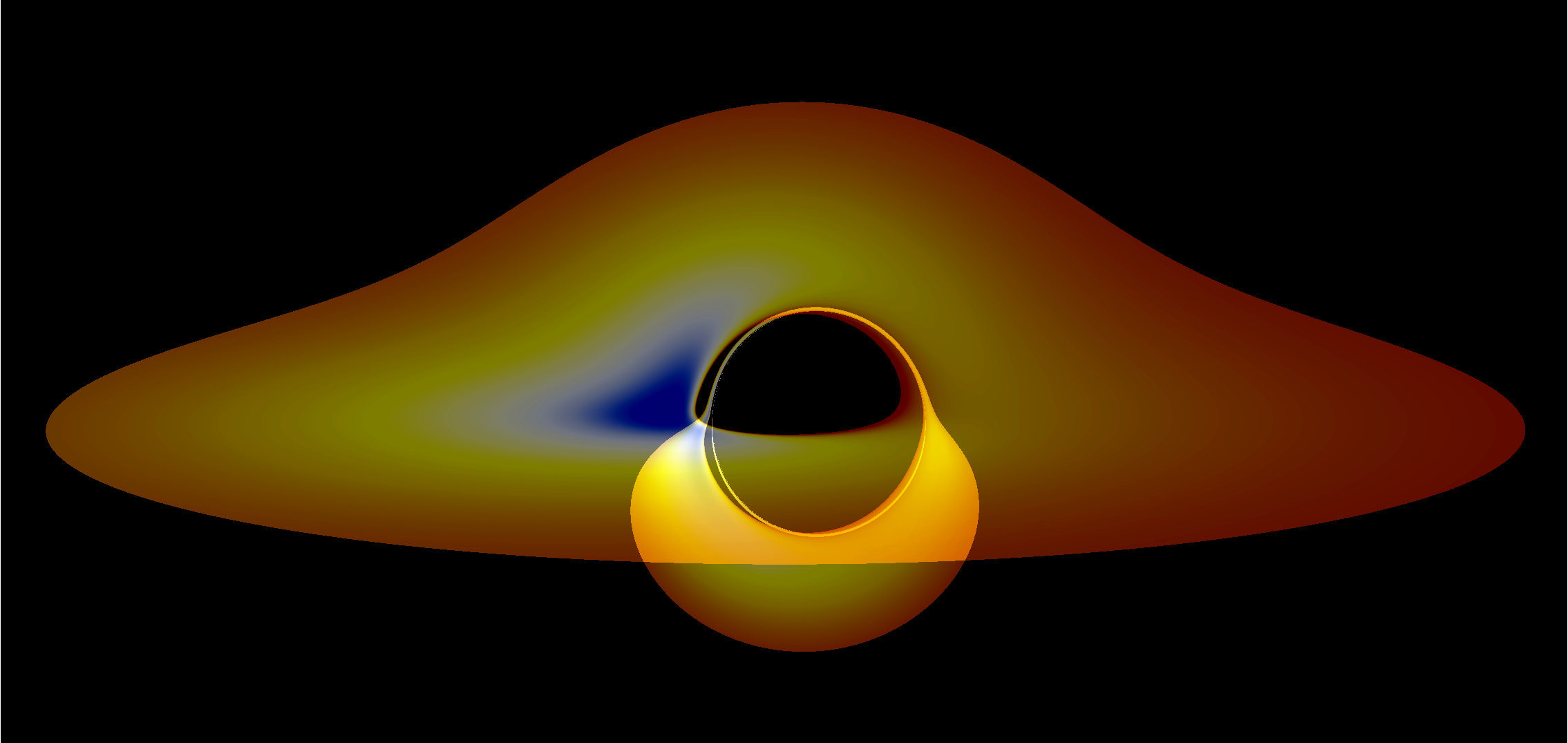}
   \caption{\small Images of a prograde geometrically thin accretion disk around a black hole with vanishing Gaussian curvature ($\kappa = 0$) and horizon radius $r_H = 0.3$, observed from a circumferential distance $\tilde r_{\text{obs}} = 200\,M$ at an inclination angle $\theta_{\text{obs}} = 80^\circ$. The images correspond to configuration \textbf{VI}$^{\,0}_{\,0.3}$ with parameters $\omega_s/\mu = 0.988000$, $M\mu = 0.320010$, and $q = 0.647791$. The color palette and panel rendering follow the same conventions as in Fig.~\ref{fig:Disk_P_OUT_I}, with the disk shown opaque in the upper panel and semi-transparent in the lower panel, revealing both direct and higher-order relativistic images. Additional parameters are listed in Table~\ref{tab_7} (Appendix A).}
    \label{fig:Disk_P_VI}
\end{figure} 

\begin{figure}[t!]
    \centering
    \includegraphics[width=0.985\textwidth]{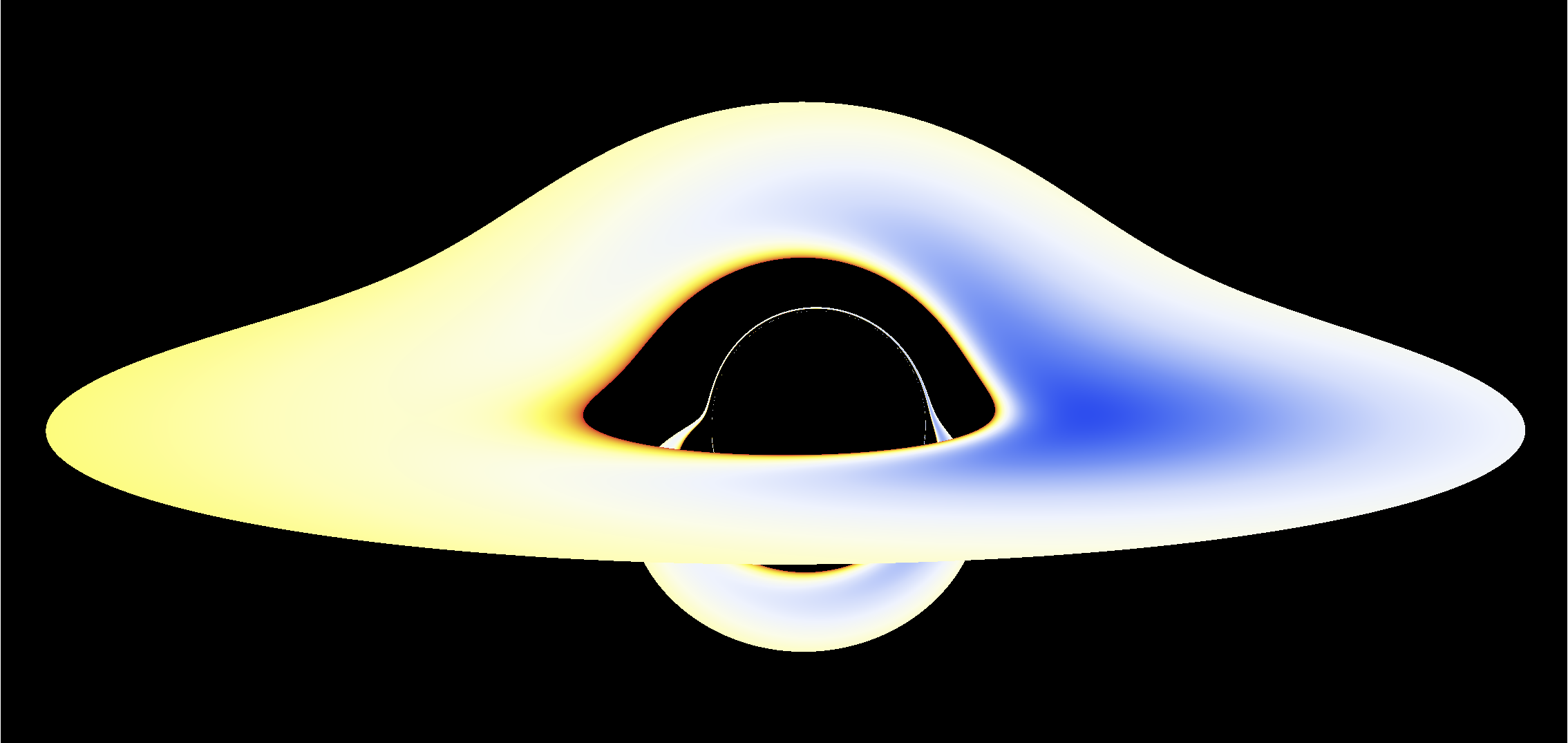}\\ \vspace{0.8mm}
    \includegraphics[width=0.985\textwidth]{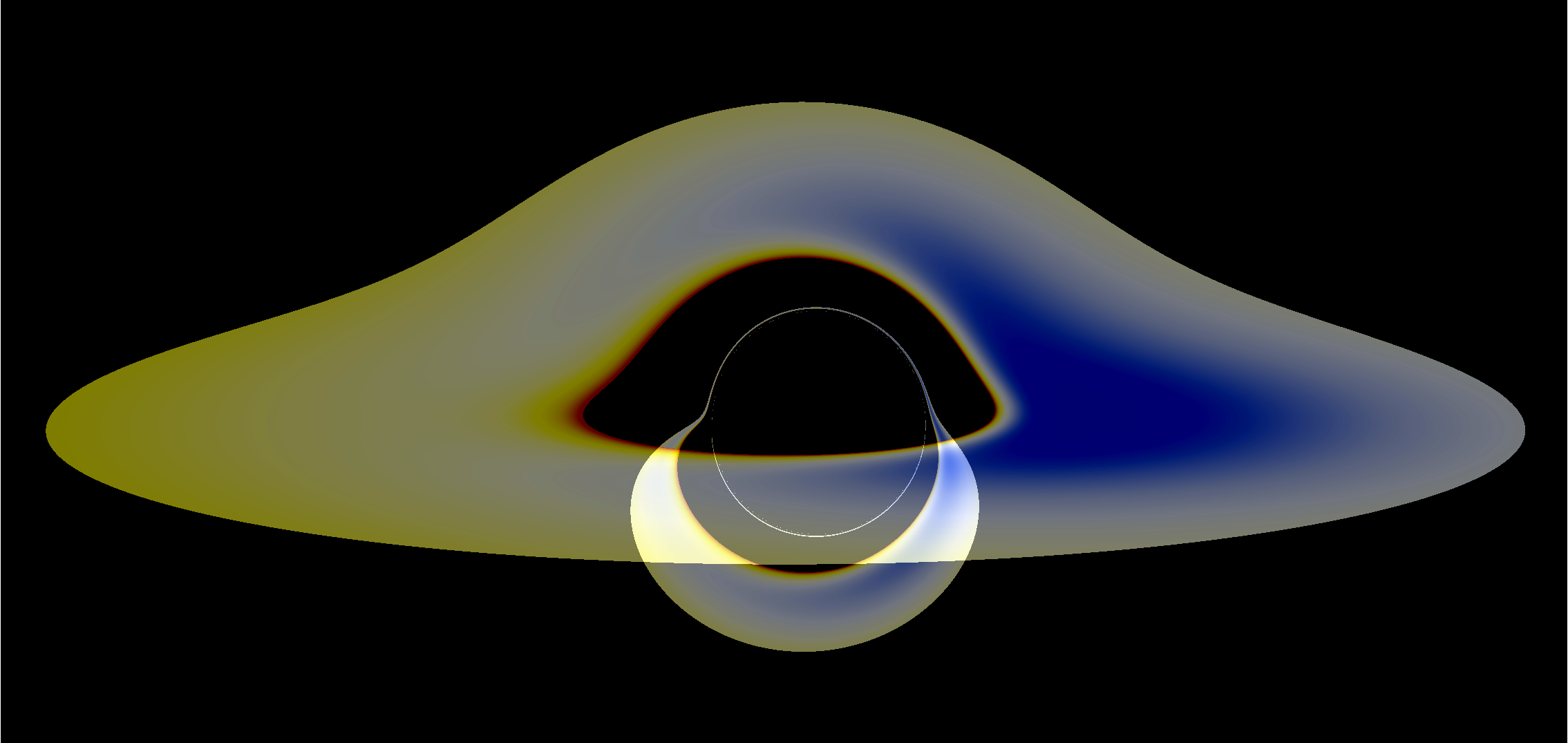}
   \caption{\small Images of a retrograde geometrically thin accretion disk around a black hole with vanishing Gaussian curvature ($\kappa = 0$) and horizon radius $r_H = 0.3$, observed from a circumferential distance $\tilde r_{\text{obs}} = 200\,M$ at an inclination angle $\theta_{\text{obs}} = 80^\circ$. The images correspond to configuration \textbf{VI}$^{\,0}_{\,0.3}$ with parameters $\omega_s/\mu = 0.988000$, $M\mu = 0.320010$, and $q = 0.647791$. The color palette and panel rendering follow the same conventions as in Fig.~\ref{fig:Disk_P_OUT_I}, with the disk shown opaque in the upper panel and semi-transparent in the lower panel, revealing both direct and higher-order relativistic images. Additional parameters are listed in Table~\ref{tab_7} (Appendix~A).}
    \label{fig:Disk_R_VI}
\end{figure}

The prograde accretion disk reaches a luminosity maximum of $F_{O}^{\,\max} \simeq 87.2932\,\dot{M}\times10^{-5}$ at $r \simeq 0.958$, accompanied by a modest blueshift of $z \simeq -0.280$ (Table \ref{tab_KerrBH_SH}). When compared with a Kerr black hole of the same spin ($|a/M| = 0.883338$, Table \ref{tab_KerrBH}), the peak flux is reduced by $25.68\%$ (Table \ref{Comparison_Kerr_BH_SH}). This reduction reflects the moderate modification of the circular-orbit structure induced by the scalar field, which slightly lowers the radiative efficiency of the innermost prograde region.

The retrograde disk attains a peak luminosity of $F_{O}^{\,\max} \simeq 2.10851\,\dot{M}\times10^{-5}$ at $r \simeq 2.955$, with a blueshift of $z \simeq -0.204$ (Table \ref{tab_KerrBH_SH}). Its flux is nearly twice that of the corresponding Kerr configuration, showing a $99.34\%$ increase (Table \ref{Comparison_Kerr_BH_SH}). This enhancement originates from a slight inward displacement of the retrograde emitting region in the weakly scalarized spacetime, which increases the local energy release and the observed flux. Nevertheless, the associated frequency shift remains almost identical to the Kerr prediction, indicating that photon propagation in the retrograde sector is only weakly affected by the scalar hair.

These results show that, although the disk morphology has largely recovered the Kerr-like structure, small but measurable deviations in the emitted flux can persist even in the weakly scalarized regime. In particular, the retrograde component remains slightly more luminous than in the Kerr spacetime, reflecting the residual modification of the underlying circular-orbit structure.

Figures \ref{fig:Disk_P_VI} and \ref{fig:Disk_R_VI} display the visual appearance of the prograde and retrograde thin accretion disks for $\mathbf{VI}^{\,0}_{\,0.3}$. The characteristic light bending and the resulting black hole shadow are clearly visible, in agreement with the shadow structures previously discussed for synchronized-hair geometries in Ref. \cite{Gyulchev2024}. 

Taken together, the sequence of configurations $\mathbf{I}^{\,0}_{\,0.01}$ -- $\mathbf{VI}^{\,0}_{\,0.3}$ reveals a clear systematic trend in the emission properties of thin accretion disks around scalarized Kerr black holes. In the strongly scalarized regime the modified circular-orbit structure gives rise to multiple emitting regions and large luminosity enhancements, particularly in the retrograde sector. As the scalar charge decreases, the disk morphology progressively approaches the Kerr configuration, and the additional radiative components disappear. Nevertheless, noticeable deviations in the emitted flux may persist even in moderately and weakly scalarized solutions, most prominently in the retrograde disk, which remains especially sensitive to the underlying orbital structure of the spacetime.

\section{Conclusion}

In this work, we have investigated the geodesic structure, light-ring configurations, and observational appearance of Kerr black holes endowed with synchronized scalar hair, with particular emphasis on the emission properties of geometrically thin Novikov--Thorne accretion disks. By combining an analysis of timelike circular orbits with backward ray tracing, we have systematically explored how the normalized scalar charge $q$ modifies orbital stability, light-ring structure, and the observable radiation from both prograde and retrograde disks.

Our results reveal that synchronized scalar hair induces a qualitatively rich strong-field structure that directly imprints on thin-disk emission. In the highly scalarized regime ($q \rightarrow 1$), the spacetime develops multiple circular-orbit regions associated with a layered system of light rings, including stable ones. While each ray-traced image corresponds to a single continuous disk component, the underlying orbital structure allows the formation of additional inner emitting regions. In this regime, both prograde and retrograde disks can exhibit substantial deviations from Kerr, including luminosity enhancements exceeding one order of magnitude in specific components. These effects arise primarily from the displacement of stable circular orbits toward the light-ring structure rather than from extreme frequency shifts alone.

A particularly robust feature of the scalarized solutions is the marked asymmetry between prograde and counter-rotating disks. Whereas prograde emission gradually approaches Kerr-like behavior as scalarization weakens, the retrograde sector remains highly sensitive to modifications of the strong-field geometry. In intermediate configurations, new inner retrograde radiative rings emerge, with peak luminosities enhanced by nearly a factor of twenty relative to Kerr. Even in the weakly scalarized limit, the retrograde flux can remain significantly elevated (approaching a factor of two in configuration $\mathbf{VI}^{\,0}_{\,0.3}$), despite frequency shifts that remain close to their Kerr values. This indicates that the dominant mechanism behind the luminosity enhancement is the modified orbital structure rather than altered photon propagation alone.

Across the sequence of solutions $\mathbf{I}^{\,0}_{\,0.01}$ -- $\mathbf{VI}^{\,0}_{\,0.3}$, a clear trend emerges. Strong scalarization generates additional emitting regions and large luminosity contrasts, particularly in the retrograde sector. As $q$ decreases, the disk morphology progressively simplifies and converges toward the Kerr paradigm, yet residual flux asymmetries persist. The transition between these regimes is continuous, indicating that scalar hair modifies the near-horizon orbital landscape in a systematic and potentially observable manner.

From an observational perspective, these findings suggest that high-resolution imaging of accretion flows may provide a sensitive probe of scalarized black hole geometries. Recent horizon-scale observations by the Event Horizon Telescope (EHT) collaboration have demonstrated the capability to resolve ring-like emission structures around supermassive black holes \cite{event2019first,event2022first}. Although long-term accretion tends to align disks in a prograde configuration through mechanisms such as the Bardeen--Petterson effect \cite{Bardeen1975,Scheuer1996}, counter-rotating flows may arise in chaotic accretion episodes, galaxy mergers, or tidal disruption events \cite{King2006Chaotic,Nixon2012Retrograde,Stone2019TDE}. Within the idealized Novikov--Thorne thin-disk framework adopted here, such configurations provide a useful setting for exploring the observational signatures of both prograde and retrograde accretion.

It is also important to emphasize that not all deviations from the Kerr geometry lead to qualitatively similar disk-emission signatures. For instance, the shadow and light-ring properties of Einstein-dilaton-Gauss-Bonnet black holes have been investigated in detail \cite{Cunha2017EdGB}. In those cases the deviations typically manifest as moderate shifts in shadow size or geodesic structure, without producing the restructuring of circular-orbit regions found here. The distinctive feature of synchronized scalar hair is the bifurcation of circular geodesic branches associated with the modified light-ring system, which directly reshapes the emissivity profile of thin accretion disks.

Overall, synchronized scalar hair leaves distinct and observable imprints on thin accretion flows. The interplay between light-ring structure, orbital stability, and disk emission provides a coherent framework in which deviations from the Kerr geometry may appear through characteristic strong-field observational signatures, with the retrograde sector emerging as a particularly sensitive diagnostic of scalarization. These results suggest that the observational properties of thin accretion disks may provide a complementary avenue for probing scalar-hair extensions of the Kerr paradigm in the strong-gravity regime.

\newpage

\begin{appendix}
\section{Physical quantities of selected solutions}\label{A1}

We provide a detailed description of the solutions used to create the black hole thin accretion disks. These solutions are taken from \cite{collodel2020rotating}, where sequences with fixed horizon radii were analyzed for zero Gaussian curvature $\kappa$.

\begin{table}[h!]
  \centering
  \footnotesize
    \setlength{\tabcolsep}{8.1pt} 
    \renewcommand{\arraystretch}{1.4} 
    \caption{\small Physical quantities of hairy black hole solutions, as shown in Fig. \ref{fig:M-Omega Space}, correspond to a zero Gaussian curvature of the target space, $\kappa=0$. Each configuration is labeled as \textbf{X}$^{\,0}_{\,v}$, where \textbf{X} denotes the configuration number, the superscript $0$ specifies the Gaussian curvature $\kappa$, and the subscript $v$ indicates the black hole horizon radius $r_{H}$.}   
    \label{tab_7}
    \vspace{1.5mm} 
  \begin{tabular}{lcccccccccc}
    \hline\hline
    Label & $\omega_{s}/\mu$ & $M\mu$ & $J\mu^2$ & $M_{BH}\mu$ & $J_{BH}\mu^2$ & $M_{\psi}\mu$ & $J_{\psi}\mu^2$ & $J_{\psi}/J$ & $J/M^2$ & $J_{BH}/M_{BH}^2$ \\
    \hline
    $\mathbf{I}^{\,0}_{\,0.01}$  & 0.6792 & 0.8820 & 0.7259 & 0.0035 & 0.0001 & 0.8785 & 0.7258 & 0.9999 & 0.9331 & 5.8772 \\
    $\mathbf{II}^{\,0}_{\,0.01}$   & 0.8353 & 0.6482 & 0.4068 & 0.0049 & 0.0011 & 0.6434 & 0.4057 & 0.9973 & 0.9680 & 46.444 \\    
    $\mathbf{III}^{\,0}_{\,0.05}$ & 0.7064 & 0.9082 & 0.8329 & 0.0252 & 0.0045 & 0.9415 & 0.8283 & 0.9945 & 0.8915 & 7.1310 \\
    $\mathbf{IV}^{\,0}_{\,0.1}$   & 0.7385 & 1.0004 & 0.8535 & 0.0870 & 0.0302 & 0.9134 & 0.8232 & 0.9645 & 0.8528 & 3.9982 \\
    $\mathbf{V}^{\,0}_{\,0.2}$    & 0.8955 & 0.8787 & 0.6790 & 0.2796 & 0.1012 & 0.5990 & 0.5776 & 0.8508 & 0.8793 & 1.2946 \\
    $\mathbf{VI}^{\,0}_{\,0.3}$   & 0.9880 & 0.3200 & 0.1260 & 0.2377 & 0.0444 & 0.0823 & 0.0816 & 0.6478 & 1.2304 & 0.7856 \\
    \hline\hline
  \end{tabular}
\end{table}

\end{appendix}

\section*{Acknowledgements}

This study is financed by the Bulgarian National Science Fund (NSF) under Grant KP-06-DV/8 within the funding programme ``VIHREN--2024''. D.D. acknowledges financial support from the Spanish Ministry of Science and Innovation through the Ram\'on y Cajal programme (grant RYC2023-042559-I), funded by MCIN/AEI/\href{https://doi.org/10.13039/501100011033}{10.13039/50110001\-1033}, from an Emmy Noether Research Group funded by the German Research Foundation (DFG) under Grant No.~DO~1771/1-1, and by the Spanish Agencia Estatal de Investigaci\'on (grant PID2024-159689NB-C21) funded by MICIU/AEI/10.13039/501100011033 and by FEDER / EU.

\bibliography{ref}

@article{event2019first,
  title={First M87 event horizon telescope results. IV. Imaging the central supermassive black hole},
  author={Event Horizon Telescope Collaboration and others},
  journal={arXiv preprint arXiv:1906.11241},
  year={2019}
}

@article{event2022first,
doi = {10.3847/2041-8213/ac6675},
url = {https://dx.doi.org/10.3847/2041-8213/ac6675},
year = {2022},
month = {may},
publisher = {The American Astronomical Society},
volume = {930},
number = {2},
pages = {L13},
author = {Event Horizon Telescope Collaboration et al.},
title = {First Sagittarius A* Event Horizon Telescope Results. II. EHT and Multiwavelength Observations, Data Processing, and Calibration},
journal = {The Astrophysical Journal Letters}
}

@article{collodel2020rotating,
  title={Rotating tensor-multiscalar black holes with two scalars},
  author={Collodel, Lucas G and Doneva, Daniela D and Yazadjiev, Stoytcho S},
  journal={Physical Review D},
  volume={102},
  number={8},
  pages={084032},
  year={2020},
  publisher={APS}
}

@article{hairysol1,
  title={Stationary scalar clouds around rotating black holes},
  author={Hod, Shahar},
  journal={Physical Review D},
  volume={86},
  number={10},
  pages={104026},
  year={2012},
  publisher={APS}
}

@article{herdeiro2014,
  title={Kerr black holes with scalar hair},
  author={Herdeiro, Carlos AR and Radu, Eugen},
  journal={Physical review letters},
  volume={112},
  number={22},
  pages={221101},
  year={2014},
  publisher={APS}
}

@article{herdeiro2015,
  title={Construction and physical properties of Kerr black holes with scalar hair},
  author={Herdeiro, Carlos and Radu, Eugen},
  journal={Classical and Quantum Gravity},
  volume={32},
  number={14},
  pages={144001},
  year={2015},
  publisher={IOP Publishing}
}

@article{horbatsch2015tensor,
  title={Tensor-multi-scalar theories: relativistic stars and 3+ 1 decomposition},
  author={Horbatsch, Michael and Silva, Hector O and Gerosa, Davide and Pani, Paolo and Berti, Emanuele and Gualtieri, Leonardo and Sperhake, Ulrich},
  journal={Classical and Quantum Gravity},
  volume={32},
  number={20},
  pages={204001},
  year={2015},
  publisher={IOP Publishing}
}

@article{doneva1,
  title={Dark compact objects in massive tensor-multi-scalar theories of gravity},
  author={Yazadjiev, Stoytcho S and Doneva, Daniela D},
  journal={Physical Review D},
  volume={99},
  number={8},
  pages={084011},
  year={2019},
  publisher={APS}
}

@article{doneva2,
  title={Rotating tensor-multiscalar solitons},
  author={Collodel, Lucas G and Doneva, Daniela D and Yazadjiev, Stoytcho S},
  journal={Physical Review D},
  volume={101},
  number={4},
  pages={044021},
  year={2020},
  publisher={APS}
}

@article{cunha2015shadows,
  title={Shadows of Kerr black holes with scalar hair},
  author={Cunha, Pedro VP and Herdeiro, Carlos AR and Radu, Eugen and R{\'u}narsson, Helgi F},
  journal={Physical review letters},
  volume={115},
  number={21},
  pages={211102},
  year={2015},
  publisher={APS}
}

@article{cunha2016shadows,
  title={Shadows of Kerr black holes with and without scalar hair},
  author={Cunha, Pedro VP and Herdeiro, Carlos AR and Radu, Eugen and Runarsson, Helgi F},
  journal={International Journal of Modern Physics D},
  volume={25},
  number={09},
  pages={1641021},
  year={2016},
  publisher={World Scientific}
}

@article{lora2022osiris,
  title={OSIRIS: A New Code for Ray Tracing Around Compact Objects},
  author={Lora-Clavijo, FD and Pimentel, OM and others},
  journal={arXiv preprint arXiv:2202.00086},
  year={2022}
}

@article{Damour:1992we,
    author = "Damour, Thibault and Esposito-Farese, Gilles",
    title = "{Tensor multiscalar theories of gravitation}",
    reportNumber = "IHES-P-91-93, CPT-91-PE-2542",
    doi = "10.1088/0264-9381/9/9/015",
    journal = "Class. Quant. Grav.",
    volume = "9",
    pages = "2093--2176",
    year = "1992"
}

@article{Gyulchev2024,
  title = {Shadows of rotating hairy Kerr black holes coupled to time periodic scalar fields with a nonflat target space},
  author = {Gyulchev, Galin N. and Roy, Ayush and Collodel, Lucas G. and Nedkova, Petya G. and Yazadjiev, Stoytcho S. and Doneva, Daniela D.},
  journal = {Phys. Rev. D},
  volume = {109},
  issue = {10},
  pages = {104051},
  numpages = {20},
  year = {2024},
  month = {May},
  publisher = {American Physical Society},
  doi = {10.1103/PhysRevD.109.104051},
  url = {https://link.aps.org/doi/10.1103/PhysRevD.109.104051}
}

@article{Delgado2022,
  title = {Equatorial timelike circular orbits around generic ultracompact objects},
  author = {Delgado, Jorge F. M. and Herdeiro, Carlos A. R. and Radu, Eugen},
  journal = {Phys. Rev. D},
  volume = {105},
  issue = {6},
  pages = {064026},
  numpages = {19},
  year = {2022},
  month = {Mar},
  publisher = {American Physical Society},
  doi = {10.1103/PhysRevD.105.064026},
  url = {https://link.aps.org/doi/10.1103/PhysRevD.105.064026}
}

@article{Collodel_2021,
doi = {10.3847/1538-4357/abe305},
url = {https://dx.doi.org/10.3847/1538-4357/abe305},
year = {2021},
month = {mar},
publisher = {The American Astronomical Society},
volume = {910},
number = {1},
pages = {52},
author = {Collodel, Lucas G. and Doneva, Daniela D. and Yazadjiev, Stoytcho S.},
title = {Circular Orbit Structure and Thin Accretion Disks around Kerr Black Holes with Scalar Hair},
journal = {The Astrophysical Journal}
}

@article{Pessah2012,
  author = {Pessah, Martin E. and Chan, Chi-Kwan},
  title = {On Angular Momentum Transport in Boundary Layers},
  journal = {The Astrophysical Journal},
  volume = {751},
  number = {1},
  pages = {48},
  year = {2012},
  doi = {10.1088/0004-637X/751/1/48},
  url = {https://arxiv.org/abs/1111.4219}
}

@article{Marshall2018,
  author = {Marshall, N. and Avara, M. J. and McKinney, J. C.},
  title = {Magnetically arrested disk state and the suppression of the magnetorotational instability},
  journal = {Monthly Notices of the Royal Astronomical Society},
  volume = {478},
  number = {2},
  pages = {1837--1852},
  year = {2018},
  doi = {10.1093/mnras/sty1175},
  url = {https://academic.oup.com/mnras/article/478/2/1837/4993537}
}

@incollection{Novikov1973,
  author    = {Novikov, I. D. and Thorne, K. S.},
  title     = {Astrophysics of black holes},
  booktitle = {Black Holes (Les Astres Occlus)},
  editor    = {C. DeWitt and B. DeWitt},
  publisher = {Gordon and Breach},
  address   = {New York},
  pages     = {343--450},
  year      = {1973}
}

@article{Page1974,
  author  = {Page, D. N. and Thorne, K. S.},
  title   = {Disk-Accretion onto a Black Hole. Time-Averaged Structure of Accretion Disk},
  journal = {Astrophysical Journal},
  volume  = {191},
  pages   = {499--506},
  year    = {1974},
  doi     = {10.1086/152990}
}

@article{Sengo_2024,
doi = {10.1088/1475-7516/2024/05/054},
url = {https://dx.doi.org/10.1088/1475-7516/2024/05/054},
year = {2024},
month = {may},
publisher = {IOP Publishing},
volume = {2024},
number = {05},
pages = {054},
author = {Sengo, Ivo and Cunha, Pedro V.P. and Herdeiro, Carlos A.R. and Radu, Eugen},
title = {The imitation game reloaded: effective shadows of dynamically robust spinning Proca stars},
journal = {Journal of Cosmology and Astroparticle Physics}
}

@article{Luminet_1979,
    author = "Luminet, J. -P.",
    title = "{Image of a spherical black hole with thin accretion disk}",
    journal = "Astron. Astrophys.",
    volume = "75",
    pages = "228--235",
    year = "1979"
}

@article{CunninghamBardeen1972,
  author  = {Cunningham, C. T. and Bardeen, J. M.},
  title   = {The optical appearance of a star orbiting an extreme Kerr black hole},
  journal = {Astrophys. J.},
  volume  = {173},
  pages   = {L137},
  year    = {1972},
  doi     = {10.1086/180927}
}

@article{Cunningham1975,
  author  = {Cunningham, C. T.},
  title   = {The effects of redshifts and focusing on the spectrum of an accretion disk around a Kerr black hole},
  journal = {Astrophys. J.},
  volume  = {202},
  pages   = {788--802},
  year    = {1975},
  doi     = {10.1086/153997}
}

@article{Fanton1997,
  author  = {Fanton, C. and Calvani, M. and de Felice, F. and Cadez, A.},
  title   = {Detecting accretion disks in active galactic nuclei},
  journal = {Publ. Astron. Soc. Pac.},
  volume  = {109},
  pages   = {706--718},
  year    = {1997},
  doi     = {10.1086/133948}
}

@article{GCandHairyKerr,
doi = {},
url = {},
year = {2026},
month = {apr},
publisher = {IOP Publishing},
volume = {},
number = {},
pages = {},
author = {Gyulchev, Galin N. and Deliyski, Valentin O. and Nedkova, Petya G. and Yazadjiev, Stoytcho S. and Doneva, Daniela D.},
title = {Gaussian Curvature Effects in Hairy Kerr Black Hole Imaging},
journal = {Journal of Physics: Conference Series}
}

@article{ThinDiskHorndeski2024,
  author  = {Zeng, X. and Wang, P. and Yang, C.},
  title   = {Thin accretion disk images of rotating hairy Horndeski black holes},
  journal = {Eur. Phys. J. C},
  volume  = {84},
  pages   = {113},
  year    = {2024},
  doi     = {10.1140/epjc/s10052-024-12513-4}
}

@article{BosonStarThinDisk2025,
  author  = {Li, Z. and Zhou, S. and Bambi, C.},
  title   = {Optical images of massive boson stars illuminated by thin accretion disks},
  journal = {Eur. Phys. J. C},
  volume  = {85},
  pages   = {102},
  year    = {2025},
  doi     = {10.1140/epjc/s10052-025-12702-1}
}

@article{WormholeDiskImaging2023,
  author  = {Hao, Chen-Hao and Su, Xin and Wang, Yong-Qiang},
  title   = {AdS Ellis wormholes with scalar field},
  journal = {Eur. Phys. J. C},
  volume  = {85},
  number  = {3},
  pages   = {348},
  year    = {2025},
  doi     = {10.1140/epjc/s10052-025-13937-0}
}

@article{Ishkaeva_2023,
  title = {Image of an accreting general {Ellis}--{Bronnikov} wormhole},
  author = {Ishkaeva, Valeria A. and Sushkov, Sergey V.},
  journal = {Phys. Rev. D},
  volume = {108},
  issue = {8},
  pages = {084054},
  numpages = {10},
  year = {2023},
  month = {Oct},
  publisher = {American Physical Society},
  doi = {10.1103/PhysRevD.108.084054},
  url = {https://link.aps.org/doi/10.1103/PhysRevD.108.084054}
}

@article{herdeiro2016a,
  title = {Kerr black holes with self-interacting scalar hair: Hairier but not heavier},
  author = {Herdeiro, Carlos A. R. and Radu, Eugen and R\'unarsson, Helgi},
  journal = {Phys. Rev. D},
  volume = {92},
  issue = {8},
  pages = {084059},
  numpages = {8},
  year = {2015},
  month = {Oct},
  publisher = {American Physical Society},
  doi = {10.1103/PhysRevD.92.084059},
  url = {https://link.aps.org/doi/10.1103/PhysRevD.92.084059}
}

@article{brihaye2016,
title = {Inside black holes with synchronized hair},
journal = {Physics Letters B},
volume = {760},
pages = {279-287},
year = {2016},
issn = {0370-2693},
doi = {https://doi.org/10.1016/j.physletb.2016.06.078},
url = {https://www.sciencedirect.com/science/article/pii/S0370269316303392},
author = {Yves Brihaye and Carlos Herdeiro and Eugen Radu}
}

@article{delgado2016,
title = {Kerr-Newman black holes with scalar hair},
journal = {Physics Letters B},
volume = {761},
pages = {234-241},
year = {2016},
issn = {0370-2693},
doi = {https://doi.org/10.1016/j.physletb.2016.08.032},
url = {https://www.sciencedirect.com/science/article/pii/S037026931630452X},
author = {Jorge F.M. Delgado and Carlos A.R. Herdeiro and Eugen Radu and Helgi R\'unarsson}
}

@article{herdeiro2025,
    author = {Nicoules, Jordan and Ferreira, Jos{\'e} and Herdeiro, Carlos A. R. and Radu, Eugen and Zilh{\~a}o, Miguel},
    title = {Splitting the Gravitational Atom: Instabilities of Black Holes with Synchronized/Resonant Hair},
    eprint = {2509.20450},
    archivePrefix = {arXiv},
    primaryClass = {gr-qc},
    month = {9},
    year = {2025},
    journal={arXiv preprint arXiv:2509.20450}
}

@article{Cunha2017EdGB,
  author = {Cunha, P. V. P. and Herdeiro, C. A. R. and Radu, E.},
  title = {Shadows of Einstein-dilaton-Gauss-Bonnet black holes},
  journal = {Phys. Rev. D},
  volume = {96},
  pages = {024039},
  year = {2017},
  doi = {10.1103/PhysRevD.96.024039}
}

@article{Bardeen1975,
  author = {Bardeen, J. M. and Petterson, J. A.},
  title = {The Lense--Thirring Effect and Accretion Disks around Kerr Black Holes},
  journal = {Astrophysical Journal Letters},
  volume = {195},
  pages = {L65--L67},
  year = {1975},
  doi = {10.1086/181711}
}

@article{Scheuer1996,
  author = {Scheuer, P. A. G. and Feiler, R.},
  title = {On the Bardeen--Petterson effect},
  journal = {Monthly Notices of the Royal Astronomical Society},
  volume = {282},
  pages = {291--296},
  year = {1996},
  doi = {10.1093/mnras/282.1.291}
}

@article{King2006Chaotic,
  author = {King, A. R. and Pringle, J. E.},
  title = {Growing supermassive black holes by chaotic accretion},
  journal = {Monthly Notices of the Royal Astronomical Society},
  volume = {373},
  pages = {L90--L92},
  year = {2006},
  doi = {10.1111/j.1745-3933.2006.00241.x}
}

@article{Nixon2012Retrograde,
  author = {Nixon, C. and King, A. and Price, D.},
  title = {Tearing up the disc: misaligned accretion on to a binary},
  journal = {Monthly Notices of the Royal Astronomical Society},
  volume = {422},
  pages = {2547--2554},
  year = {2012},
  doi = {10.1111/j.1365-2966.2012.20819.x}
}

@article{Stone2019TDE,
  author = {Stone, N. C. and Vasiliev, E. and Kesden, M. and Rossi, E.},
  title = {Rates of stellar tidal disruption as probes of the supermassive black hole mass function},
  journal = {Space Science Reviews},
  volume = {215},
  pages = {26},
  year = {2019},
  doi = {10.1007/s11214-019-0591-0}
}

\end{document}